\documentclass[useAMS,usenatbib,usegraphicx]{mn2e}
\usepackage{color}
\definecolor{AliceBlue}{rgb}{0.94,0.97,1.00}
\definecolor{AntiqueWhite1}{rgb}{1.00,0.94,0.86}
\definecolor{AntiqueWhite2}{rgb}{0.93,0.87,0.80}
\definecolor{AntiqueWhite3}{rgb}{0.80,0.75,0.69}
\definecolor{AntiqueWhite4}{rgb}{0.55,0.51,0.47}
\definecolor{AntiqueWhite}{rgb}{0.98,0.92,0.84}
\definecolor{BlanchedAlmond}{rgb}{1.00,0.92,0.80}
\definecolor{BlueViolet}{rgb}{0.54,0.17,0.89}
\definecolor{CadetBlue1}{rgb}{0.60,0.96,1.00}
\definecolor{CadetBlue2}{rgb}{0.56,0.90,0.93}
\definecolor{CadetBlue3}{rgb}{0.48,0.77,0.80}
\definecolor{CadetBlue4}{rgb}{0.33,0.53,0.55}
\definecolor{CadetBlue}{rgb}{0.37,0.62,0.63}
\definecolor{CornflowerBlue}{rgb}{0.39,0.58,0.93}
\definecolor{DarkBlue}{rgb}{0.00,0.00,0.55}
\definecolor{DarkCyan}{rgb}{0.00,0.55,0.55}
\definecolor{DarkGoldenrod1}{rgb}{1.00,0.73,0.06}
\definecolor{DarkGoldenrod2}{rgb}{0.93,0.68,0.05}
\definecolor{DarkGoldenrod3}{rgb}{0.80,0.58,0.05}
\definecolor{DarkGoldenrod4}{rgb}{0.55,0.40,0.03}
\definecolor{DarkGoldenrod}{rgb}{0.72,0.53,0.04}
\definecolor{DarkGray}{rgb}{0.66,0.66,0.66}
\definecolor{DarkGreen}{rgb}{0.00,0.39,0.00}
\definecolor{DarkGrey}{rgb}{0.66,0.66,0.66}
\definecolor{DarkKhaki}{rgb}{0.74,0.72,0.42}
\definecolor{DarkMagenta}{rgb}{0.55,0.00,0.55}
\definecolor{DarkOliveGreen1}{rgb}{0.79,1.00,0.44}
\definecolor{DarkOliveGreen2}{rgb}{0.74,0.93,0.41}
\definecolor{DarkOliveGreen3}{rgb}{0.64,0.80,0.35}
\definecolor{DarkOliveGreen4}{rgb}{0.43,0.55,0.24}
\definecolor{DarkOliveGreen}{rgb}{0.33,0.42,0.18}
\definecolor{DarkOrange1}{rgb}{1.00,0.50,0.00}
\definecolor{DarkOrange2}{rgb}{0.93,0.46,0.00}
\definecolor{DarkOrange3}{rgb}{0.80,0.40,0.00}
\definecolor{DarkOrange4}{rgb}{0.55,0.27,0.00}
\definecolor{DarkOrange}{rgb}{1.00,0.55,0.00}
\definecolor{DarkOrchid1}{rgb}{0.75,0.24,1.00}
\definecolor{DarkOrchid2}{rgb}{0.70,0.23,0.93}
\definecolor{DarkOrchid3}{rgb}{0.60,0.20,0.80}
\definecolor{DarkOrchid4}{rgb}{0.41,0.13,0.55}
\definecolor{DarkOrchid}{rgb}{0.60,0.20,0.80}
\definecolor{DarkRed}{rgb}{0.55,0.00,0.00}
\definecolor{DarkSalmon}{rgb}{0.91,0.59,0.48}
\definecolor{DarkSeaGreen1}{rgb}{0.76,1.00,0.76}
\definecolor{DarkSeaGreen2}{rgb}{0.71,0.93,0.71}
\definecolor{DarkSeaGreen3}{rgb}{0.61,0.80,0.61}
\definecolor{DarkSeaGreen4}{rgb}{0.41,0.55,0.41}
\definecolor{DarkSeaGreen}{rgb}{0.56,0.74,0.56}
\definecolor{DarkSlateBlue}{rgb}{0.28,0.24,0.55}
\definecolor{DarkSlateGray1}{rgb}{0.59,1.00,1.00}
\definecolor{DarkSlateGray2}{rgb}{0.55,0.93,0.93}
\definecolor{DarkSlateGray3}{rgb}{0.47,0.80,0.80}
\definecolor{DarkSlateGray4}{rgb}{0.32,0.55,0.55}
\definecolor{DarkSlateGray}{rgb}{0.18,0.31,0.31}
\definecolor{DarkSlateGrey}{rgb}{0.18,0.31,0.31}
\definecolor{DarkTurquoise}{rgb}{0.00,0.81,0.82}
\definecolor{DarkViolet}{rgb}{0.58,0.00,0.83}
\definecolor{DeepPink1}{rgb}{1.00,0.08,0.58}
\definecolor{DeepPink2}{rgb}{0.93,0.07,0.54}
\definecolor{DeepPink3}{rgb}{0.80,0.06,0.46}
\definecolor{DeepPink4}{rgb}{0.55,0.04,0.31}
\definecolor{DeepPink}{rgb}{1.00,0.08,0.58}
\definecolor{DeepSkyBlue1}{rgb}{0.00,0.75,1.00}
\definecolor{DeepSkyBlue2}{rgb}{0.00,0.70,0.93}
\definecolor{DeepSkyBlue3}{rgb}{0.00,0.60,0.80}
\definecolor{DeepSkyBlue4}{rgb}{0.00,0.41,0.55}
\definecolor{DeepSkyBlue}{rgb}{0.00,0.75,1.00}
\definecolor{DimGray}{rgb}{0.41,0.41,0.41}
\definecolor{DimGrey}{rgb}{0.41,0.41,0.41}
\definecolor{DodgerBlue1}{rgb}{0.12,0.56,1.00}
\definecolor{DodgerBlue2}{rgb}{0.11,0.53,0.93}
\definecolor{DodgerBlue3}{rgb}{0.09,0.45,0.80}
\definecolor{DodgerBlue4}{rgb}{0.06,0.31,0.55}
\definecolor{DodgerBlue}{rgb}{0.12,0.56,1.00}
\definecolor{FloralWhite}{rgb}{1.00,0.98,0.94}
\definecolor{ForestGreen}{rgb}{0.13,0.55,0.13}
\definecolor{GhostWhite}{rgb}{0.97,0.97,1.00}
\definecolor{GreenYellow}{rgb}{0.68,1.00,0.18}
\definecolor{HotPink1}{rgb}{1.00,0.43,0.71}
\definecolor{HotPink2}{rgb}{0.93,0.42,0.65}
\definecolor{HotPink3}{rgb}{0.80,0.38,0.56}
\definecolor{HotPink4}{rgb}{0.55,0.23,0.38}
\definecolor{HotPink}{rgb}{1.00,0.41,0.71}
\definecolor{IndianRed1}{rgb}{1.00,0.42,0.42}
\definecolor{IndianRed2}{rgb}{0.93,0.39,0.39}
\definecolor{IndianRed3}{rgb}{0.80,0.33,0.33}
\definecolor{IndianRed4}{rgb}{0.55,0.23,0.23}
\definecolor{IndianRed}{rgb}{0.80,0.36,0.36}
\definecolor{LavenderBlush1}{rgb}{1.00,0.94,0.96}
\definecolor{LavenderBlush2}{rgb}{0.93,0.88,0.90}
\definecolor{LavenderBlush3}{rgb}{0.80,0.76,0.77}
\definecolor{LavenderBlush4}{rgb}{0.55,0.51,0.53}
\definecolor{LavenderBlush}{rgb}{1.00,0.94,0.96}
\definecolor{LawnGreen}{rgb}{0.49,0.99,0.00}
\definecolor{LemonChiffon1}{rgb}{1.00,0.98,0.80}
\definecolor{LemonChiffon2}{rgb}{0.93,0.91,0.75}
\definecolor{LemonChiffon3}{rgb}{0.80,0.79,0.65}
\definecolor{LemonChiffon4}{rgb}{0.55,0.54,0.44}
\definecolor{LemonChiffon}{rgb}{1.00,0.98,0.80}
\definecolor{LightBlue1}{rgb}{0.75,0.94,1.00}
\definecolor{LightBlue2}{rgb}{0.70,0.87,0.93}
\definecolor{LightBlue3}{rgb}{0.60,0.75,0.80}
\definecolor{LightBlue4}{rgb}{0.41,0.51,0.55}
\definecolor{LightBlue}{rgb}{0.68,0.85,0.90}
\definecolor{LightCoral}{rgb}{0.94,0.50,0.50}
\definecolor{LightCyan1}{rgb}{0.88,1.00,1.00}
\definecolor{LightCyan2}{rgb}{0.82,0.93,0.93}
\definecolor{LightCyan3}{rgb}{0.71,0.80,0.80}
\definecolor{LightCyan4}{rgb}{0.48,0.55,0.55}
\definecolor{LightCyan}{rgb}{0.88,1.00,1.00}
\definecolor{LightGoldenrod1}{rgb}{1.00,0.93,0.55}
\definecolor{LightGoldenrod2}{rgb}{0.93,0.86,0.51}
\definecolor{LightGoldenrod3}{rgb}{0.80,0.75,0.44}
\definecolor{LightGoldenrod4}{rgb}{0.55,0.51,0.30}
\definecolor{LightGoldenrodYellow}{rgb}{0.98,0.98,0.82}
\definecolor{LightGoldenrod}{rgb}{0.93,0.87,0.51}
\definecolor{LightGray}{rgb}{0.83,0.83,0.83}
\definecolor{LightGreen}{rgb}{0.56,0.93,0.56}
\definecolor{LightGrey}{rgb}{0.83,0.83,0.83}
\definecolor{LightPink1}{rgb}{1.00,0.68,0.73}
\definecolor{LightPink2}{rgb}{0.93,0.64,0.68}
\definecolor{LightPink3}{rgb}{0.80,0.55,0.58}
\definecolor{LightPink4}{rgb}{0.55,0.37,0.40}
\definecolor{LightPink}{rgb}{1.00,0.71,0.76}
\definecolor{LightSalmon1}{rgb}{1.00,0.63,0.48}
\definecolor{LightSalmon2}{rgb}{0.93,0.58,0.45}
\definecolor{LightSalmon3}{rgb}{0.80,0.51,0.38}
\definecolor{LightSalmon4}{rgb}{0.55,0.34,0.26}
\definecolor{LightSalmon}{rgb}{1.00,0.63,0.48}
\definecolor{LightSeaGreen}{rgb}{0.13,0.70,0.67}
\definecolor{LightSkyBlue1}{rgb}{0.69,0.89,1.00}
\definecolor{LightSkyBlue2}{rgb}{0.64,0.83,0.93}
\definecolor{LightSkyBlue3}{rgb}{0.55,0.71,0.80}
\definecolor{LightSkyBlue4}{rgb}{0.38,0.48,0.55}
\definecolor{LightSkyBlue}{rgb}{0.53,0.81,0.98}
\definecolor{LightSlateBlue}{rgb}{0.52,0.44,1.00}
\definecolor{LightSlateGray}{rgb}{0.47,0.53,0.60}
\definecolor{LightSlateGrey}{rgb}{0.47,0.53,0.60}
\definecolor{LightSteelBlue1}{rgb}{0.79,0.88,1.00}
\definecolor{LightSteelBlue2}{rgb}{0.74,0.82,0.93}
\definecolor{LightSteelBlue3}{rgb}{0.64,0.71,0.80}
\definecolor{LightSteelBlue4}{rgb}{0.43,0.48,0.55}
\definecolor{LightSteelBlue}{rgb}{0.69,0.77,0.87}
\definecolor{LightYellow1}{rgb}{1.00,1.00,0.88}
\definecolor{LightYellow2}{rgb}{0.93,0.93,0.82}
\definecolor{LightYellow3}{rgb}{0.80,0.80,0.71}
\definecolor{LightYellow4}{rgb}{0.55,0.55,0.48}
\definecolor{LightYellow}{rgb}{1.00,1.00,0.88}
\definecolor{LimeGreen}{rgb}{0.20,0.80,0.20}
\definecolor{MediumAquamarine}{rgb}{0.40,0.80,0.67}
\definecolor{MediumBlue}{rgb}{0.00,0.00,0.80}
\definecolor{MediumOrchid1}{rgb}{0.88,0.40,1.00}
\definecolor{MediumOrchid2}{rgb}{0.82,0.37,0.93}
\definecolor{MediumOrchid3}{rgb}{0.71,0.32,0.80}
\definecolor{MediumOrchid4}{rgb}{0.48,0.22,0.55}
\definecolor{MediumOrchid}{rgb}{0.73,0.33,0.83}
\definecolor{MediumPurple1}{rgb}{0.67,0.51,1.00}
\definecolor{MediumPurple2}{rgb}{0.62,0.47,0.93}
\definecolor{MediumPurple3}{rgb}{0.54,0.41,0.80}
\definecolor{MediumPurple4}{rgb}{0.36,0.28,0.55}
\definecolor{MediumPurple}{rgb}{0.58,0.44,0.86}
\definecolor{MediumSeaGreen}{rgb}{0.24,0.70,0.44}
\definecolor{MediumSlateBlue}{rgb}{0.48,0.41,0.93}
\definecolor{MediumSpringGreen}{rgb}{0.00,0.98,0.60}
\definecolor{MediumTurquoise}{rgb}{0.28,0.82,0.80}
\definecolor{MediumVioletRed}{rgb}{0.78,0.08,0.52}
\definecolor{MidnightBlue}{rgb}{0.10,0.10,0.44}
\definecolor{MintCream}{rgb}{0.96,1.00,0.98}
\definecolor{MistyRose1}{rgb}{1.00,0.89,0.88}
\definecolor{MistyRose2}{rgb}{0.93,0.84,0.82}
\definecolor{MistyRose3}{rgb}{0.80,0.72,0.71}
\definecolor{MistyRose4}{rgb}{0.55,0.49,0.48}
\definecolor{MistyRose}{rgb}{1.00,0.89,0.88}
\definecolor{NavajoWhite1}{rgb}{1.00,0.87,0.68}
\definecolor{NavajoWhite2}{rgb}{0.93,0.81,0.63}
\definecolor{NavajoWhite3}{rgb}{0.80,0.70,0.55}
\definecolor{NavajoWhite4}{rgb}{0.55,0.47,0.37}
\definecolor{NavajoWhite}{rgb}{1.00,0.87,0.68}
\definecolor{NavyBlue}{rgb}{0.00,0.00,0.50}
\definecolor{OldLace}{rgb}{0.99,0.96,0.90}
\definecolor{OliveDrab1}{rgb}{0.75,1.00,0.24}
\definecolor{OliveDrab2}{rgb}{0.70,0.93,0.23}
\definecolor{OliveDrab3}{rgb}{0.60,0.80,0.20}
\definecolor{OliveDrab4}{rgb}{0.41,0.55,0.13}
\definecolor{OliveDrab}{rgb}{0.42,0.56,0.14}
\definecolor{OrangeRed1}{rgb}{1.00,0.27,0.00}
\definecolor{OrangeRed2}{rgb}{0.93,0.25,0.00}
\definecolor{OrangeRed3}{rgb}{0.80,0.22,0.00}
\definecolor{OrangeRed4}{rgb}{0.55,0.15,0.00}
\definecolor{OrangeRed}{rgb}{1.00,0.27,0.00}
\definecolor{PaleGoldenrod}{rgb}{0.93,0.91,0.67}
\definecolor{PaleGreen1}{rgb}{0.60,1.00,0.60}
\definecolor{PaleGreen2}{rgb}{0.56,0.93,0.56}
\definecolor{PaleGreen3}{rgb}{0.49,0.80,0.49}
\definecolor{PaleGreen4}{rgb}{0.33,0.55,0.33}
\definecolor{PaleGreen}{rgb}{0.60,0.98,0.60}
\definecolor{PaleTurquoise1}{rgb}{0.73,1.00,1.00}
\definecolor{PaleTurquoise2}{rgb}{0.68,0.93,0.93}
\definecolor{PaleTurquoise3}{rgb}{0.59,0.80,0.80}
\definecolor{PaleTurquoise4}{rgb}{0.40,0.55,0.55}
\definecolor{PaleTurquoise}{rgb}{0.69,0.93,0.93}
\definecolor{PaleVioletRed1}{rgb}{1.00,0.51,0.67}
\definecolor{PaleVioletRed2}{rgb}{0.93,0.47,0.62}
\definecolor{PaleVioletRed3}{rgb}{0.80,0.41,0.54}
\definecolor{PaleVioletRed4}{rgb}{0.55,0.28,0.36}
\definecolor{PaleVioletRed}{rgb}{0.86,0.44,0.58}
\definecolor{PapayaWhip}{rgb}{1.00,0.94,0.84}
\definecolor{PeachPuff1}{rgb}{1.00,0.85,0.73}
\definecolor{PeachPuff2}{rgb}{0.93,0.80,0.68}
\definecolor{PeachPuff3}{rgb}{0.80,0.69,0.58}
\definecolor{PeachPuff4}{rgb}{0.55,0.47,0.40}
\definecolor{PeachPuff}{rgb}{1.00,0.85,0.73}
\definecolor{PowderBlue}{rgb}{0.69,0.88,0.90}
\definecolor{RosyBrown1}{rgb}{1.00,0.76,0.76}
\definecolor{RosyBrown2}{rgb}{0.93,0.71,0.71}
\definecolor{RosyBrown3}{rgb}{0.80,0.61,0.61}
\definecolor{RosyBrown4}{rgb}{0.55,0.41,0.41}
\definecolor{RosyBrown}{rgb}{0.74,0.56,0.56}
\definecolor{RoyalBlue1}{rgb}{0.28,0.46,1.00}
\definecolor{RoyalBlue2}{rgb}{0.26,0.43,0.93}
\definecolor{RoyalBlue3}{rgb}{0.23,0.37,0.80}
\definecolor{RoyalBlue4}{rgb}{0.15,0.25,0.55}
\definecolor{RoyalBlue}{rgb}{0.25,0.41,0.88}
\definecolor{SaddleBrown}{rgb}{0.55,0.27,0.07}
\definecolor{SandyBrown}{rgb}{0.96,0.64,0.38}
\definecolor{SeaGreen1}{rgb}{0.33,1.00,0.62}
\definecolor{SeaGreen2}{rgb}{0.31,0.93,0.58}
\definecolor{SeaGreen3}{rgb}{0.26,0.80,0.50}
\definecolor{SeaGreen4}{rgb}{0.18,0.55,0.34}
\definecolor{SeaGreen}{rgb}{0.18,0.55,0.34}
\definecolor{SkyBlue1}{rgb}{0.53,0.81,1.00}
\definecolor{SkyBlue2}{rgb}{0.49,0.75,0.93}
\definecolor{SkyBlue3}{rgb}{0.42,0.65,0.80}
\definecolor{SkyBlue4}{rgb}{0.29,0.44,0.55}
\definecolor{SkyBlue}{rgb}{0.53,0.81,0.92}
\definecolor{SlateBlue1}{rgb}{0.51,0.44,1.00}
\definecolor{SlateBlue2}{rgb}{0.48,0.40,0.93}
\definecolor{SlateBlue3}{rgb}{0.41,0.35,0.80}
\definecolor{SlateBlue4}{rgb}{0.28,0.24,0.55}
\definecolor{SlateBlue}{rgb}{0.42,0.35,0.80}
\definecolor{SlateGray1}{rgb}{0.78,0.89,1.00}
\definecolor{SlateGray2}{rgb}{0.73,0.83,0.93}
\definecolor{SlateGray3}{rgb}{0.62,0.71,0.80}
\definecolor{SlateGray4}{rgb}{0.42,0.48,0.55}
\definecolor{SlateGray}{rgb}{0.44,0.50,0.56}
\definecolor{SlateGrey}{rgb}{0.44,0.50,0.56}
\definecolor{SpringGreen1}{rgb}{0.00,1.00,0.50}
\definecolor{SpringGreen2}{rgb}{0.00,0.93,0.46}
\definecolor{SpringGreen3}{rgb}{0.00,0.80,0.40}
\definecolor{SpringGreen4}{rgb}{0.00,0.55,0.27}
\definecolor{SpringGreen}{rgb}{0.00,1.00,0.50}
\definecolor{SteelBlue1}{rgb}{0.39,0.72,1.00}
\definecolor{SteelBlue2}{rgb}{0.36,0.67,0.93}
\definecolor{SteelBlue3}{rgb}{0.31,0.58,0.80}
\definecolor{SteelBlue4}{rgb}{0.21,0.39,0.55}
\definecolor{SteelBlue}{rgb}{0.27,0.51,0.71}
\definecolor{VioletRed1}{rgb}{1.00,0.24,0.59}
\definecolor{VioletRed2}{rgb}{0.93,0.23,0.55}
\definecolor{VioletRed3}{rgb}{0.80,0.20,0.47}
\definecolor{VioletRed4}{rgb}{0.55,0.13,0.32}
\definecolor{VioletRed}{rgb}{0.82,0.13,0.56}
\definecolor{WhiteSmoke}{rgb}{0.96,0.96,0.96}
\definecolor{YellowGreen}{rgb}{0.60,0.80,0.20}
\definecolor{aliceblue}{rgb}{0.94,0.97,1.00}
\definecolor{antiquewhite}{rgb}{0.98,0.92,0.84}
\definecolor{aquamarine1}{rgb}{0.50,1.00,0.83}
\definecolor{aquamarine2}{rgb}{0.46,0.93,0.78}
\definecolor{aquamarine3}{rgb}{0.40,0.80,0.67}
\definecolor{aquamarine4}{rgb}{0.27,0.55,0.45}
\definecolor{aquamarine}{rgb}{0.50,1.00,0.83}
\definecolor{azure1}{rgb}{0.94,1.00,1.00}
\definecolor{azure2}{rgb}{0.88,0.93,0.93}
\definecolor{azure3}{rgb}{0.76,0.80,0.80}
\definecolor{azure4}{rgb}{0.51,0.55,0.55}
\definecolor{azure}{rgb}{0.94,1.00,1.00}
\definecolor{beige}{rgb}{0.96,0.96,0.86}
\definecolor{bisque1}{rgb}{1.00,0.89,0.77}
\definecolor{bisque2}{rgb}{0.93,0.84,0.72}
\definecolor{bisque3}{rgb}{0.80,0.72,0.62}
\definecolor{bisque4}{rgb}{0.55,0.49,0.42}
\definecolor{bisque}{rgb}{1.00,0.89,0.77}
\definecolor{black}{rgb}{0.00,0.00,0.00}
\definecolor{blanchedalmond}{rgb}{1.00,0.92,0.80}
\definecolor{blue1}{rgb}{0.00,0.00,1.00}
\definecolor{blue2}{rgb}{0.00,0.00,0.93}
\definecolor{blue3}{rgb}{0.00,0.00,0.80}
\definecolor{blue4}{rgb}{0.00,0.00,0.55}
\definecolor{blueviolet}{rgb}{0.54,0.17,0.89}
\definecolor{blue}{rgb}{0.00,0.00,1.00}
\definecolor{brown1}{rgb}{1.00,0.25,0.25}
\definecolor{brown2}{rgb}{0.93,0.23,0.23}
\definecolor{brown3}{rgb}{0.80,0.20,0.20}
\definecolor{brown4}{rgb}{0.55,0.14,0.14}
\definecolor{brown}{rgb}{0.65,0.16,0.16}
\definecolor{burlywood1}{rgb}{1.00,0.83,0.61}
\definecolor{burlywood2}{rgb}{0.93,0.77,0.57}
\definecolor{burlywood3}{rgb}{0.80,0.67,0.49}
\definecolor{burlywood4}{rgb}{0.55,0.45,0.33}
\definecolor{burlywood}{rgb}{0.87,0.72,0.53}
\definecolor{cadetblue}{rgb}{0.37,0.62,0.63}
\definecolor{chartreuse1}{rgb}{0.50,1.00,0.00}
\definecolor{chartreuse2}{rgb}{0.46,0.93,0.00}
\definecolor{chartreuse3}{rgb}{0.40,0.80,0.00}
\definecolor{chartreuse4}{rgb}{0.27,0.55,0.00}
\definecolor{chartreuse}{rgb}{0.50,1.00,0.00}
\definecolor{chocolate1}{rgb}{1.00,0.50,0.14}
\definecolor{chocolate2}{rgb}{0.93,0.46,0.13}
\definecolor{chocolate3}{rgb}{0.80,0.40,0.11}
\definecolor{chocolate4}{rgb}{0.55,0.27,0.07}
\definecolor{chocolate}{rgb}{0.82,0.41,0.12}
\definecolor{coral1}{rgb}{1.00,0.45,0.34}
\definecolor{coral2}{rgb}{0.93,0.42,0.31}
\definecolor{coral3}{rgb}{0.80,0.36,0.27}
\definecolor{coral4}{rgb}{0.55,0.24,0.18}
\definecolor{coral}{rgb}{1.00,0.50,0.31}
\definecolor{cornflowerblue}{rgb}{0.39,0.58,0.93}
\definecolor{cornsilk1}{rgb}{1.00,0.97,0.86}
\definecolor{cornsilk2}{rgb}{0.93,0.91,0.80}
\definecolor{cornsilk3}{rgb}{0.80,0.78,0.69}
\definecolor{cornsilk4}{rgb}{0.55,0.53,0.47}
\definecolor{cornsilk}{rgb}{1.00,0.97,0.86}
\definecolor{cyan1}{rgb}{0.00,1.00,1.00}
\definecolor{cyan2}{rgb}{0.00,0.93,0.93}
\definecolor{cyan3}{rgb}{0.00,0.80,0.80}
\definecolor{cyan4}{rgb}{0.00,0.55,0.55}
\definecolor{cyan}{rgb}{0.00,1.00,1.00}
\definecolor{darkblue}{rgb}{0.00,0.00,0.55}
\definecolor{darkcyan}{rgb}{0.00,0.55,0.55}
\definecolor{darkgoldenrod}{rgb}{0.72,0.53,0.04}
\definecolor{darkgray}{rgb}{0.66,0.66,0.66}
\definecolor{darkgreen}{rgb}{0.00,0.39,0.00}
\definecolor{darkgrey}{rgb}{0.66,0.66,0.66}
\definecolor{darkkhaki}{rgb}{0.74,0.72,0.42}
\definecolor{darkmagenta}{rgb}{0.55,0.00,0.55}
\definecolor{darkolive}{rgb}{0.33,0.42,0.18}
\definecolor{darkorange}{rgb}{1.00,0.55,0.00}
\definecolor{darkorchid}{rgb}{0.60,0.20,0.80}
\definecolor{darkred}{rgb}{0.55,0.00,0.00}
\definecolor{darksalmon}{rgb}{0.91,0.59,0.48}
\definecolor{darksea}{rgb}{0.56,0.74,0.56}
\definecolor{darkslate}{rgb}{0.18,0.31,0.31}
\definecolor{darkslate}{rgb}{0.18,0.31,0.31}
\definecolor{darkslate}{rgb}{0.28,0.24,0.55}
\definecolor{darkturquoise}{rgb}{0.00,0.81,0.82}
\definecolor{darkviolet}{rgb}{0.58,0.00,0.83}
\definecolor{deeppink}{rgb}{1.00,0.08,0.58}
\definecolor{deepsky}{rgb}{0.00,0.75,1.00}
\definecolor{dimgray}{rgb}{0.41,0.41,0.41}
\definecolor{dimgrey}{rgb}{0.41,0.41,0.41}
\definecolor{dodgerblue}{rgb}{0.12,0.56,1.00}
\definecolor{firebrick1}{rgb}{1.00,0.19,0.19}
\definecolor{firebrick2}{rgb}{0.93,0.17,0.17}
\definecolor{firebrick3}{rgb}{0.80,0.15,0.15}
\definecolor{firebrick4}{rgb}{0.55,0.10,0.10}
\definecolor{firebrick}{rgb}{0.70,0.13,0.13}
\definecolor{floralwhite}{rgb}{1.00,0.98,0.94}
\definecolor{forestgreen}{rgb}{0.13,0.55,0.13}
\definecolor{gainsboro}{rgb}{0.86,0.86,0.86}
\definecolor{ghostwhite}{rgb}{0.97,0.97,1.00}
\definecolor{gold1}{rgb}{1.00,0.84,0.00}
\definecolor{gold2}{rgb}{0.93,0.79,0.00}
\definecolor{gold3}{rgb}{0.80,0.68,0.00}
\definecolor{gold4}{rgb}{0.55,0.46,0.00}
\definecolor{goldenrod1}{rgb}{1.00,0.76,0.15}
\definecolor{goldenrod2}{rgb}{0.93,0.71,0.13}
\definecolor{goldenrod3}{rgb}{0.80,0.61,0.11}
\definecolor{goldenrod4}{rgb}{0.55,0.41,0.08}
\definecolor{goldenrod}{rgb}{0.85,0.65,0.13}
\definecolor{gold}{rgb}{1.00,0.84,0.00}
\definecolor{gray0}{rgb}{0.00,0.00,0.00}
\definecolor{gray100}{rgb}{1.00,1.00,1.00}
\definecolor{gray10}{rgb}{0.10,0.10,0.10}
\definecolor{gray11}{rgb}{0.11,0.11,0.11}
\definecolor{gray12}{rgb}{0.12,0.12,0.12}
\definecolor{gray13}{rgb}{0.13,0.13,0.13}
\definecolor{gray14}{rgb}{0.14,0.14,0.14}
\definecolor{gray15}{rgb}{0.15,0.15,0.15}
\definecolor{gray16}{rgb}{0.16,0.16,0.16}
\definecolor{gray17}{rgb}{0.17,0.17,0.17}
\definecolor{gray18}{rgb}{0.18,0.18,0.18}
\definecolor{gray19}{rgb}{0.19,0.19,0.19}
\definecolor{gray1}{rgb}{0.01,0.01,0.01}
\definecolor{gray20}{rgb}{0.20,0.20,0.20}
\definecolor{gray21}{rgb}{0.21,0.21,0.21}
\definecolor{gray22}{rgb}{0.22,0.22,0.22}
\definecolor{gray23}{rgb}{0.23,0.23,0.23}
\definecolor{gray24}{rgb}{0.24,0.24,0.24}
\definecolor{gray25}{rgb}{0.25,0.25,0.25}
\definecolor{gray26}{rgb}{0.26,0.26,0.26}
\definecolor{gray27}{rgb}{0.27,0.27,0.27}
\definecolor{gray28}{rgb}{0.28,0.28,0.28}
\definecolor{gray29}{rgb}{0.29,0.29,0.29}
\definecolor{gray2}{rgb}{0.02,0.02,0.02}
\definecolor{gray30}{rgb}{0.30,0.30,0.30}
\definecolor{gray31}{rgb}{0.31,0.31,0.31}
\definecolor{gray32}{rgb}{0.32,0.32,0.32}
\definecolor{gray33}{rgb}{0.33,0.33,0.33}
\definecolor{gray34}{rgb}{0.34,0.34,0.34}
\definecolor{gray35}{rgb}{0.35,0.35,0.35}
\definecolor{gray36}{rgb}{0.36,0.36,0.36}
\definecolor{gray37}{rgb}{0.37,0.37,0.37}
\definecolor{gray38}{rgb}{0.38,0.38,0.38}
\definecolor{gray39}{rgb}{0.39,0.39,0.39}
\definecolor{gray3}{rgb}{0.03,0.03,0.03}
\definecolor{gray40}{rgb}{0.40,0.40,0.40}
\definecolor{gray41}{rgb}{0.41,0.41,0.41}
\definecolor{gray42}{rgb}{0.42,0.42,0.42}
\definecolor{gray43}{rgb}{0.43,0.43,0.43}
\definecolor{gray44}{rgb}{0.44,0.44,0.44}
\definecolor{gray45}{rgb}{0.45,0.45,0.45}
\definecolor{gray46}{rgb}{0.46,0.46,0.46}
\definecolor{gray47}{rgb}{0.47,0.47,0.47}
\definecolor{gray48}{rgb}{0.48,0.48,0.48}
\definecolor{gray49}{rgb}{0.49,0.49,0.49}
\definecolor{gray4}{rgb}{0.04,0.04,0.04}
\definecolor{gray50}{rgb}{0.50,0.50,0.50}
\definecolor{gray51}{rgb}{0.51,0.51,0.51}
\definecolor{gray52}{rgb}{0.52,0.52,0.52}
\definecolor{gray53}{rgb}{0.53,0.53,0.53}
\definecolor{gray54}{rgb}{0.54,0.54,0.54}
\definecolor{gray55}{rgb}{0.55,0.55,0.55}
\definecolor{gray56}{rgb}{0.56,0.56,0.56}
\definecolor{gray57}{rgb}{0.57,0.57,0.57}
\definecolor{gray58}{rgb}{0.58,0.58,0.58}
\definecolor{gray59}{rgb}{0.59,0.59,0.59}
\definecolor{gray5}{rgb}{0.05,0.05,0.05}
\definecolor{gray60}{rgb}{0.60,0.60,0.60}
\definecolor{gray61}{rgb}{0.61,0.61,0.61}
\definecolor{gray62}{rgb}{0.62,0.62,0.62}
\definecolor{gray63}{rgb}{0.63,0.63,0.63}
\definecolor{gray64}{rgb}{0.64,0.64,0.64}
\definecolor{gray65}{rgb}{0.65,0.65,0.65}
\definecolor{gray66}{rgb}{0.66,0.66,0.66}
\definecolor{gray67}{rgb}{0.67,0.67,0.67}
\definecolor{gray68}{rgb}{0.68,0.68,0.68}
\definecolor{gray69}{rgb}{0.69,0.69,0.69}
\definecolor{gray6}{rgb}{0.06,0.06,0.06}
\definecolor{gray70}{rgb}{0.70,0.70,0.70}
\definecolor{gray71}{rgb}{0.71,0.71,0.71}
\definecolor{gray72}{rgb}{0.72,0.72,0.72}
\definecolor{gray73}{rgb}{0.73,0.73,0.73}
\definecolor{gray74}{rgb}{0.74,0.74,0.74}
\definecolor{gray75}{rgb}{0.75,0.75,0.75}
\definecolor{gray76}{rgb}{0.76,0.76,0.76}
\definecolor{gray77}{rgb}{0.77,0.77,0.77}
\definecolor{gray78}{rgb}{0.78,0.78,0.78}
\definecolor{gray79}{rgb}{0.79,0.79,0.79}
\definecolor{gray7}{rgb}{0.07,0.07,0.07}
\definecolor{gray80}{rgb}{0.80,0.80,0.80}
\definecolor{gray81}{rgb}{0.81,0.81,0.81}
\definecolor{gray82}{rgb}{0.82,0.82,0.82}
\definecolor{gray83}{rgb}{0.83,0.83,0.83}
\definecolor{gray84}{rgb}{0.84,0.84,0.84}
\definecolor{gray85}{rgb}{0.85,0.85,0.85}
\definecolor{gray86}{rgb}{0.86,0.86,0.86}
\definecolor{gray87}{rgb}{0.87,0.87,0.87}
\definecolor{gray88}{rgb}{0.88,0.88,0.88}
\definecolor{gray89}{rgb}{0.89,0.89,0.89}
\definecolor{gray8}{rgb}{0.08,0.08,0.08}
\definecolor{gray90}{rgb}{0.90,0.90,0.90}
\definecolor{gray91}{rgb}{0.91,0.91,0.91}
\definecolor{gray92}{rgb}{0.92,0.92,0.92}
\definecolor{gray93}{rgb}{0.93,0.93,0.93}
\definecolor{gray94}{rgb}{0.94,0.94,0.94}
\definecolor{gray95}{rgb}{0.95,0.95,0.95}
\definecolor{gray96}{rgb}{0.96,0.96,0.96}
\definecolor{gray97}{rgb}{0.97,0.97,0.97}
\definecolor{gray98}{rgb}{0.98,0.98,0.98}
\definecolor{gray99}{rgb}{0.99,0.99,0.99}
\definecolor{gray9}{rgb}{0.09,0.09,0.09}
\definecolor{gray}{rgb}{0.75,0.75,0.75}
\definecolor{green1}{rgb}{0.00,1.00,0.00}
\definecolor{green2}{rgb}{0.00,0.93,0.00}
\definecolor{green3}{rgb}{0.00,0.80,0.00}
\definecolor{green4}{rgb}{0.00,0.55,0.00}
\definecolor{greenyellow}{rgb}{0.68,1.00,0.18}
\definecolor{green}{rgb}{0.00,1.00,0.00}
\definecolor{grey0}{rgb}{0.00,0.00,0.00}
\definecolor{grey100}{rgb}{1.00,1.00,1.00}
\definecolor{grey10}{rgb}{0.10,0.10,0.10}
\definecolor{grey11}{rgb}{0.11,0.11,0.11}
\definecolor{grey12}{rgb}{0.12,0.12,0.12}
\definecolor{grey13}{rgb}{0.13,0.13,0.13}
\definecolor{grey14}{rgb}{0.14,0.14,0.14}
\definecolor{grey15}{rgb}{0.15,0.15,0.15}
\definecolor{grey16}{rgb}{0.16,0.16,0.16}
\definecolor{grey17}{rgb}{0.17,0.17,0.17}
\definecolor{grey18}{rgb}{0.18,0.18,0.18}
\definecolor{grey19}{rgb}{0.19,0.19,0.19}
\definecolor{grey1}{rgb}{0.01,0.01,0.01}
\definecolor{grey20}{rgb}{0.20,0.20,0.20}
\definecolor{grey21}{rgb}{0.21,0.21,0.21}
\definecolor{grey22}{rgb}{0.22,0.22,0.22}
\definecolor{grey23}{rgb}{0.23,0.23,0.23}
\definecolor{grey24}{rgb}{0.24,0.24,0.24}
\definecolor{grey25}{rgb}{0.25,0.25,0.25}
\definecolor{grey26}{rgb}{0.26,0.26,0.26}
\definecolor{grey27}{rgb}{0.27,0.27,0.27}
\definecolor{grey28}{rgb}{0.28,0.28,0.28}
\definecolor{grey29}{rgb}{0.29,0.29,0.29}
\definecolor{grey2}{rgb}{0.02,0.02,0.02}
\definecolor{grey30}{rgb}{0.30,0.30,0.30}
\definecolor{grey31}{rgb}{0.31,0.31,0.31}
\definecolor{grey32}{rgb}{0.32,0.32,0.32}
\definecolor{grey33}{rgb}{0.33,0.33,0.33}
\definecolor{grey34}{rgb}{0.34,0.34,0.34}
\definecolor{grey35}{rgb}{0.35,0.35,0.35}
\definecolor{grey36}{rgb}{0.36,0.36,0.36}
\definecolor{grey37}{rgb}{0.37,0.37,0.37}
\definecolor{grey38}{rgb}{0.38,0.38,0.38}
\definecolor{grey39}{rgb}{0.39,0.39,0.39}
\definecolor{grey3}{rgb}{0.03,0.03,0.03}
\definecolor{grey40}{rgb}{0.40,0.40,0.40}
\definecolor{grey41}{rgb}{0.41,0.41,0.41}
\definecolor{grey42}{rgb}{0.42,0.42,0.42}
\definecolor{grey43}{rgb}{0.43,0.43,0.43}
\definecolor{grey44}{rgb}{0.44,0.44,0.44}
\definecolor{grey45}{rgb}{0.45,0.45,0.45}
\definecolor{grey46}{rgb}{0.46,0.46,0.46}
\definecolor{grey47}{rgb}{0.47,0.47,0.47}
\definecolor{grey48}{rgb}{0.48,0.48,0.48}
\definecolor{grey49}{rgb}{0.49,0.49,0.49}
\definecolor{grey4}{rgb}{0.04,0.04,0.04}
\definecolor{grey50}{rgb}{0.50,0.50,0.50}
\definecolor{grey51}{rgb}{0.51,0.51,0.51}
\definecolor{grey52}{rgb}{0.52,0.52,0.52}
\definecolor{grey53}{rgb}{0.53,0.53,0.53}
\definecolor{grey54}{rgb}{0.54,0.54,0.54}
\definecolor{grey55}{rgb}{0.55,0.55,0.55}
\definecolor{grey56}{rgb}{0.56,0.56,0.56}
\definecolor{grey57}{rgb}{0.57,0.57,0.57}
\definecolor{grey58}{rgb}{0.58,0.58,0.58}
\definecolor{grey59}{rgb}{0.59,0.59,0.59}
\definecolor{grey5}{rgb}{0.05,0.05,0.05}
\definecolor{grey60}{rgb}{0.60,0.60,0.60}
\definecolor{grey61}{rgb}{0.61,0.61,0.61}
\definecolor{grey62}{rgb}{0.62,0.62,0.62}
\definecolor{grey63}{rgb}{0.63,0.63,0.63}
\definecolor{grey64}{rgb}{0.64,0.64,0.64}
\definecolor{grey65}{rgb}{0.65,0.65,0.65}
\definecolor{grey66}{rgb}{0.66,0.66,0.66}
\definecolor{grey67}{rgb}{0.67,0.67,0.67}
\definecolor{grey68}{rgb}{0.68,0.68,0.68}
\definecolor{grey69}{rgb}{0.69,0.69,0.69}
\definecolor{grey6}{rgb}{0.06,0.06,0.06}
\definecolor{grey70}{rgb}{0.70,0.70,0.70}
\definecolor{grey71}{rgb}{0.71,0.71,0.71}
\definecolor{grey72}{rgb}{0.72,0.72,0.72}
\definecolor{grey73}{rgb}{0.73,0.73,0.73}
\definecolor{grey74}{rgb}{0.74,0.74,0.74}
\definecolor{grey75}{rgb}{0.75,0.75,0.75}
\definecolor{grey76}{rgb}{0.76,0.76,0.76}
\definecolor{grey77}{rgb}{0.77,0.77,0.77}
\definecolor{grey78}{rgb}{0.78,0.78,0.78}
\definecolor{grey79}{rgb}{0.79,0.79,0.79}
\definecolor{grey7}{rgb}{0.07,0.07,0.07}
\definecolor{grey80}{rgb}{0.80,0.80,0.80}
\definecolor{grey81}{rgb}{0.81,0.81,0.81}
\definecolor{grey82}{rgb}{0.82,0.82,0.82}
\definecolor{grey83}{rgb}{0.83,0.83,0.83}
\definecolor{grey84}{rgb}{0.84,0.84,0.84}
\definecolor{grey85}{rgb}{0.85,0.85,0.85}
\definecolor{grey86}{rgb}{0.86,0.86,0.86}
\definecolor{grey87}{rgb}{0.87,0.87,0.87}
\definecolor{grey88}{rgb}{0.88,0.88,0.88}
\definecolor{grey89}{rgb}{0.89,0.89,0.89}
\definecolor{grey8}{rgb}{0.08,0.08,0.08}
\definecolor{grey90}{rgb}{0.90,0.90,0.90}
\definecolor{grey91}{rgb}{0.91,0.91,0.91}
\definecolor{grey92}{rgb}{0.92,0.92,0.92}
\definecolor{grey93}{rgb}{0.93,0.93,0.93}
\definecolor{grey94}{rgb}{0.94,0.94,0.94}
\definecolor{grey95}{rgb}{0.95,0.95,0.95}
\definecolor{grey96}{rgb}{0.96,0.96,0.96}
\definecolor{grey97}{rgb}{0.97,0.97,0.97}
\definecolor{grey98}{rgb}{0.98,0.98,0.98}
\definecolor{grey99}{rgb}{0.99,0.99,0.99}
\definecolor{grey9}{rgb}{0.09,0.09,0.09}
\definecolor{grey}{rgb}{0.75,0.75,0.75}
\definecolor{honeydew1}{rgb}{0.94,1.00,0.94}
\definecolor{honeydew2}{rgb}{0.88,0.93,0.88}
\definecolor{honeydew3}{rgb}{0.76,0.80,0.76}
\definecolor{honeydew4}{rgb}{0.51,0.55,0.51}
\definecolor{honeydew}{rgb}{0.94,1.00,0.94}
\definecolor{hotpink}{rgb}{1.00,0.41,0.71}
\definecolor{indianred}{rgb}{0.80,0.36,0.36}
\definecolor{ivory1}{rgb}{1.00,1.00,0.94}
\definecolor{ivory2}{rgb}{0.93,0.93,0.88}
\definecolor{ivory3}{rgb}{0.80,0.80,0.76}
\definecolor{ivory4}{rgb}{0.55,0.55,0.51}
\definecolor{ivory}{rgb}{1.00,1.00,0.94}
\definecolor{khaki1}{rgb}{1.00,0.96,0.56}
\definecolor{khaki2}{rgb}{0.93,0.90,0.52}
\definecolor{khaki3}{rgb}{0.80,0.78,0.45}
\definecolor{khaki4}{rgb}{0.55,0.53,0.31}
\definecolor{khaki}{rgb}{0.94,0.90,0.55}
\definecolor{lavenderblush}{rgb}{1.00,0.94,0.96}
\definecolor{lavender}{rgb}{0.90,0.90,0.98}
\definecolor{lawngreen}{rgb}{0.49,0.99,0.00}
\definecolor{lemonchiffon}{rgb}{1.00,0.98,0.80}
\definecolor{lightblue}{rgb}{0.68,0.85,0.90}
\definecolor{lightcoral}{rgb}{0.94,0.50,0.50}
\definecolor{lightcyan}{rgb}{0.88,1.00,1.00}
\definecolor{lightgoldenrod}{rgb}{0.93,0.87,0.51}
\definecolor{lightgoldenrod}{rgb}{0.98,0.98,0.82}
\definecolor{lightgray}{rgb}{0.83,0.83,0.83}
\definecolor{lightgreen}{rgb}{0.56,0.93,0.56}
\definecolor{lightgrey}{rgb}{0.83,0.83,0.83}
\definecolor{lightpink}{rgb}{1.00,0.71,0.76}
\definecolor{lightsalmon}{rgb}{1.00,0.63,0.48}
\definecolor{lightsea}{rgb}{0.13,0.70,0.67}
\definecolor{lightsky}{rgb}{0.53,0.81,0.98}
\definecolor{lightslate}{rgb}{0.47,0.53,0.60}
\definecolor{lightslate}{rgb}{0.47,0.53,0.60}
\definecolor{lightslate}{rgb}{0.52,0.44,1.00}
\definecolor{lightsteel}{rgb}{0.69,0.77,0.87}
\definecolor{lightyellow}{rgb}{1.00,1.00,0.88}
\definecolor{limegreen}{rgb}{0.20,0.80,0.20}
\definecolor{linen}{rgb}{0.98,0.94,0.90}
\definecolor{magenta1}{rgb}{1.00,0.00,1.00}
\definecolor{magenta2}{rgb}{0.93,0.00,0.93}
\definecolor{magenta3}{rgb}{0.80,0.00,0.80}
\definecolor{magenta4}{rgb}{0.55,0.00,0.55}
\definecolor{magenta}{rgb}{1.00,0.00,1.00}
\definecolor{maroon1}{rgb}{1.00,0.20,0.70}
\definecolor{maroon2}{rgb}{0.93,0.19,0.65}
\definecolor{maroon3}{rgb}{0.80,0.16,0.56}
\definecolor{maroon4}{rgb}{0.55,0.11,0.38}
\definecolor{maroon}{rgb}{0.69,0.19,0.38}
\definecolor{mediumaquamarine}{rgb}{0.40,0.80,0.67}
\definecolor{mediumblue}{rgb}{0.00,0.00,0.80}
\definecolor{mediumorchid}{rgb}{0.73,0.33,0.83}
\definecolor{mediumpurple}{rgb}{0.58,0.44,0.86}
\definecolor{mediumsea}{rgb}{0.24,0.70,0.44}
\definecolor{mediumslate}{rgb}{0.48,0.41,0.93}
\definecolor{mediumspring}{rgb}{0.00,0.98,0.60}
\definecolor{mediumturquoise}{rgb}{0.28,0.82,0.80}
\definecolor{mediumviolet}{rgb}{0.78,0.08,0.52}
\definecolor{midnightblue}{rgb}{0.10,0.10,0.44}
\definecolor{mintcream}{rgb}{0.96,1.00,0.98}
\definecolor{mistyrose}{rgb}{1.00,0.89,0.88}
\definecolor{moccasin}{rgb}{1.00,0.89,0.71}
\definecolor{navajowhite}{rgb}{1.00,0.87,0.68}
\definecolor{navyblue}{rgb}{0.00,0.00,0.50}
\definecolor{navy}{rgb}{0.00,0.00,0.50}
\definecolor{oldlace}{rgb}{0.99,0.96,0.90}
\definecolor{olivedrab}{rgb}{0.42,0.56,0.14}
\definecolor{orange1}{rgb}{1.00,0.65,0.00}
\definecolor{orange2}{rgb}{0.93,0.60,0.00}
\definecolor{orange3}{rgb}{0.80,0.52,0.00}
\definecolor{orange4}{rgb}{0.55,0.35,0.00}
\definecolor{orangered}{rgb}{1.00,0.27,0.00}
\definecolor{orange}{rgb}{1.00,0.65,0.00}
\definecolor{orchid1}{rgb}{1.00,0.51,0.98}
\definecolor{orchid2}{rgb}{0.93,0.48,0.91}
\definecolor{orchid3}{rgb}{0.80,0.41,0.79}
\definecolor{orchid4}{rgb}{0.55,0.28,0.54}
\definecolor{orchid}{rgb}{0.85,0.44,0.84}
\definecolor{palegoldenrod}{rgb}{0.93,0.91,0.67}
\definecolor{palegreen}{rgb}{0.60,0.98,0.60}
\definecolor{paleturquoise}{rgb}{0.69,0.93,0.93}
\definecolor{paleviolet}{rgb}{0.86,0.44,0.58}
\definecolor{papayawhip}{rgb}{1.00,0.94,0.84}
\definecolor{peachpuff}{rgb}{1.00,0.85,0.73}
\definecolor{peru}{rgb}{0.80,0.52,0.25}
\definecolor{pink1}{rgb}{1.00,0.71,0.77}
\definecolor{pink2}{rgb}{0.93,0.66,0.72}
\definecolor{pink3}{rgb}{0.80,0.57,0.62}
\definecolor{pink4}{rgb}{0.55,0.39,0.42}
\definecolor{pink}{rgb}{1.00,0.75,0.80}
\definecolor{plum1}{rgb}{1.00,0.73,1.00}
\definecolor{plum2}{rgb}{0.93,0.68,0.93}
\definecolor{plum3}{rgb}{0.80,0.59,0.80}
\definecolor{plum4}{rgb}{0.55,0.40,0.55}
\definecolor{plum}{rgb}{0.87,0.63,0.87}
\definecolor{powderblue}{rgb}{0.69,0.88,0.90}
\definecolor{purple1}{rgb}{0.61,0.19,1.00}
\definecolor{purple2}{rgb}{0.57,0.17,0.93}
\definecolor{purple3}{rgb}{0.49,0.15,0.80}
\definecolor{purple4}{rgb}{0.33,0.10,0.55}
\definecolor{purple}{rgb}{0.63,0.13,0.94}
\definecolor{red1}{rgb}{1.00,0.00,0.00}
\definecolor{red2}{rgb}{0.93,0.00,0.00}
\definecolor{red3}{rgb}{0.80,0.00,0.00}
\definecolor{red4}{rgb}{0.55,0.00,0.00}
\definecolor{red}{rgb}{1.00,0.00,0.00}
\definecolor{rosybrown}{rgb}{0.74,0.56,0.56}
\definecolor{royalblue}{rgb}{0.25,0.41,0.88}
\definecolor{saddlebrown}{rgb}{0.55,0.27,0.07}
\definecolor{salmon1}{rgb}{1.00,0.55,0.41}
\definecolor{salmon2}{rgb}{0.93,0.51,0.38}
\definecolor{salmon3}{rgb}{0.80,0.44,0.33}
\definecolor{salmon4}{rgb}{0.55,0.30,0.22}
\definecolor{salmon}{rgb}{0.98,0.50,0.45}
\definecolor{sandybrown}{rgb}{0.96,0.64,0.38}
\definecolor{seagreen}{rgb}{0.18,0.55,0.34}
\definecolor{seashell1}{rgb}{1.00,0.96,0.93}
\definecolor{seashell2}{rgb}{0.93,0.90,0.87}
\definecolor{seashell3}{rgb}{0.80,0.77,0.75}
\definecolor{seashell4}{rgb}{0.55,0.53,0.51}
\definecolor{seashell}{rgb}{1.00,0.96,0.93}
\definecolor{sienna1}{rgb}{1.00,0.51,0.28}
\definecolor{sienna2}{rgb}{0.93,0.47,0.26}
\definecolor{sienna3}{rgb}{0.80,0.41,0.22}
\definecolor{sienna4}{rgb}{0.55,0.28,0.15}
\definecolor{sienna}{rgb}{0.63,0.32,0.18}
\definecolor{skyblue}{rgb}{0.53,0.81,0.92}
\definecolor{slateblue}{rgb}{0.42,0.35,0.80}
\definecolor{slategray}{rgb}{0.44,0.50,0.56}
\definecolor{slategrey}{rgb}{0.44,0.50,0.56}
\definecolor{snow1}{rgb}{1.00,0.98,0.98}
\definecolor{snow2}{rgb}{0.93,0.91,0.91}
\definecolor{snow3}{rgb}{0.80,0.79,0.79}
\definecolor{snow4}{rgb}{0.55,0.54,0.54}
\definecolor{snow}{rgb}{1.00,0.98,0.98}
\definecolor{springgreen}{rgb}{0.00,1.00,0.50}
\definecolor{steelblue}{rgb}{0.27,0.51,0.71}
\definecolor{tan1}{rgb}{1.00,0.65,0.31}
\definecolor{tan2}{rgb}{0.93,0.60,0.29}
\definecolor{tan3}{rgb}{0.80,0.52,0.25}
\definecolor{tan4}{rgb}{0.55,0.35,0.17}
\definecolor{tan}{rgb}{0.82,0.71,0.55}
\definecolor{thistle1}{rgb}{1.00,0.88,1.00}
\definecolor{thistle2}{rgb}{0.93,0.82,0.93}
\definecolor{thistle3}{rgb}{0.80,0.71,0.80}
\definecolor{thistle4}{rgb}{0.55,0.48,0.55}
\definecolor{thistle}{rgb}{0.85,0.75,0.85}
\definecolor{tomato1}{rgb}{1.00,0.39,0.28}
\definecolor{tomato2}{rgb}{0.93,0.36,0.26}
\definecolor{tomato3}{rgb}{0.80,0.31,0.22}
\definecolor{tomato4}{rgb}{0.55,0.21,0.15}
\definecolor{tomato}{rgb}{1.00,0.39,0.28}
\definecolor{turquoise1}{rgb}{0.00,0.96,1.00}
\definecolor{turquoise2}{rgb}{0.00,0.90,0.93}
\definecolor{turquoise3}{rgb}{0.00,0.77,0.80}
\definecolor{turquoise4}{rgb}{0.00,0.53,0.55}
\definecolor{turquoise}{rgb}{0.25,0.88,0.82}
\definecolor{violetred}{rgb}{0.82,0.13,0.56}
\definecolor{violet}{rgb}{0.93,0.51,0.93}
\definecolor{wheat1}{rgb}{1.00,0.91,0.73}
\definecolor{wheat2}{rgb}{0.93,0.85,0.68}
\definecolor{wheat3}{rgb}{0.80,0.73,0.59}
\definecolor{wheat4}{rgb}{0.55,0.49,0.40}
\definecolor{wheat}{rgb}{0.96,0.87,0.70}
\definecolor{whitesmoke}{rgb}{0.96,0.96,0.96}
\definecolor{white}{rgb}{1.00,1.00,1.00}
\definecolor{yellow1}{rgb}{1.00,1.00,0.00}
\definecolor{yellow2}{rgb}{0.93,0.93,0.00}
\definecolor{yellow3}{rgb}{0.80,0.80,0.00}
\definecolor{yellow4}{rgb}{0.55,0.55,0.00}
\definecolor{yellowgreen}{rgb}{0.60,0.80,0.20}
\definecolor{yellow}{rgb}{1.00,1.00,0.00}

\usepackage{amsmath}
\usepackage{amssymb}
\usepackage{graphicx}
\usepackage{eso-pic}
\usepackage{amsmath,amsfonts,amssymb,aas_macros}

\usepackage[draft]{hyperref}

\pdfoutput=1
\usepackage[normalem]{ulem} 

\usepackage{capt-of}

\newcommand{\mice}{\textsc{mice}}
\newcommand{\halogen}{\textsc{halogen}}
\newcommand{\nb}{$N$-Body}

\newcommand{\mpch}{{\ifmmode{h^{-1}{\rm Mpc}}\else{$h^{-1}$Mpc}\fi}}
\newcommand{\hMpc}{{\ifmmode{h^{-1}{\rm Mpc}}\else{$h^{-1}$Mpc}\fi}}
\newcommand{\hkpc}{{\ifmmode{h^{-1}{\rm kpc}}\else{$h^{-1}$kpc}\fi}}
\newcommand{\degree}{^{\circ}}

\newcommand{\Eq}[1]{Equation~\ref{#1}}
\newcommand{\Fig}[1]{Figure~\ref{#1}}

\newcommand{\Sec}[1]{Section~\ref{#1}}
\newcommand{\Tab}[1]{Table~\ref{#1}}

\newcommand{\Cls}{DES-BAO-$\ell$-METHOD}
\newcommand{\main}{DES-BAO-MAIN}
\newcommand{\comov}{DES-BAO-$s_\perp$-METHOD}
\newcommand{\templates}{DES-BAO-$\theta$-METHOD}
\newcommand{\sample}{DES-BAO-SAMPLE}
\newcommand{\photoz}{DES-BAO-PHOTOZ}

\newlength{\figwidth}
\newlength{\figtable}
\newlength{\figtripple}
\newlength{\resplot}
\title[Y1-DES mocks]
{\vskip 0 in
\begin{minipage}{7.03 in}
\begin{flushright}
{\rm \small FERMILAB-PUB-17-587}
\end{flushright}
\end{minipage}\\
\begin{minipage}{7.03 in}
\vskip -0.2 in
\begin{flushright}
{\rm \small DES-2017-0292}
\end{flushright}
\end{minipage}\\
\vskip -0.0in
\begin{minipage}{7.03 in}
\vskip -0.4 in
\begin{flushright}
{\rm \small IFT-UAM/CSIC-17-124}
\end{flushright}
\end{minipage}\\
\vskip 0.0in
Dark Energy Survey Year 1 Results: galaxy mock catalogues for BAO
}

\begin{document}

\author[Avila et al.]{
\parbox{\textwidth}{
\Large
S.~Avila$^{1,2,3}$\thanks{e-mail: santiagoavilaperez@gmail.com}, 
M.~Crocce$^{4}$, 
A.~J.~Ross$^5$,
J.~Garc\'ia-Bellido$^{2,3}$, 
W.~J.~Percival$^{1}$, 
N. ~Banik$^{6,7,8,9}$,  
H.~Camacho$^{10,11}$, 
N.~Kokron$^{10,11,12,13}$, 
K. ~C.~Chan$^{4,14}$,
F. Andrade-Oliveira$^{11,15}$, 
R. Gomes$^{10,11}$,  
D. Gomes$^{10,11}$, 
M. Lima$^{10,11}$, 
R. Rosenfeld$^{11,12}$, 
A.~I.~Salvador$^{2,3}$
O.~Friedrich$^{16,17}$, 
F.~B.~Abdalla$^{18,19}$, 
J.~Annis$^{7}$, 
A.~Benoit-L{\'e}vy$^{18,20,21}$, 
E.~Bertin$^{20,21}$,
D.~Brooks$^{18}$, 
M.~Carrasco~Kind$^{22,23}$, 
J.~Carretero$^{24}$, 
F.~J.~Castander$^{4}$, 
C.~E.~Cunha$^{25}$,
L.~N.~da Costa$^{11,26}$, 
C.~Davis$^{25}$, 
J.~De~Vicente$^{27}$, 
P.~Doel$^{18}$, 
P.~Fosalba$^{4}$,
J.~Frieman$^{7,28}$, 
D.~W.~Gerdes$^{29,30}$, 
D.~Gruen$^{25,31}$, 
R.~A.~Gruendl$^{22,23}$, 
G.~Gutierrez$^{7}$,
W.~G.~Hartley$^{18,32}$, 
D.~Hollowood$^{33}$, 
K.~Honscheid$^{5,34}$, 
D.~J.~James$^{35}$, 
K.~Kuehn$^{36}$,
N.~Kuropatkin$^{7}$, 
R.~Miquel$^{24,37}$, 
A.~A.~Plazas$^{38}$, 
E.~Sanchez$^{27}$, 
V.~Scarpine$^{7}$, 
R.~Schindler$^{31}$, 
M.~Schubnell$^{30}$, 
I.~Sevilla-Noarbe$^{27}$,  
M.~Smith$^{39}$, 
F.~Sobreira$^{11,40}$,        
E.~Suchyta$^{41}$, 
M.~E.~C.~Swanson$^{23}$, 
G.~Tarle$^{30}$, 
D.~Thomas$^{1}$, 
A.~R.~Walker$^{42}$      
\begin{center} (The Dark Energy Survey Collaboration) \end{center}
}
\vspace{0.4cm}
\\
\parbox{\textwidth}{
\scriptsize
  $^{1}$ Institute of Cosmology \& Gravitation, Dennis Sciama Building, University of Portsmouth, Portsmouth, PO1 3FX, UK\\
  $^{2}$ Departamento de F\'isica Te\'orica, M\'odulo C-15, Facultad de Ciencias, Universidad Aut\'onoma de Madrid, 28049 Cantoblanco, Madrid, Spain\\
  $^{3}$ Instituto de F\'isica Te\'orica, UAM-CSIC, Universidad Autonoma de Madrid, 28049 Cantoblanco, Madrid, Spain\\
  $^{4}$ Institut de Ci\`encies de l'Espai, IEEC-CSIC, Campus UAB, Facultat de Ci\`encies, Torre C5 par-2, Barcelona 08193, Spain \\
  $^{5}$ Center for Cosmology and AstroParticle Physics, The Ohio State University, Columbus, OH 43210, USA \\
  $^{6}$ Department of Physics, University of Florida, Gainesville, Florida 32611, USA\\
  $^{7}$ Fermi National Accelerator Laboratory, Batavia, Illinois 60510, USA\\
  $^{8}$ GRAPPA, Institute of Theoretical Physics, University of Amsterdam, Science Park 904, 1090 GL Amsterdam \\
  $^{9}$ Lorentz Institute, Leiden University, Niels Bohrweg 2, Leiden, NL-2333 CA, The Netherlands \\
  $^{10}$ Departamento de F\'{\i}sica Matem\'atica, Instituto de F\'{\i}sica, Universidade de S\~ao Paulo,CP 66318, S\~ao Paulo, SP, 05314-970, Brazil\\
  $^{11}$ Laborat\'orio Interinstitucional de e-Astronomia, Rua General Jos\'e Cristino, 77, S\~ao Crist\'ov\~ao, Rio de Janeiro, RJ, 20921-400, Brazil \\
  $^{12}$ ICTP South American Institute for Fundamental Research \& Instituto de F\'isica Te\'orica, Universidade Estadual Paulista, S\~ao Paulo, Brazil \\
  $^{13}$ Department of Physics, Stanford University, 382 Via Pueblo Mall, Stanford, CA 94305, USA \\
  $^{14}$ School of Physics and Astronomy, Sun Yat-Sen University, Guangzhou 510275, China \\
  $^{15}$ Instituto de F\'isica Te\'orica, Universidade Estadual Paulista, S\~ao Paulo, Brazil \\
  $^{16}$ Max Planck Institute for Extraterrestrial Physics, Giessenbachstrasse, 85748 Garching, Germany\\
  $^{17}$ Universit\"ats-Sternwarte, Fakult\"at f\"ur Physik, Ludwig-Maximilians Universit\"at M\"unchen, Scheinerstr. 1, 81679 M\"unchen, Germany\\
  $^{18}$ Department of Physics \& Astronomy, University College London, Gower Street, London, WC1E 6BT, UK\\
  $^{19}$ Department of Physics and Electronics, Rhodes University, PO Box 94, Grahamstown, 6140, South Africa\\
  $^{20}$ CNRS, UMR 7095, Institut d'Astrophysique de Paris, F-75014, Paris, France\\
  $^{21}$ Sorbonne Universit\'es, UPMC Univ Paris 06, UMR 7095, Institut d'Astrophysique de Paris, F-75014, Paris, France\\
  $^{22}$ Department of Astronomy, University of Illinois, 1002 W. Green Street, Urbana, IL 61801, USA\\
  $^{23}$ National Center for Supercomputing Applications, 1205 West Clark St., Urbana, IL 61801, USA\\
  $^{24}$ Institut de F\'{\i}sica d'Altes Energies (IFAE), The Barcelona Institute of Science and Technology, Campus UAB, 08193 Bellaterra (Barcelona) Spain\\
  $^{25}$ Kavli Institute for Particle Astrophysics \& Cosmology, P. O. Box 2450, Stanford University, Stanford, CA 94305, USA\\
  $^{26}$ Observat\'orio Nacional, Rua Gal. Jos\'e Cristino 77, Rio de Janeiro, RJ - 20921-400, Brazil\\
  $^{27}$ Centro de Investigaciones Energ\'eticas, Medioambientales y Tecnol\'ogicas (CIEMAT), Madrid, Spain\\
  $^{28}$ Kavli Institute for Cosmological Physics, University of Chicago, Chicago, IL 60637, USA\\
  $^{29}$ Department of Astronomy, University of Michigan, Ann Arbor, MI 48109, USA\\
  $^{30}$ Department of Physics, University of Michigan, Ann Arbor, MI 48109, USA\\
  $^{31}$ SLAC National Accelerator Laboratory, Menlo Park, CA 94025, USA\\
  $^{32}$ Department of Physics, ETH Zurich, Wolfgang-Pauli-Strasse 16, CH-8093 Zurich, Switzerland\\
  $^{33}$ Santa Cruz Institute for Particle Physics, Santa Cruz, CA 95064, USA\\
  $^{34}$ Department of Physics, The Ohio State University, Columbus, OH 43210, USA\\
  $^{35}$ Astronomy Department, University of Washington, Box 351580, Seattle, WA 98195, USA\\
  $^{36}$ Australian Astronomical Observatory, North Ryde, NSW 2113, Australia\\
  $^{37}$ Instituci\'o Catalana de Recerca i Estudis Avan\c{c}ats, E-08010 Barcelona, Spain\\
  $^{38}$ Jet Propulsion Laboratory, California Institute of Technology, 4800 Oak Grove Dr., Pasadena, CA 91109, USA\\
  $^{39}$ School of Physics and Astronomy, University of Southampton,  Southampton, SO17 1BJ, UK\\
  $^{40}$ Instituto de F\'isica Gleb Wataghin, Universidade Estadual de Campinas, 13083-859, Campinas, SP, Brazil\\
  $^{41}$ Computer Science and Mathematics Division, Oak Ridge National Laboratory, Oak Ridge, TN 37831\\
  $^{42}$ Cerro Tololo Inter-American Observatory, National Optical Astronomy Observatory, Casilla 603, La Serena, Chile\\
}
}

\date{\today}

\pagerange{\pageref{firstpage}--\pageref{lastpage}} \pubyear{2017}\volume{0000}

\maketitle
\label{firstpage}

\newpage

\newpage

\clearpage

\begin{abstract}
Mock catalogues are a crucial tool in the analysis of galaxy surveys data, both for the accurate computation of covariance matrices, and for the optimisation of analysis methodology and validation of data sets. In this paper, we present a set of 1800 galaxy mock catalogues designed to match the Dark Energy Survey Year-1 BAO sample \citep{sample} in abundance, observational volume, redshift distribution and uncertainty, and redshift dependent clustering. 
The simulated samples were built upon \halogen\ \citep{halogen} halo catalogues, based on a $2LPT$ density field with an empirical halo bias. For each of them, a lightcone is constructed by the superposition of snapshots in the redshift range $0.45<z<1.4$.
Uncertainties introduced by so-called photometric redshifts estimators were modelled with a \textit{double-skewed-Gaussian} curve fitted to the data. We populate halos with galaxies by introducing a hybrid Halo Occupation Distribution - Halo Abundance Matching model with two free parameters. These are adjusted to achieve a galaxy bias evolution $b(z_{\rm ph})$ that matches the data at the 1-$\sigma$ level in the range $0.6<z_{\rm ph}<1.0$. We further analyse the galaxy mock catalogues and compare their clustering to the data using the angular correlation function $	w(\theta)$, the comoving transverse separation clustering $\xi_{\mu<0.8}(s_{\perp})$ and the angular power spectrum $C_\ell$, finding them in agreement. This is the first large set of three-dimensional \{ra,dec,$z$\} galaxy mock catalogues able to simultaneously accurately reproduce the photometric redshift uncertainties and the galaxy clustering. 
\end{abstract}


\section{Introduction} \label{sec:introduction}

The Large Scale Structure (LSS) of the Universe has proven to be a very powerful tool to study Cosmology. In particular, distance measurements of the Baryonic Acoustic Oscillation (BAO) scale \citep{thBAO1,thBAO2} can be used to infer the expansion history of the Universe and, hence, to constrain dark energy properties. Whereas most BAO detections have been performed by spectroscopic galaxy surveys, able to estimate radial positions with great accuracy \citep{BAO1,BAO2,BAO3,BAO4,BAO5,BAO6,BAO7}, the size and depth of the Dark Energy Survey\footnote{\url{www.darkenergysurvey.org}} (DES) gives us the opportunity to measure the BAO angular distance $D_A(z)$ with competing constraining power using only photometry. Photometric galaxy surveys provide moderately accurate estimates of the redshift of galaxies from the magnitudes observed through a number of filters (5, in the case of DES) \citep{photoz}, making it more difficult to obtain BAO  measurements. 
However, the fidelity with which the BAO is observed is boosted by the photometric survey's capability to explore larger areas of the sky ($1318$ deg$^2$ for the BAO DES sample from data taken in the first year, $\sim5000\ $deg$^2$ for the complete survey) and larger number density of galaxies, reducing the shot noise. The current Year-1 (Y1) DES data already allow us to probe a range $0.6<z<1$ poorly explored with BAO physics. 

This paper is released within a series of studies devoted to the measurement of
the BAO scale with the DES Y1 data. The main results are presented 
in \citet{Y1BAO} (here after \main), including a $\sim 4\%$ precision $D_A$ BAO measurement. \citet{sample} defines the sample selection optimised for BAO analysis (hereafter \sample). A photometric redshift validation over the sample is performed 
in Gazta{\~n}aga et al. (in preparation) (hereafter \photoz). A method to extract the BAO from angular clustering in tomographic redshift bins is presented in Chan et al. (2018) (from now \templates). \citet{3DBAO} (\comov\ in the remainder) explains a method to extract the BAO information from the comoving transverse distance clustering. Camacho et al. (2018) (\Cls\ from now), presents a method to extract the BAO scale from the angular power spectrum. This paper will be devoted to the simulations used in the analysis.

In order to analyse the data we need an adequate theoretical framework. Even though there are analytic models that can help us understand the structure formation of the Universe 
\citep{PS,peak_split,EPS,halomodel,ZA,Kaiser,2lpt_0}, most realistic models are based on numerical simulations. Simulations have the additional advantages that they allow us to easily include observational effects such as masks and redshift uncertainties and can realistically mimic how these couple with other sources of uncertainty such as cosmic variance or shot noise. For the estimation of the covariance matrices of our measurements we need a number of the order of hundreds to thousands of simulations, depending on the size of the data vector analysed \citep{cov1}, 
in order that the uncertainty in the covariance matrices is subdominant for the final results. As full $N$-Body simulations require considerable computing resources, running that number of $N$-Body simulations is unfeasible.
Approximate mock catalogues are an alternative to simulate our data set in a much more computationally efficient way \citep{lognormal,peakpatch,pthalos,mocks1,mocks2,mocks3,halogen,cola,pinocchio_2,qpm,nifty,monaco}. 
These methods are limited in accuracy at small scales, however, these method have been shown to reproduced accurately the large scales \citep{nifty}. 
Alternatively, we can use a lower number of mock catalogues combining them with theory using hybrid methods \citep{cov2,cov3,cov4,cov5}, or methods that can re-sample $N$-Body simulations (e.g. \citealt{schneider}).
These alternatives would still rely on $\sim 100-200$ simulations (see \templates, or \citealt{schneider}), in order to have subdominant noise in the covariance.

Galaxy mock catalogues are important in Large Scale Structure studies not only for the computation of covariance matrices, but also crucial when optimising the  methodology and understanding the significance of any particularity found in the data itself, and learn how to interpret/deal with it (see, for example, Appendix A in \main).

In this paper we present a set of 1800 mock catalogues designed to statistically match those properties of the DES Y1-BAO sample.
The main properties from the simulations that we need to match to the data in order to correctly reproduce the covariance are: the galaxy abundance, the galaxy bias evolution, the redshift uncertainties and the shape of the sampled volume (angular mask and redshift range).
The definition of the reference sample is summarised in \Sec{sec:sample}. 
As a first step, we use the halo generator method called \halogen\ \citep[summarised in \Sec{sec:halogen}]{halogen}, to create dark matter halo catalogues in Cartesian coordinates and fixed redshift. We then generate a lightcone (\Sec{sec:lightcone}) by transforming our catalogues to observational coordinates $\{ {\rm ra},{\rm dec}, z_{\rm sp} \}$, accounting for redshift evolution, and implement the survey mask (\Sec{sec:mask}). 
In \Sec{sec:photoz} we model and implement the redshift uncertainties introduced in the sample by the photometric redshift techniques.
The galaxy clustering model is described in \Sec{sec:hod}, where we introduce a redshift evolving hybrid Halo Occupation Distribution (HOD) - Halo Abundance Matching (HAM) model. Finally, in \Sec{sec:results} we analyse the set of mock catalogues, comparing their covariance matrices with the theoretical model in \templates, and we compare the clustering measurements in angular configuration space ($w_i(\theta)$), three-dimensional configuration space ($\xi_{<\mu0.8}(s_{\perp})$) and angular harmonic space ($C_{l}^i$) of our mock catalogues with the data and theoretical models. We conclude in \Sec{sec:conclusions}.

\section{The Reference Data}
\label{sec:sample}

The aim of this paper is to reproduce in a cosmological simulation all the properties relevant for BAO analysis of the DES Y1-BAO sample. We describe how this sample is selected below in \Sec{sec:Y1BAO}, 
and how the redshifts of that sample are obtained in \Sec{sec:photometry}.
We also describe how we compute correlation functions from data or simulations in \Sec{sec:cute}.

\subsection{The Y1-BAO Sample}
\label{sec:Y1BAO}

The Y1-BAO sample is a subsample of the Gold Catalogue \citep{gold} obtained from the first Year of DES observations \citep{survey}. The Gold Catalogue provides `clean' galaxy catalogues and photometry as described in \citet{gold}. 
A footprint quantified using a  Healpix \citep{healpix} map with $n_{\rm side}=4096$   
is provided with all the areas with at least $90s$ of exposure time in all the filters $g$,$r$,$i$ and $z$, summing up to $\sim 1800 {\rm deg}^2$. 
After vetoing bright stars and the Large Magellanic Cloud, the area is reduced to $\sim 1500 {\rm deg}^2$.

The Y1-BAO sample selection procedure was optimised to obtain precise BAO measurements at high redshift and is fully described in \sample. The Y1-BAO sample is obtained applying three main selection criteria:

\begin{eqnarray}
\label{eq:sample}
\begin{aligned}
& 17.5 < i_{\rm auto} < 19.0 + 3.0\ z_{\rm BPZ-MA}\\
& (i_{\rm auto} - z_{\rm auto}) + 2.0(r_{\rm auto} - i_{\rm auto}) > 1.7 \\
& 0.6 < z_{\rm ph} < 1.0\\
\end{aligned}
\end{eqnarray}
\noindent
with $X_{\rm auto}$ being the \textsc{mag\_auto} magnitude in the band $X$, and 
$z_{\rm BPZ-MA}$ the photometric redshift obtained by BPZ \citep{bpz1} using \textsc{mag\_auto} photometry, and $z_{\rm ph}$ being the photometric redshift (either $z_{\rm BPZ-MA}$ or $z_{\rm DNF-MOF}$ see below). 
Apart from the three main cuts in \Eq{eq:sample}, we remove outliers in color space and perform a star-galaxy separation. 
Further veto masks are applied to the Y1-BAO sample guaranteeing at least a $80\%$ coverage of each pixel in the four bands, requiring sufficient depth-limit in different bands and removing `bad regions'. The final Y1-BAO sample after all the veto masks have been applied covers an effective area of $1318\ {\rm deg^2}$ (see more details in \sample).

\subsection{Photometric redshifts}
\label{sec:photometry}

The redshift estimation for each galaxy is based on the magnitude observed in each filter. For this paper, we will use two combinations of photometry and photo-z code, respectively: \textsc{mag\_auto} with BPZ (BPZ-MA), and MOF  with DNF (DNF-MOF). 

\textsc{mag\_auto} photometry is derived from the flux of the coadded image, as measured by the SExtractor software \citep{MA} from each of the bands. On the other hand, the MOF approach (Multi-Object Fitting, \citealt{gold}) makes a multi-epoch, multi-band fit to the shape of the object instead of on the coadded image as well as subtracting the light of neighbouring objects. The flux is fit with this common shape for each band separately.

A thorough description and comparison of both photometric redshift methods utilised here can be found in \citet{photoz}. First, we have BPZ (Bayesian Photometric Redshift, \citealt{bpz1,bpz2}), which
is a method based on synthetic templates of spectra convolved with the DES filters, and makes use of Bayesian inference. On the other hand, we have DNF (Directional Neighbourhood Fitting, \citealt{DNF}), which is a training-based method. 

Both methods take the results from the chosen photometry in the four bands and give a Probability-Distribution-Function (PDF) for the redshift of each galaxy: $P(z)$. As a full PDF for each galaxy would build up a very large dataset to transfer and work with, here we take two quantities from each PDF: the mean $z_{\rm ph}\equiv\langle P(z) \rangle $, and a random draw from the distribution $z_{\rm mc}$. We will explain in \Sec{sec:photoz} how we will model the effect of photometric redshift in our simulations. 

By default, the reference data will use the DNF-MOF redshift $z_{\rm DNF-MOF}$, since this is the one used in \main\ for the fiducial results.
We will only include $z_{\rm BPZ-MA}$ in \Sec{sec:photoz}, since it was the reference redshift when part of the methodology presented in that section was designed. 

\subsection{2-point Correlation Functions}
\label{sec:cute}

Throughout this paper we analyse 2-point correlation functions in repeated occasions. 
In all cases we use the Landy-Szalay estimator \citep{LS}:

\begin{equation}
\Psi(\textbf{x})=\frac{DD(\textbf{x})-2DR(\textbf{x})+RR(\textbf{x})}{RR(\textbf{x})}
\label{eq:LS}
\end{equation}
with DD, DR and RR being respectively the number of Data-Data, Data-Random and Random-Random pairs separated by a distance $\textbf{x}$. The correlation $\Psi$ refers to either an angular correlation 
denoted by $w$, or a three-dimensional correlation $\xi$. The variable $\textbf{x}$ may correspond to the angular separation $\theta$ projected on to the sky, or the three-dimensional comoving separation $r$. In the three-dimensional case, we will sometimes study the anisotropic correlation, distinguishing between 
the distance parallel to the line-of-sight and perpendicular to it, having 
$\textbf{x} =\{ s_{\parallel},s_{\perp}\}$.

The data D may refer to observed data or simulated data. Random catalogues R are produced by populating
the same sampled volume as the data with randomly distributed points. 
All the correlation function presented here were computed with the public code \textsc{cute} \citep{cute}\footnote{\url{https://github.com/damonge/CUTE}}.


\section{Halo lightcone catalogues} \label{sec:halo}

Prior to the generation of the galaxy catalogues, we need to construct the field of dark matter
halos. 
For this, we will use the technique called \halogen\ \citep{halogen}, a technique that produces halo catalogues with Cartesian coordinates embedded in a cube and at a given time slice (\textit{snapshot}). By superposing a series of \halogen\ snapshots, we construct an observational catalogue with angular coordinates and redshift $\{{\rm ra}$,${\rm dec}$,$z_{\rm sp}\}$: a lightcone halo catalogue.
Finally, we describe how we implement the survey mask in the mock catalogues in order to statistically reproduce the angular distribution of the data.

\subsection{HALOGEN}
\label{sec:halogen}

\halogen\footnote{\url{https://github.com/savila/HALOGEN}} is a fast approximate method to generate halo mock catalogues. It was designed and described in \citet{halogen}, and compared with other methods in \citet{nifty}.
We summarise it here as 4 major steps:	

\begin{enumerate} 
	\item Generate a distribution of dark matter particles with $2^{\rm nd}$-order Perturbation Theory \citep[2LPT]{2lpt_0,bouchet} at fixed redshift in a box of size $L_{\rm box}$. Distribute those particles onto a grid with cells of size $l_{\rm cell}$.
	\item Produce a list of halo masses $M_h$ from a theoretical/empirical Halo Mass Function (HMF).
	\item Place the halos at the position of particles with a probability dependent on the cell density and halo mass as $P_{\rm cell} \propto \rho_{\rm cell} ^{\alpha(M_h)}$. Within cells we choose random particles, while imposing an exclusion criterion to avoid halo overlap (using the $R_{200,\ {\rm crit}}$ derived from the halo mass). Mass conservation is ensured within cells by not allowing more halos once the mass of the halos surpasses the original dark matter mass. 
	\item Assign the velocities of the selected particles to the halos rescaled through a factor: $\mathbf{v}_{\rm halo}=f_{\rm vel}(M_h)\cdot \mathbf{v}_{\rm part}$
\end{enumerate}

There are one parameter, and  two functions of halo mass that have been introduced in the method and need to to be set for each run. We set the cell size $l_{\rm cell}=5\mpch$ as in \citet{halogen}.
The parameter $\alpha(M_h)$ controls the halo bias and is fitted to a reference $N$-Body simulation to match the mass-dependent clustering. The factor $f_{\rm vel}(M_h)$ is also calibrated against an \nb\ simulation in order to reproduce the variance of the halo velocities, crucial for the redshift-space distortions. 
For this study we use the \mice\ simulation as a reference for this calibration. 

The MICE Grand Challenge simulation, described in \citep{mice1,mice2,mice3}, is based on a cosmology with parameters:
$\Omega_M =0.25$, $\Omega_\Lambda =0.75$, $\Omega_b=0.044$, $h=0.7$, $\sigma_8=0.8$, $n_s =0.95$ matching early WMAP data \citep{wmap3}. We will use this fiducial cosmology throughout the paper. The box size of the simulation is $L_{\rm box}=3072$\hMpc\ and made use 
of $4096^3$ particles. The \halogen\ catalogues use a lower mass resolution with $1280^3$ particles in order to reduce the required computing resources.  We use the same box size and cosmology for \halogen.

For the calibration we used the same phases of the initial conditions as the \nb\ simulation and fitted 
the \halogen\ parameters with the snapshots at $z_{\rm snap}=0,0.5,1.0,1.5$. 
We input to \halogen\ a hybrid Halo Mass Function, using the HMF directly 
measured from \mice\ catalogues at low masses, while using an analytic 
expression \citep{watson} generated with \textsc{hmfcalc}\footnote{\url{http://hmf.icrar.org}} \citep{hmfcalc}	 for the large masses, where the HMF from the \mice\ catalogues are noise dominated.

\begin{figure}
\includegraphics[height=1.1\linewidth,angle=270]{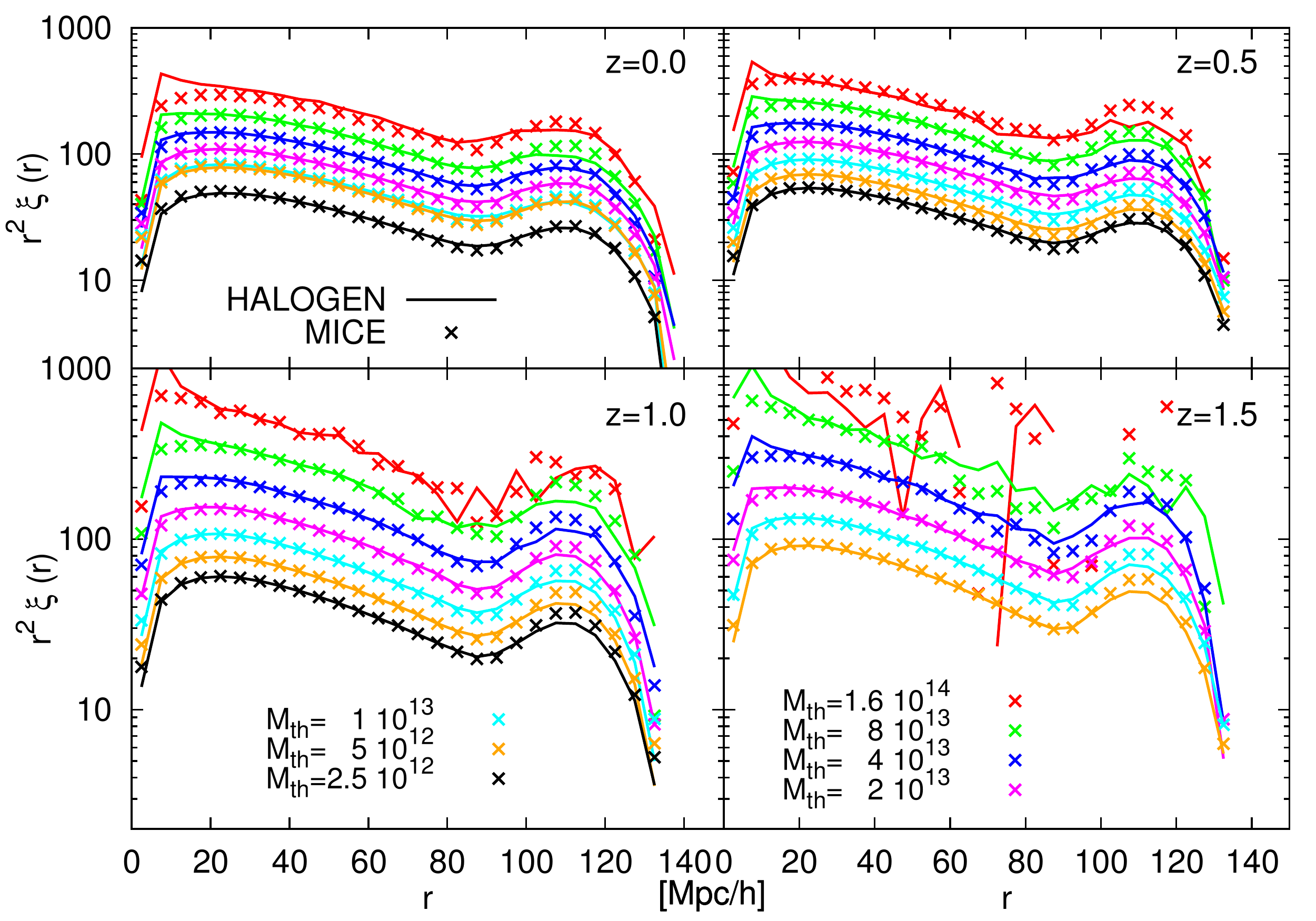}
  \caption{2-point correlation function of \mice\ vs. \halogen\ halos in the simulation box at the snapshots $z=0.0$, $0.5$, $1.0$ and $1.5$ as labelled.
  We display the different mass thresholds $M_{\rm th}$  used during the fit (finding higher correlations for higher $M_{\rm th}$). Note that correlations have been multiplied by $r^2$ to highlight the large scales.}
  \label{fig:MICE_box}
\end{figure}

We fitted the clustering of the halos in logarithmic mass bins (with factor 2 in mass threshold), this being a slight variation with respect to the method used in \citet{halogen}. The minimum halo mass that we 
probe is $M_h=2.5\times 10^{12} M_{\odot}/h$ for snaphots at $z\leq 1.0$, whereas we use a minimum mass of $M_h=5.0\times 10^{12} M_{\odot}/h$ for higher redshift snaphots. 
Once the parameter calibration is finished, we find a good agreement between \mice\ and \halogen\ correlation functions as a function of redshift and mass, as shown in \Fig{fig:MICE_box}.

\subsection{Lightcone}
\label{sec:lightcone}

We place the observer at the origin (i.e., one corner of the box), so that we can simulate one octant of the sky, and transform to spherical coordinates.
We use the notation $z_{\rm sp}$ for the redshift of a halo or galaxy as it would be observed with a spectroscopic survey (i.e., with negligible uncertainty).  
We have that

\begin{equation}
 \begin{aligned}
      &{z_{\rm sp}} = z(r) + \frac{1+z(r)}{c} \vec{u} \cdot \hat{r}\ ,\\
 \end{aligned}
 \label{eq:zsp}
\end{equation}
with $\vec{r}=\{X,Y,Z\}$ the comoving position,  $\vec{u}$ the comoving velocity, $r=\lvert \vec{r} \lvert$, $\hat{r}=\vec{r}/r$ and $z(r)$ the inverse of 

\begin{equation}
\label{eq:rz}
 r(z) = c \int_0^{z}  \frac{dz'}{H(z')} \ .
\end{equation}

The first term in \Eq{eq:zsp} corresponds to the redshift due to the Hubble expansion, whereas the second term is the contribution from the peculiar velocity of the galaxy.

Given the redshift range covered by DES, we need to allow for redshift-dependent clustering, and hence we will let the \halogen\ parameters ($\alpha$, $f_{\rm vel}$, and the HMF)
vary as a function of redshift.

For that, we interpolate $\alpha$, $f_{\rm vel}$ and ${\rm log}n$ (the logarithm of the HMF) using cubic splines from the reference redshifts $z_{\rm snap}=0,0.5,1,1.5$ at which parameters were fitted to our output redshifts $z_{{\rm snap}, i}=0.3$,
$0.55$, $0.625$, $0.675$, $0.725$, $0.775$, $0.825$, $0.875$, $0.925$, $0.975$, $1.05$ and $1.3$, then
\halogen\ was run at those redshifts. 
We build the lightcone from the superposition of spherical $z_{\rm sp}$-shells drawn from the snapshots by setting the edges at the intermediate redshifts. So each snapshot $i$ contributes galaxies whose redshift is in the interval $z_{\rm sp}\in [\frac{z_{{\rm snap}, i-1}+z_{{\rm snap}, i}}{2},\frac{z_{{\rm snap}, i}+z_{{\rm snap}, i+1}}{2}]$, also imposing the edges of the lightcone at $z_{\rm sp}=0.1$ and $z_{\rm sp}=1.42$ (which is the maximum redshift reachable given the chosen geometry and cosmology). \textit{A priori}, there will be a relatively sharp transition of the clustering properties at the edges of the $z_{\rm sp}$-shells, but, once we have introduced the redshift uncertainties in \Sec{sec:photoz}, those transitions will be smoothed.
Throughout Sections \ref{sec:halo},\ref{sec:photoz},\ref{sec:hod} we will focus the analysis in 8 photometric redshift bins with width $\Delta z_{\rm ph}=0.05$ between $z_{\rm ph}=0.6$ and $z_{\rm ph}=1.0$. When dealing with true redshift space, we will need to extend the boundaries to the range $0.45<z_{\rm sp}<1.4$ (see \Sec{sec:photoz}).

\begin{figure}
  \centering
  \includegraphics[height=1\linewidth,angle=270]{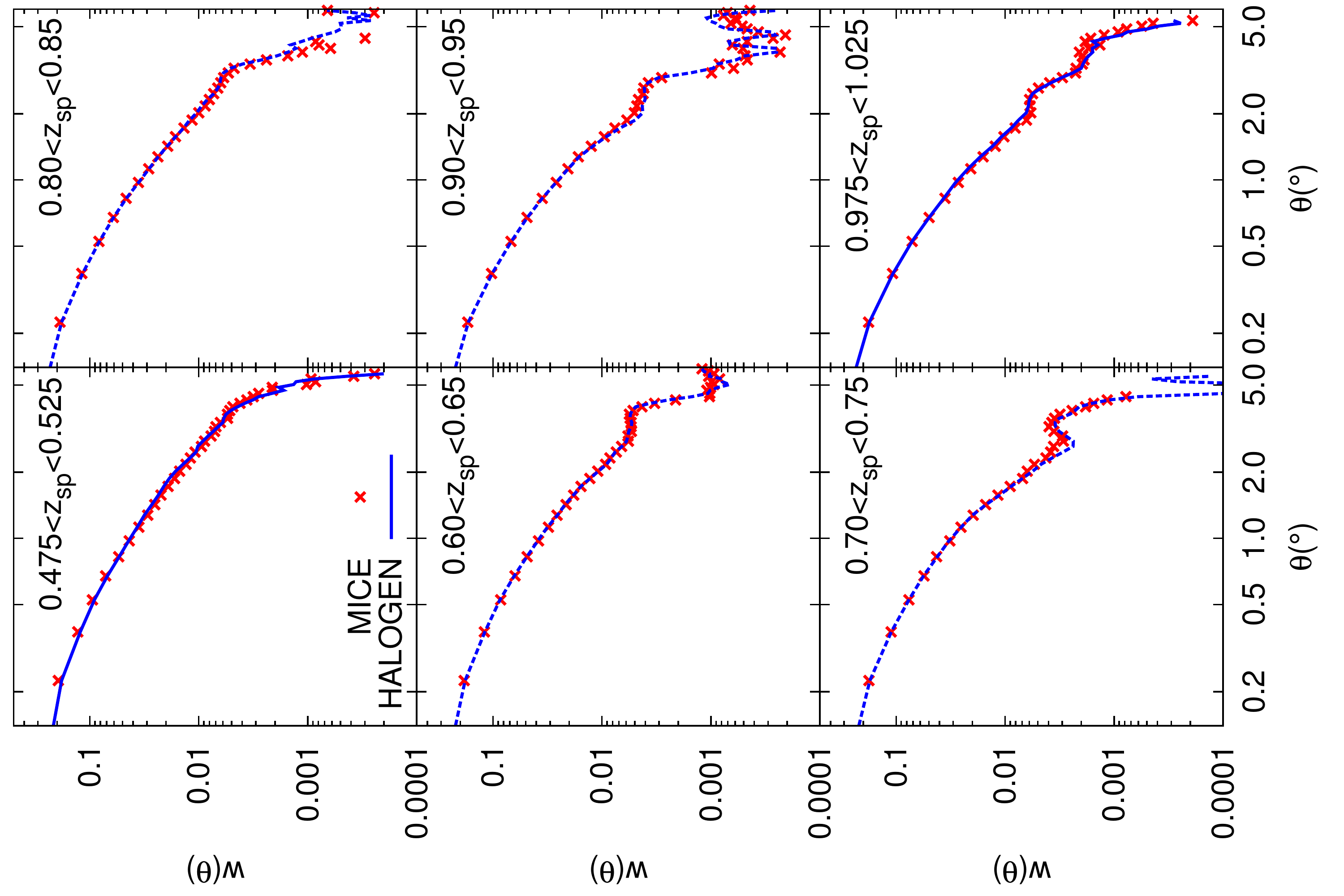}
  \caption{Angular correlation function of halos from the \mice\ (crosses) and \halogen\ (lines) lightcones. 
  The different panels correspond to different redshift bins, as labelled, with width $\Delta z_{\rm sp}=0.05$.
  We mark in solid lines the redshifts at which \halogen\ parameters were fitted, and in dashed lines results from interpolated parameters. Correlation functions shown correspond to mass cuts at $M_h=2.5\times 10^{12} M_{\odot}/h$ .}
  \label{fig:MICE_cone}
\end{figure}

Finally, we compare the resulting \halogen\ lightcone with the halo lightcone generated by \mice\ in \Fig{fig:MICE_cone}.
Note that the \mice\ simulated lightcone is constructed from fine timeslices ($\Delta z=0.005-0.025$) built on-the-fly from a full \nb\ simulation and using the velocity
of the particles to extrapolate their positions at the precise moment they cross the lightcone \citep{mice0}. 
Remarkably, despite the large differences in the methodology, the
angular correlation functions from both lightcones show very good agreement at all redshifts, independently of whether the \halogen\ parameters were fitted or interpolated. 
At large scales sampling variance becomes dominant, modifying stochastically the shape of the correlation function, but since we imposed the same phases of the initial conditions, it enables us to make a one-to-one comparison. 

\subsection{Angular selection}
\label{sec:mask}

In \Sec{sec:lightcone} we placed the observer at one corner, providing a sample covering an octant of the sky. In \Sec{sec:sample} we described how, we obtained a footprint covering the effectively observed area of the sky. This mask spreads over a large fraction of the sky and cannot be fit into a single octant. However, 
using the periodic boundary conditions of the box, we can put 8 replicas of the box together to build a larger cube, and extract a full sky lightcone catalogue, albeit with a repeating pattern of galaxies.  

\begin{figure}
  \centering
  \includegraphics[width=1\linewidth,trim={0 2cm 0 0},clip]{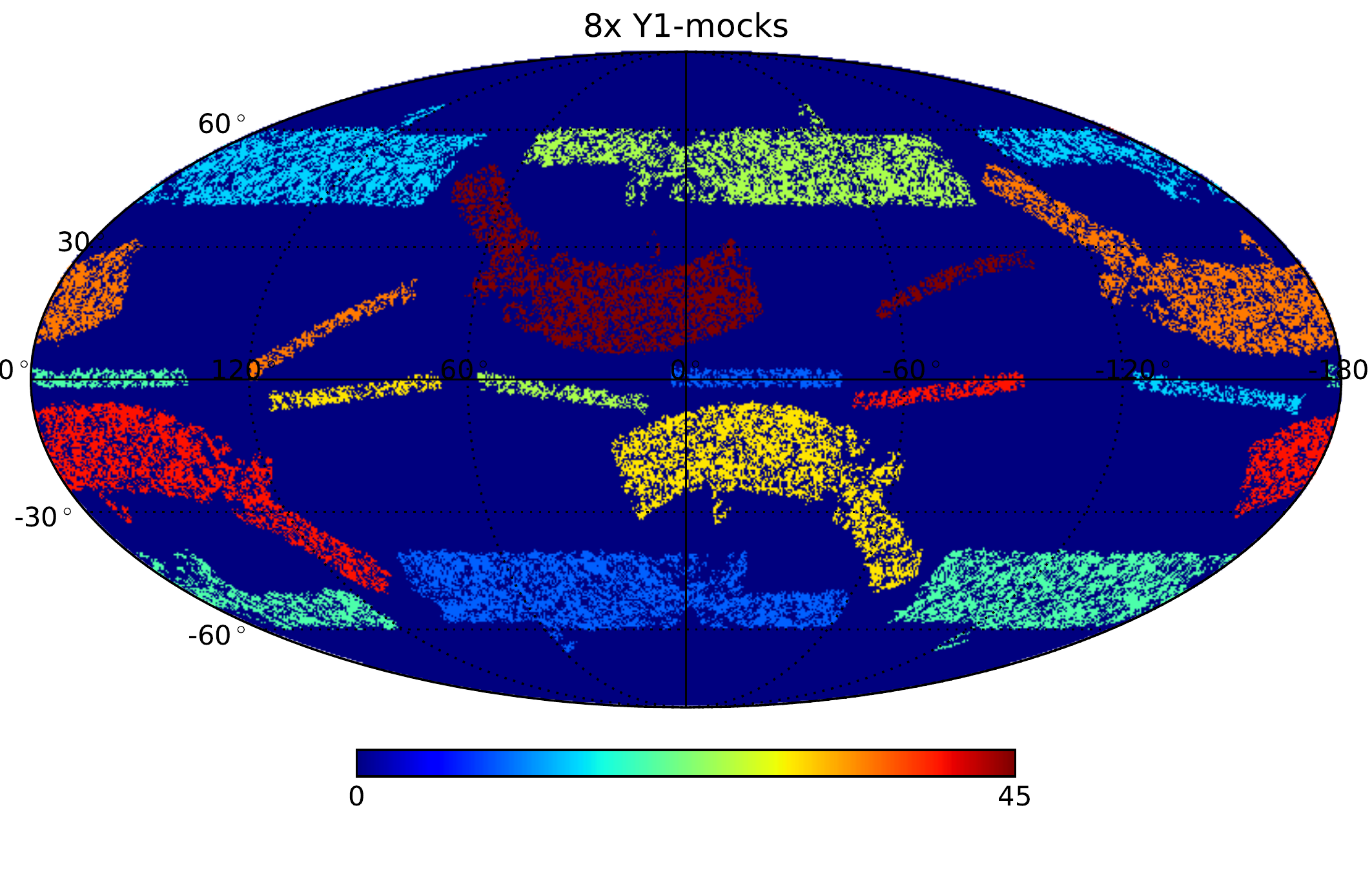}
     \caption[h]{Mask Rotations. Representation of the full sky with the eight rotations of the mask used to generate the mock catalogues. 
     Each mask, with two disjoint parts, is represented in a different colour, being the original data mask the one passing through $\{0\degree,0\degree\}$ and $\{0\degree,-60\degree\}$ in $\{$ra,dec$\}$. The mock galaxies selected by the rotated masks are, then, rotated to the position of the original mask. Having the correct angular selection is one of the key ingredients to ensure that the mocks will give us the correct covariance matrices.     \label{fig:mask}}
  \includegraphics[width=1\linewidth]{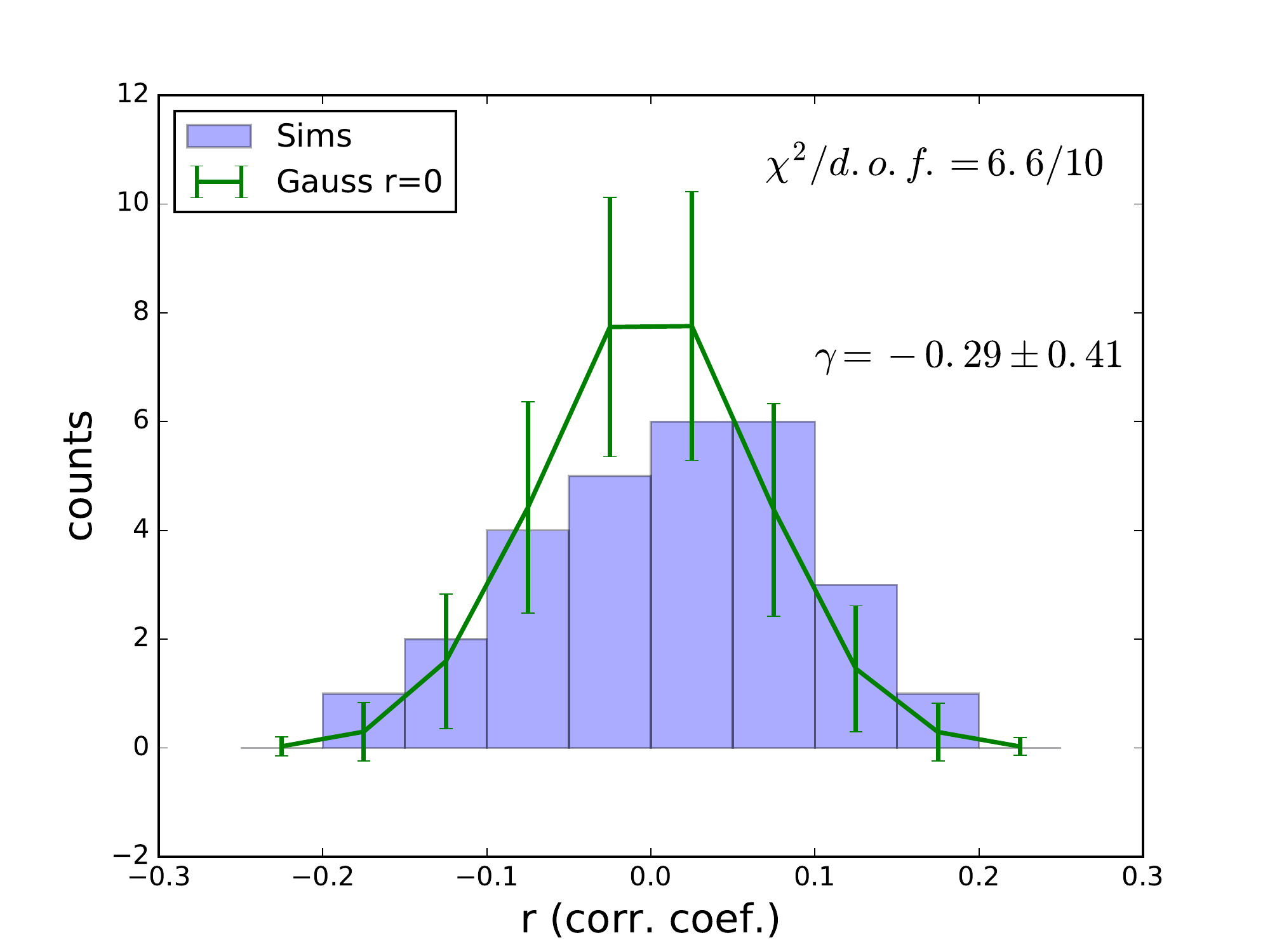} 
  \caption[h]{Distribution of correlation coefficients of $\alpha$ (see text) between different mask rotations for the \halogen\ mock catalogues (`Sims'). 
  We compare it to the expected distribution of correlations if $\alpha$ followed a Gaussian distribution and results from different masks were uncorrelated (`Gauss r=0'). 
  We find that this hypothesis (with no free parameter) to be compatible with our simulations. We also find that the observed skewness in the distribution $\gamma$ is within the expected statistical noise level under the same hypothesis.  \label{fig:corr} }

\end{figure}

In \Fig{fig:mask} we show how we can draw 8 mock catalogues with the Y1 footprint from the full sky catalogue by performing rotations on the sphere. We see that the footprint has a complicated shape with two disjoint areas: one passing by $\{{\rm ra}=0\degree,{\rm dec}=0\degree\}$ known as Stripe-82 (that overlaps with many other surveys), and another passing by $\{{\rm ra}=0\degree,{\rm dec}=-60\degree\}$, known as SPT-region (due to the overlap with the South Pole Telescope observations). While designing the rotations depicted in \Fig{fig:mask}, we made sure that every pair of footprints would not overlap and that the two disjoint areas are separated by more than the maximum scale of interest ($\sim6\degree$, see \Sec{sec:ang}).

Including this angular selection to the mock catalogues will be essential since, as shown in \Sec{sec:results} it has an important effect in the covariance matrices. 

Given the repetition of boxes, one could be concerned about the effect it could have in our measurements. Qualitatively, we do not expect this to be important for a series of reasons. First, the repetition occurs at very large scales ($L=3072\mpch$) and we only use eight replicas. This makes difficult for structures to be observed more than once, and if they are, it will always be done from a different orientation and at a different redshift. On top of that, there are three stochastic processes that will make the hypothetically repeated structure appear differently: the halo biasing (since the structure would be at different redshift, it would be drawn from a different snapshot, see \Sec{sec:halogen}), the redshift uncertainties (\Sec{sec:photoz}) and the galaxy assignment to halos (\Sec{sec:hod}).

More quantitatively, we study the correlation coefficients between the measured BAO scale $\alpha$ (defined in Equation 20 of \main) from mocks coming from the same box but different mask rotation. The distribution of the $28$ ($=7\times8/2$) correlation coefficients indicate very small correlation ranging from $r=-0.2$ to $r=0.2$ as shown (as an histogram) in \Fig{fig:corr}. In order to study if these correlations $r$ represent any significance given the number of mocks we used ($N_{\rm mocks}=1800$), we generate $N_{\rm mocks}$ Gaussian realisations of $\alpha$, distribute them in 8 groups and compute the correlation coefficients between them. We repeat this process $N_{\rm rep}=1000$ times, computing for each realisation the distribution of $r$. The mean (and 1-$\sigma$ error bar) is also shown in that panel. We compute the goodness of the model using the covariance between the $N_{\rm rep}$ realisations and  find $\chi^2/d.o.f.=6.6/10$, showing that the distribution of $r$ in our simulations is completely consistent with the null hypothesis of $\alpha$ being uncorrelated. 
We note the $r$ distribution of the simulations is skewed toward positive values: $\gamma=-0.29$. Nevertheless, this $\gamma$ value is compatible with simply being a statistical fluctuation, since its absolute value is lower than the standard deviation $\Delta\gamma=0.41$ of our $N_{\rm rep}$ realisations.

\section{Photometric redshift modelling} \label{sec:photoz}

A photometric survey like DES can determine the redshift of a galaxy with a limited precision. Typically $\sigma_z/(1+z)\approx0.017-0.05$ \citep{redmagic,gold}, 
depending on the sample and the algorithm. 
This will have a major impact in the observed clustering and has to be included in the model. 
The aim of this section is to apply the effect of the redshift uncertainties to the mock catalogues.
In particular, we model the number density of galaxies as a function of \textit{true} redshift $z_{\rm sp}$ and \textit{observed} redshift $z_{\rm ph}$: $n(z_{\rm ph},z_{\rm sp})$.

In \Sec{sec:photometry}, we explained that we will work with two sets of photometric redshifts with different properties: BPZ-MA and DNF-MOF.
As we do not know the true redshift $z_{\rm sp}$ of the galaxies in the data, we need to be careful when estimating  $n(z_{\rm ph},z_{\rm sp})$ from the data.
For both sets of photometric redshifts we will follow the same methodology starting from the PDF (or $P(z_{\rm sp})$) of each galaxy.

As we mentioned in \Sec{sec:photometry}, we take the  \textit{observed} redshift $z_{\rm ph}$  as the mean of the PDF $z_{\rm ph}=\langle P(z_{\rm sp}) \rangle $, and we also select a Monte Carlo random value from the distribution $P(z_{\rm sp})\rightarrow z_{\rm mc}$, which will be used to estimate $n(z_{\rm ph},z_{\rm sp})$. 
This way, stacking the PDFs of a large number of galaxies will be statistically equivalent to taking the normalised histogram of their $z_{\rm mc}$. Hence, we can use

\begin{equation}
 n(z_{\rm ph},z_{\rm sp}) = n(z_{\rm ph},z_{\rm mc}) \ ,
 \label{eq:nzz}
\end{equation}
being able to estimate the right-hand-side from the data and apply the left-hand-side to the simulations.

One could alternatively use the training sample to estimate directly $n(z_{\rm ph},z_{\rm sp})$. However, the technique explained above has the advantage of dividing the problem into two distinct steps: one in which the photo-$z$ codes are calibrated and validated, and one in which the science analysis from the photo-$z$ products is performed. 
We discuss at the end of this section the validation of the chosen method performed in \photoz. 
Note also that, for the mocks design, the actual definition of $z_{\rm ph}$ is not relevant as long as the $P(z_{\rm ph}|z_{\rm sp})$ is known.

We select from the data thin bins of width $\Delta z_{\rm mc} = 0.01$ and measure $\partial N/\partial z_{\rm ph}$. We denote $N$ as the total number of observed galaxies in our sample (or equivalently in our mock catalogues) and $n=dN/dV$ as the number density or abundance.
Equivalently to \Eq{eq:nzz}, we can use:

\begin{equation}
 \begin{aligned}
& \frac{\partial N}{\partial z_{\rm ph}}\Bigr|_{\substack{z_{\rm sp}}} = \frac{\partial N}{\partial z_{\rm ph}}\Bigr|_{\substack{z_{\rm mc}}} 
 \end{aligned}
 \label{eq:Npart}
\end{equation}

From now we drop the $z_{\rm mc}$ notation, since we will always be looking at distributions of $z_{\rm mc}$ (never individual values), which are equivalent to the distributions of $z_{\rm sp}$. Additionally, the focus of this paper is the simulations for which we will only have $z_{\rm sp}$.

\begin{figure}
 \centering
  \includegraphics[width=1\linewidth]{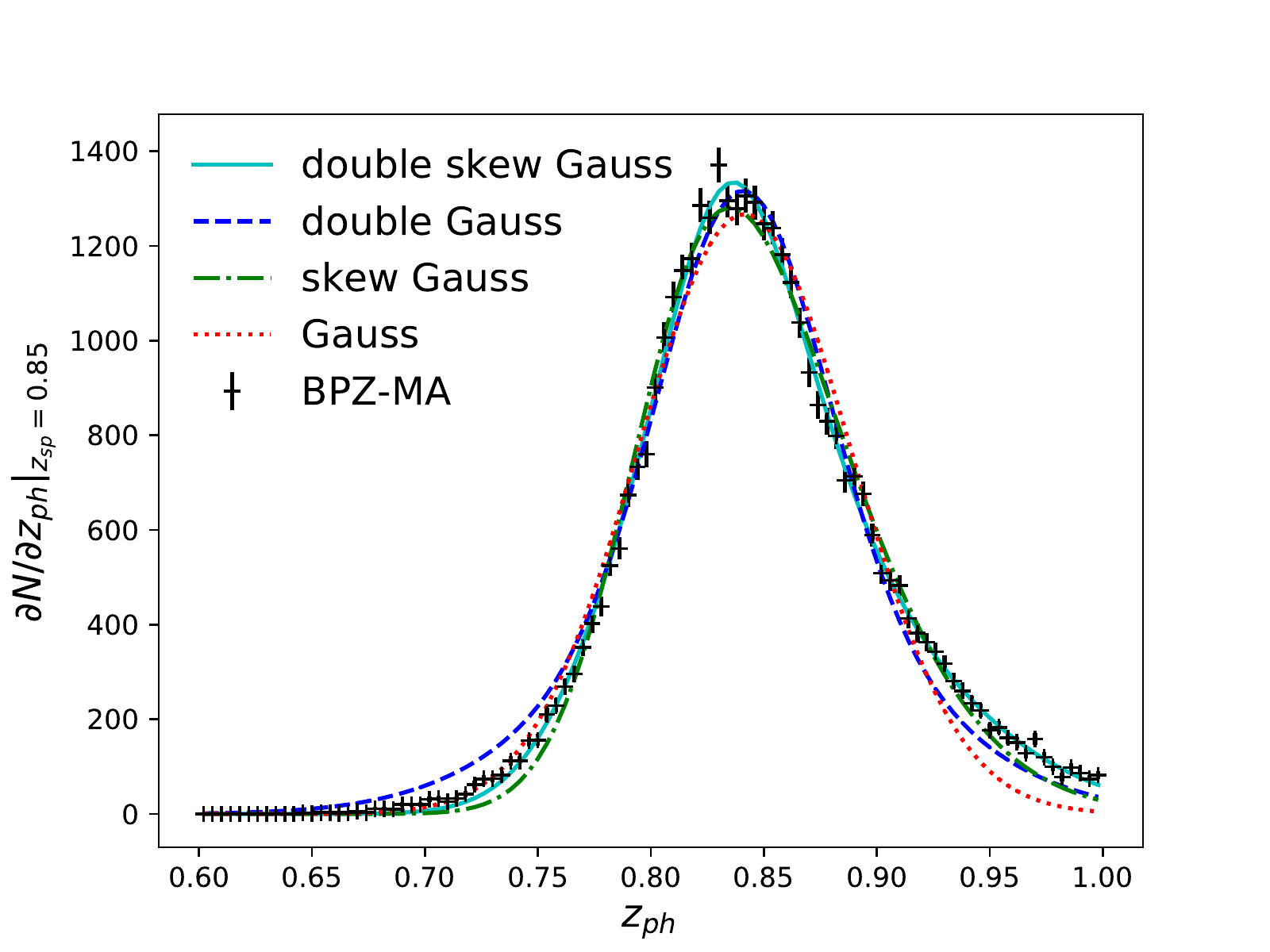}
  \includegraphics[width=1\linewidth]{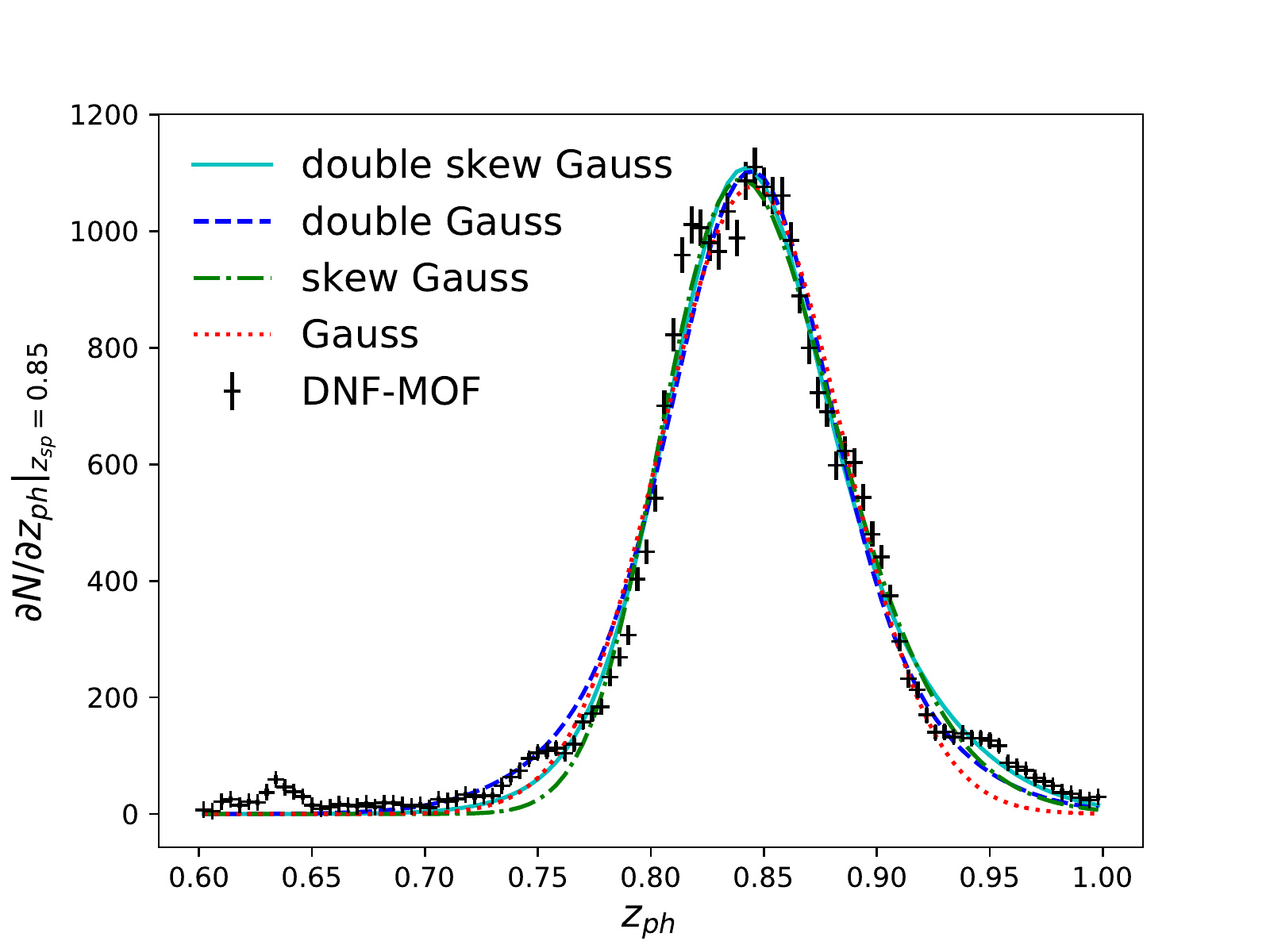}
  \caption{Abundance of galaxies at fixed true redshift $z_{\rm sp}$ as a function of photometric redshift $z_{\rm ph}$ (estimated from the joint  $\{ z_{\rm mc}, z_{\rm ph}\}$ distribution, see text and \Eq{eq:Npart}). Comparison of different fitting functions from \Eq{eq:dbsk} (lines) to the data (error bars representing the Poison noise). In the top panel we fit the BPZ-MA data, whereas in the bottom panel we fit DNF-MOF (default) data.}
  \label{fig:fit_shape}
\end{figure}

In the top panel of \Fig{fig:fit_shape} we present different fits to the BPZ-MA data for the case $\frac{\partial N}{\partial z_{\rm ph}}\Bigr|_{\substack{z_{\rm sp}=0.85}}$.
The fitting functions can be described by

\begin{table}
\begin{center}
\begin{tabular}{|l|r|r|}
\hline
 Curve & $\chi^2_{\rm BPZ-MA}/\chi^2_{\rm ref}$ &  $\chi^2_{\rm DNF-MOF}/\chi^2_{\rm ref}$ \\
\hline
Gaussian & 7.7 & 5.9 \\
Double Gaussian & 4.9 & 5.0 \\
Skewed Gaussian & 3.0 & 4.1 \\
Skewed Double Gaussian & 1 & 3.7 \\
\hline
\end{tabular}
\end{center}
\captionof{table}{Relative goodness of the fits shown in \Fig{fig:fit_shape}. The reduced $\chi^2$ were computed only taking into account Poisson noise of the histrograms. Hence, their absolute value are not significant, but their relative values (normalised to the minimum in the table), give us an idea of the improvement in the fits when addding more parameters.  For BPZ-MA the improvement is significant when including more degrees-of-freedom, whereas for DNF-MOF the improvement is less pronounced.}
\label{tab:chi}
\end{table}

\begin{equation}
 \begin{aligned}
& \frac{\partial N}{\partial z_{\rm ph}}\Bigr|_{\substack{z_{\rm sp}}} = A \cdot P(z_{\rm ph}|z_{\rm sp})    \ \ \ {\rm with} \\
& P(z_{\rm ph}|z_{\rm sp}) = \frac{1-r}{\sqrt{2\pi\sigma_1^2}}e^{-(z_{\rm ph}-\mu)^2/(2\sigma_1^2)} \ \ +  \\
& + \ \frac{r}{\sqrt{2\pi\sigma_2^2}}e^{-(z_{\rm ph}-\mu)^2/(2\sigma_2^2)}\cdot {\rm erf}\Bigg(\frac{\gamma-\mu}{\sqrt{2}\cdot \sigma_2}\Bigg), \\
& r={0,0.5,1}
 \end{aligned}
 \label{eq:dbsk}
\end{equation}
with different choices of parameters. All $A$, $\mu$, $\sigma_1$, $\sigma_2$, $\gamma$ and $r$ depend  implicitly on redshift $z_{\rm sp}$.

\begin{figure}
  \centering
  \includegraphics[height=1\linewidth,angle=270]{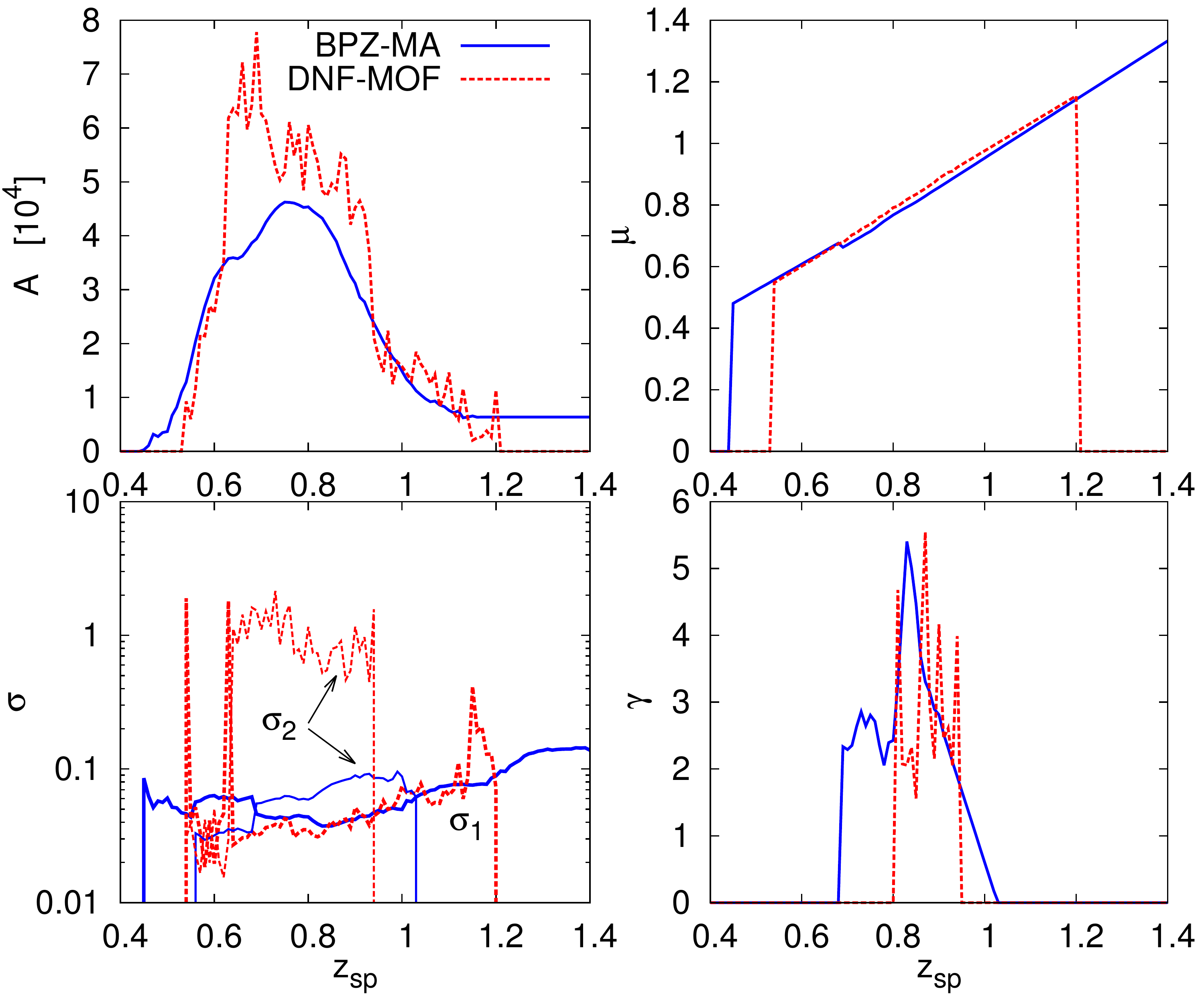}
  \caption{Best fit parameters of the fitting function described in \Eq{eq:dbsk} as a function of true redshift $z_{\rm sp}$ for both data sets. See text for details.}
  \label{fig:photoz_params}
\end{figure}

The  simplest case is a Gaussian ($\gamma=0,\sigma_2=0, r=0$). In order to include a term of kurtosis to the fit, we can 
extend the curve to a double Gaussian ($\gamma=0,\sigma_2\neq0, r=0.5$). We can also introduce skewness with the \textit{skewed} Gaussian ($\gamma\neq0,\sigma_2=0, r=1$).
Finally, the most general case considered here is the \textit{skewed double} Gaussian ($\gamma\neq0,\sigma_2\neq0, r=0.5$).  Note that for skewed curves, the parameter $\mu$ does not represent the mean.

\Fig{fig:fit_shape} and \Tab{tab:chi} show that the skewed double Gaussian does make a significant improvement to the goodness of fit for the BPZ-MA data (improving a factor of $\sim3$ to $\sim7$). However, the improvement is less significant for the DNF-MOF data, which shows more small-scale structure in the curves and is consequently more difficult to model. We discuss at the end of the section the origin of this structure.

We repeat the fit performed in \Fig{fig:fit_shape} for $z_{\rm sp}=0.85$ at all refshifts $z_{\rm sp}$. In some cases, the degeneracy between parameters makes two adjacent $z_{\rm sp}$-bins have quite different set of best fit parameters, even if the shape of the curves are similar. In order to mitigate that, we reduce the degrees of freedom in the fits when possible. We restrict the values of $r$ to $0$, $0.5$ and $1.0$ according to the values of $\sigma_1$ and $\sigma_2$. 
We also parametrise the evolution with redshift of the parameters and fix the values of these additional parameters.
For example, we fix $\mu(z_{\rm sp})$ to a straight line for most of the $z_{\rm sp}$ range, and we also set $\gamma=0$ or $\sigma_2=0$ where they stop improving the fits.
\Fig{fig:photoz_params} shows the evolution of the fitted parameters that we obtain. 
For BPZ-MA, the fits converge more easily and we can find relatively smooth evolution. For DNF-MOF, the curves have more small-scale structure, and the evolution of the best fit parameters inherits that structure.

\begin{figure}
  \centering
  \includegraphics[width=1\linewidth]{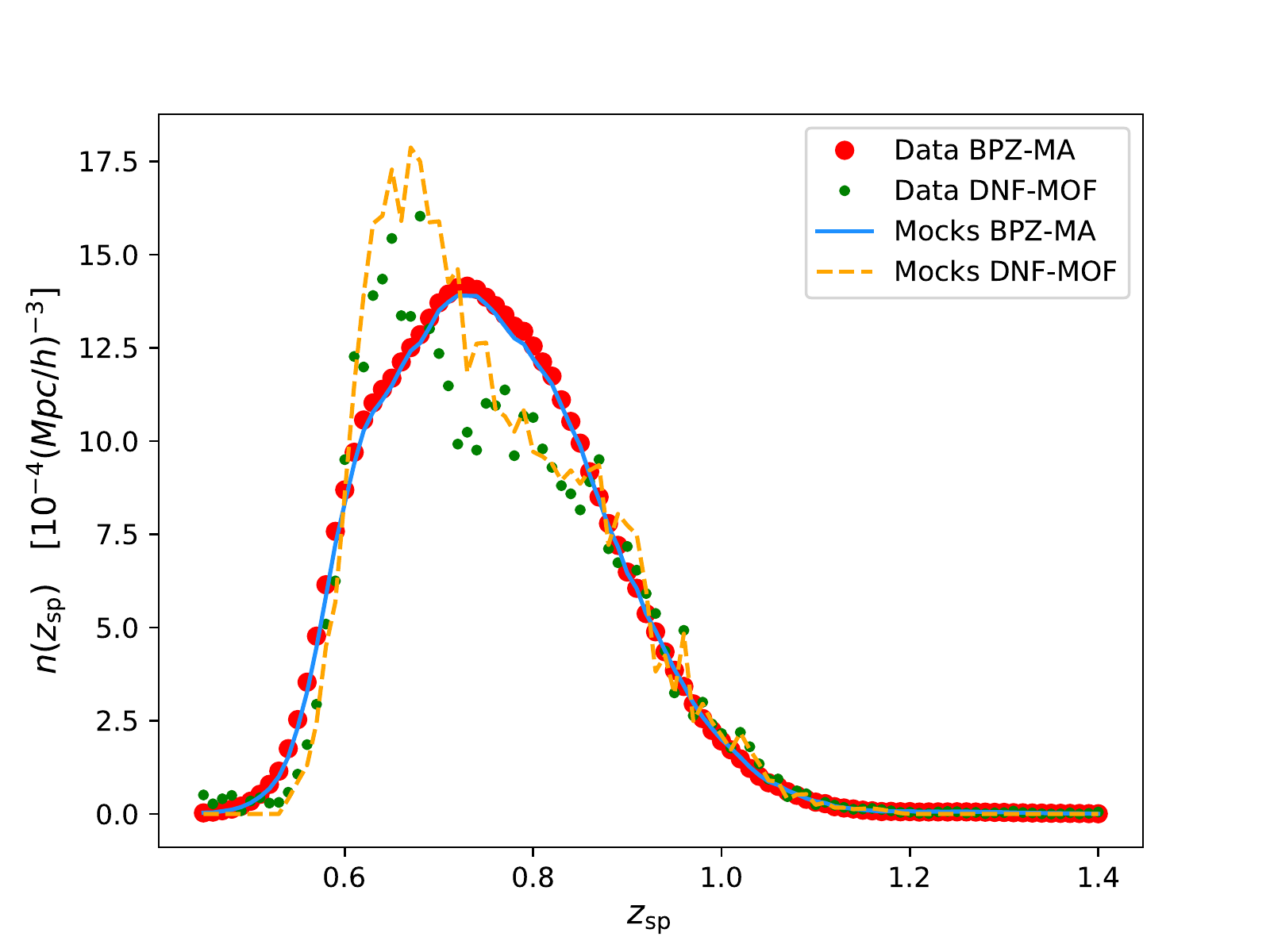}
  \includegraphics[width=1\linewidth]{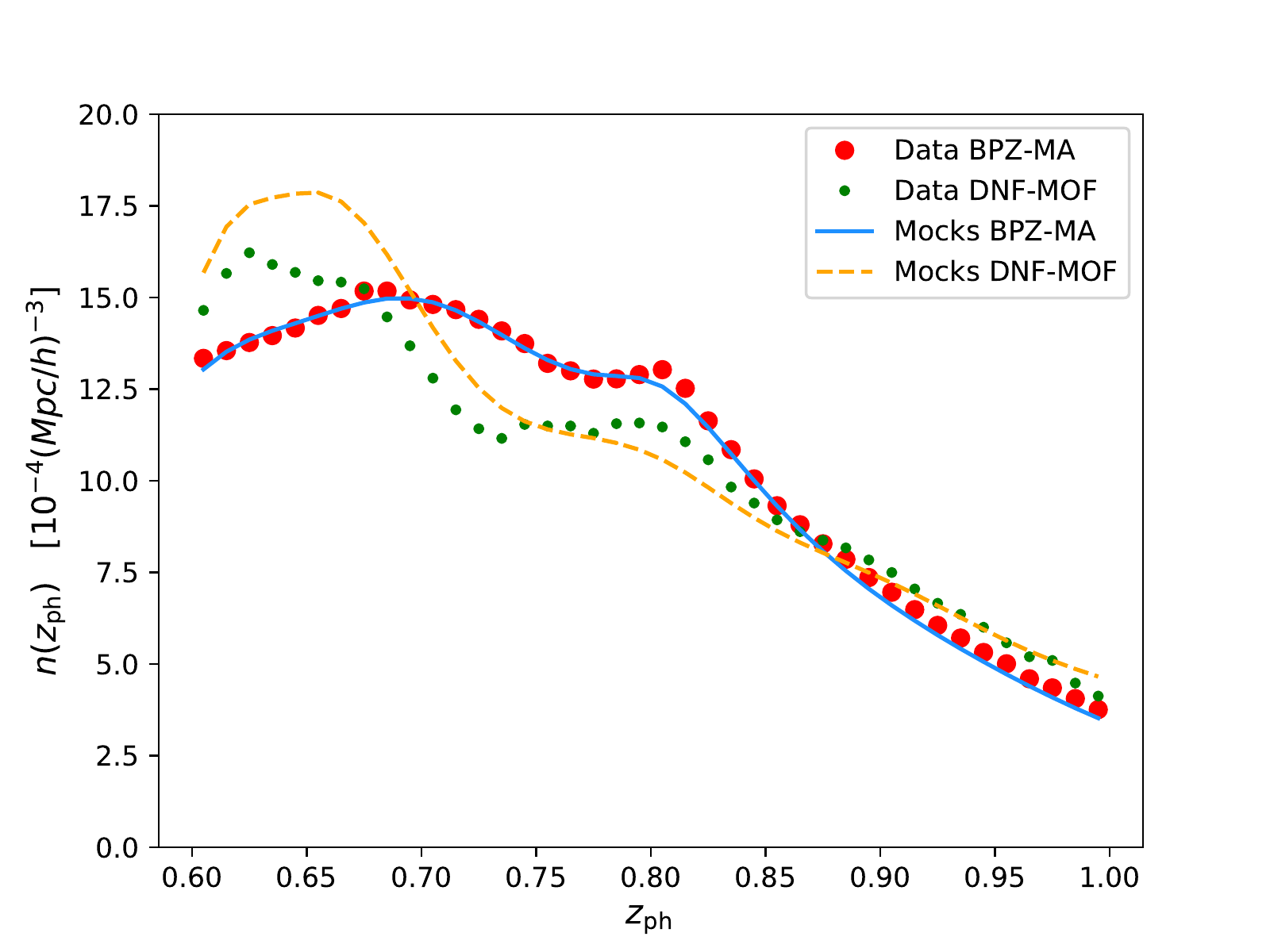}
  \caption{Number density of galaxies as a function of true redshift $z_{\rm sp}$ (Top) and photometric redshift $z_{\rm ph}$ (Bottom). 
  We compare the results from the mocks (averaged over 1800 realisations) against their reference data, for the two different data sets. This is one of the key ingredients to ensure that the mocks will give us the correct covariance matrices.}
  \label{fig:nz}
\end{figure}

The amplitude $A(z_{\rm sp})$ in \Eq{eq:dbsk} gives us the abundance of objects by simply dividing by the survey area. 

\begin{equation}
\frac{dn_A(z_{\rm sp})}{dz}=\frac{1}{\Omega_{\rm sur}} A(z_{\rm sp})
\label{eq:nz}
\end{equation}
with $n_A$ being the number of objects per unit area. Note that $dn_A(z_{\rm sp})/dz$ contains the same information as the volume number density $n$, given that the cosmology is known.

At this stage,  we can set mass thresholds $M_{\rm th}(z_{\rm sp})$ to our halo catalogues to obtain the abundance specified by the amplitude $A(z_{\rm sp})$ of the fits and \Eq{eq:nz}. 
In \Sec{sec:hod} we will equivalently select galaxies by setting luminosity thresholds $l_{\rm th}(z_{\rm sp})$, but when looking at abundance and redshift distributions, this methodology yields the same results for halos or galaxies. 
Once we fix those thresholds and apply them to our mock catalogues, we can also apply the redshift uncertainty by Monte-Carlo sampling the $P(z_{\rm ph}|z_{\rm sp})$ distribution (from \Eq{eq:dbsk}). This give us the $z_{\rm ph}$ for each halo/galaxy, and then we select the halos/galaxies in the range $0.6 < z_{\rm ph} < 1.0$. Although the binning in $z_{\rm sp}$ was already small compared to the typical redshift uncertainties, we
interpolate the value of $\mu(z_{\rm sp})$ to carry the information for $z_{\rm sp}$ beyond the precision of $0.01$. 

The resulting catalogues have an abundance of halos/galaxies as shown in \Fig{fig:nz}. 
The abundance as a function of true redshift $n(z_{\rm sp})$ (top panel)  matches the data by construction, as we have set thresholds to force it to satisfy $n(z_{\rm sp})$ derived from \Eq{eq:nz}. 
The small differences found in $n(z_{\rm sp})$, come simply from the uncertainty in the fit to \Eq{eq:dbsk}. 
Note that the shape of $A(z_{\rm sp})$ and $n(z_{\rm sp})$ are similar, but still differ due to the cuts imposed in $z_{\rm ph}$. 
When analysing the abundance in $z_{\rm ph}$ space (bottom panel), we also find an overall good agreement, 
with some differences due to \Eq{eq:dbsk} not capturing completely the $\partial N/\partial z_{\rm ph}$ distributions from the data, this effect being more pronounced for the DNF-MOF data.

\begin{figure*}
  \centering
  \includegraphics[width=\linewidth]{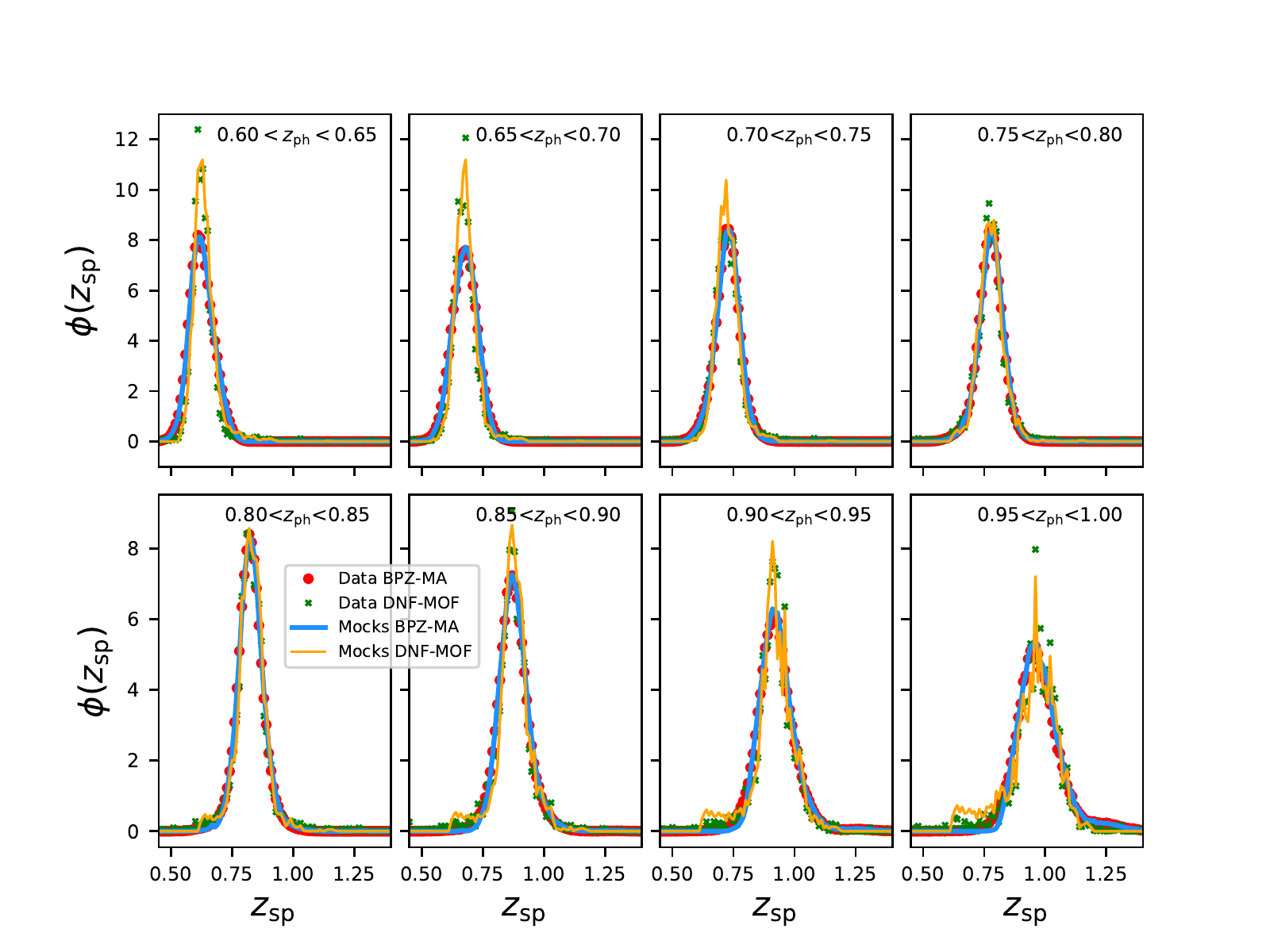}
  \caption{True redshift $z_{\rm sp}$ distribution in each of the 8 $z_{\rm ph}$ bins in the interval $0.6<z_{\rm ph}<1.0$ of data (points) versus mocks (line) for the two different datasets. Curves are normalised to have an integral of unity(\Eq{eq:phi}). This is one of the key ingredients to ensure that the mocks will give us the correct covariance matrices.  
  }
  \label{fig:zdist}
\end{figure*}

From the point of view of angular clustering analysis, the most relevant way of quantifying the photometric redshift uncertainty is to study the redshift distribution $\phi(z_{\rm sp})$ in $z_{\rm ph}$-bins $i$, which we define as:

\begin{equation}
\phi_i(z_{\rm sp})=\frac{1}{N^{\rm tot}_i}\frac{dN_i}{dz_{\rm sp}}
\label{eq:phi}
\end{equation}
with $N^{\rm tot}_i$ being the total number of galaxies in the redshift bin so that the integral of $\phi$ equals unity.

We show in \Fig{fig:zdist} the $z_{\rm sp}$ distribution for 8 equally spaced $z_{\rm ph}$-bins for both the data and mock catalogues. 
We find that data and mocks approximately agree in their $\phi_i(z_{\rm sp})$ for both data-sets.

In \photoz, we study the $\phi_i(z_{\rm sp})$ distributions of the DES Y1-BAO sample,
by comparing the results estimated from $z_{\rm mc}$ against the distribution directly observed with spectroscopic data from the COSMOS field (and corrected for sample variance). We found that results from BPZ-MA were biased and underestimated the redshift uncertainties, whereas DNF-MOF was found to give accurate $\phi(z_{\rm sp})$ distributions from the $z_{\rm mc}$ estimations. Hence, the BAO analysis in \main\ uses the DNF-MOF data set for the main results.
For that reason,
in the following sections, where we model and analyse the clustering of the mock catalogues, we will only show results for DNF-MOF (although similar clustering fits were obtained for BPZ-MA).

In order to quantify the level of agreement in the redshift uncertainties modelling, we repeated the BAO analysis performed in  \main\ (with DNF-MOF),  but assuming $\phi(z_{\rm sp})$ from the mocks instead of the $\phi(z_{\rm mc})$ from the data (the results for $z_{\rm mc}$ are denoted as ``$z$ uncal'' in Table 5 of that paper). We obtain the same best-fit value and uncertainty for the BAO scale and only $\Delta\chi^2/{\rm dof}=+1/43$, indicating that the accuracy achieved in $\phi(z_{\rm sp})$ is excellent for the purposes of our analysis.  
We will also show at the end of next section (\Fig{fig:bias}) the effect of the small differences in $\phi(z_{\rm sp})$ in the amplitude of the clustering, finding them small compared to the error bars. 

The functional form of the fits for $n(z_{\rm ph},z_{\rm sp})$ from the data presented here had originally been optimised for the shape of the $n(z_{\rm ph},z_{\rm sp})$ of BPZ-MA data. That should be taken into account when comparing the fits of both.
BPZ-MA appears to give smoother distributions easier to fit with smooth curves (\Eq{eq:dbsk}). However, given the poorer performance of the photo-z validation by BPZ-MA, we believe that the smoothness in the top panel of \Fig{fig:fit_shape} is not realistic, but rather an oversimplification. Galaxy spectra contain features that translates into structure in the redshift distribution when redshifts are estimated with photometric redshift codes using broad-band filters. These features are well captured by DNF $z_{\rm mc}$, as demonstrated by the lines in \Fig{fig:fit_shape}. These curves can also be approximately fitted with \Eq{eq:dbsk}, but with some small structures on top of the smooth curve. We leave the modelling of these structures for a future study: the level of agreement in $\phi(z_{\rm sp})$ (shown in \Fig{fig:zdist}) suggests that fitting any of the two data-sets with the \Eq{eq:dbsk} is a good approximation, and avoids the subtle problem of overfitting the data.


\section{Galaxy Clustering modelling} \label{sec:hod}

In \Sec{sec:halo} we described how to generate halo catalogues that sample the same volume as the DES Y1-BAO sample in the observational coordinates:  {$\{{\rm ra}$, ${\rm dec}$, $z_{\rm sp}\}$}.
In \Sec{sec:photoz} we described how we introduce the uncertainty in the estimation of redshift, leading to a catalogue that reproduces the abundance as a function of true and estimated redshift $\partial^2n/\partial z_{\rm ph}/\partial z_{\rm sp}$ of the data.
The next and last step, which is the focus of this section, is to produce a galaxy catalogue, able to reproduce the data clustering. For this, we will introduce hybrid HOD-HAM modelling. 

\paragraph*{Halo Abundance Matching (HAM).}
So far, all the clustering measurements shown throughout this paper were obtained from halo catalogues at a given mass threshold. 
But observed clustering is typically measured from galaxy catalogues with a magnitude-limited sample with the associated selection effects
and, more generally, with redshift-dependent colour and magnitude cuts. 

The basis of the HAM model is to assume that the most massive halo in a simulation would correspond to the most luminous galaxy in the observations and that we could do a one-to-one mapping in rank order. This is certainly very optimistic and realistic models need to add a scatter in the Luminosity-Mass relation ($L-M$) that will decrease the clustering for a magnitude-limited sample \citep{HAM_1,HAM_2,HAM_3,HAM_4,HOD_vs_HAM}.

\halogen\ was designed to only deal with main halos, neglecting subhalos. 
This limits the potential of HAM, as we can not use its natural extension to subhalos SHAM, where there is 
more freedom in the modelling by treating separately satellite and central galaxies (see e.g. \citealt{favole15}).

Nevertheless, substructure can be easily added to a main halo catalogue using a Halo Occupation Distribution technique.  

\paragraph*{Halo Occupation Distribution (HOD).}
We know that halos can host more than one galaxy, especially massive halos which match galaxy clusters. If we attribute a 
number of galaxies $N_{\rm gal}$ that is an increasing function of the halo mass ($M_h$) to a halo mock catalogue, the clustering will be enhanced,
since massive halos will be over-represented (as occurs in reality for a magnitude-limited sample).
This is the basis of the HOD methods \citep{HOD_0,HOD_1,HOD_2,HOD_3,HOD_4,HOD_zehavi,MICE_HOD,HOD_sergio,Skibba2009}. \\

The details of the HOD and HAM need to be matched to observations via parameter fitting. This process can be particularly difficult 
if one aims at having a general model that serves
for any sample with any magnitude and colour cut at any redshift (e.g. \citealt{MICE_HOD}), with the added difficulty in our case that the redshift uncertainties would also vary with colour and magnitude. 
Additionally, the HOD implementation will determine the small scale clustering
corresponding to the correlation between galaxies of the same halo \citep{halomodel}. However, this is beyond the scope of this paper
and we will only aim to match the large scale clustering of the Y1-BAO sample. 

In this paper we combine the two processes: first we add substructure with a HOD model, and in a later step we select the galaxies that enter into our sample following a HAM prescription.

With regards to the HOD model we assign to each halo one central galaxy

\begin{equation}
\label{eq:hod}
N_{\rm cent} = 1 \ ,
\end{equation}
 and $N_{\rm sat}$ satellite galaxies given by a Poisson distribution with mean

\begin{equation}
\label{eq:hod}
\langle N_{\rm sat} \rangle = \frac{M_h}{M_1}\ ,
\end{equation}
where $M_h$ is the mass of the halo, and $M_1$ is a free HOD parameter.
Note that $N_{\rm cen}$ and $N_{\rm sat}$ above do not correspond to the final occupation distribution of the halos, after applying the HAM (selecting only a subsample of these galaxies) in the later step. 

\begin{figure*}
  \centering
  \includegraphics[height=0.5\linewidth,angle=270]{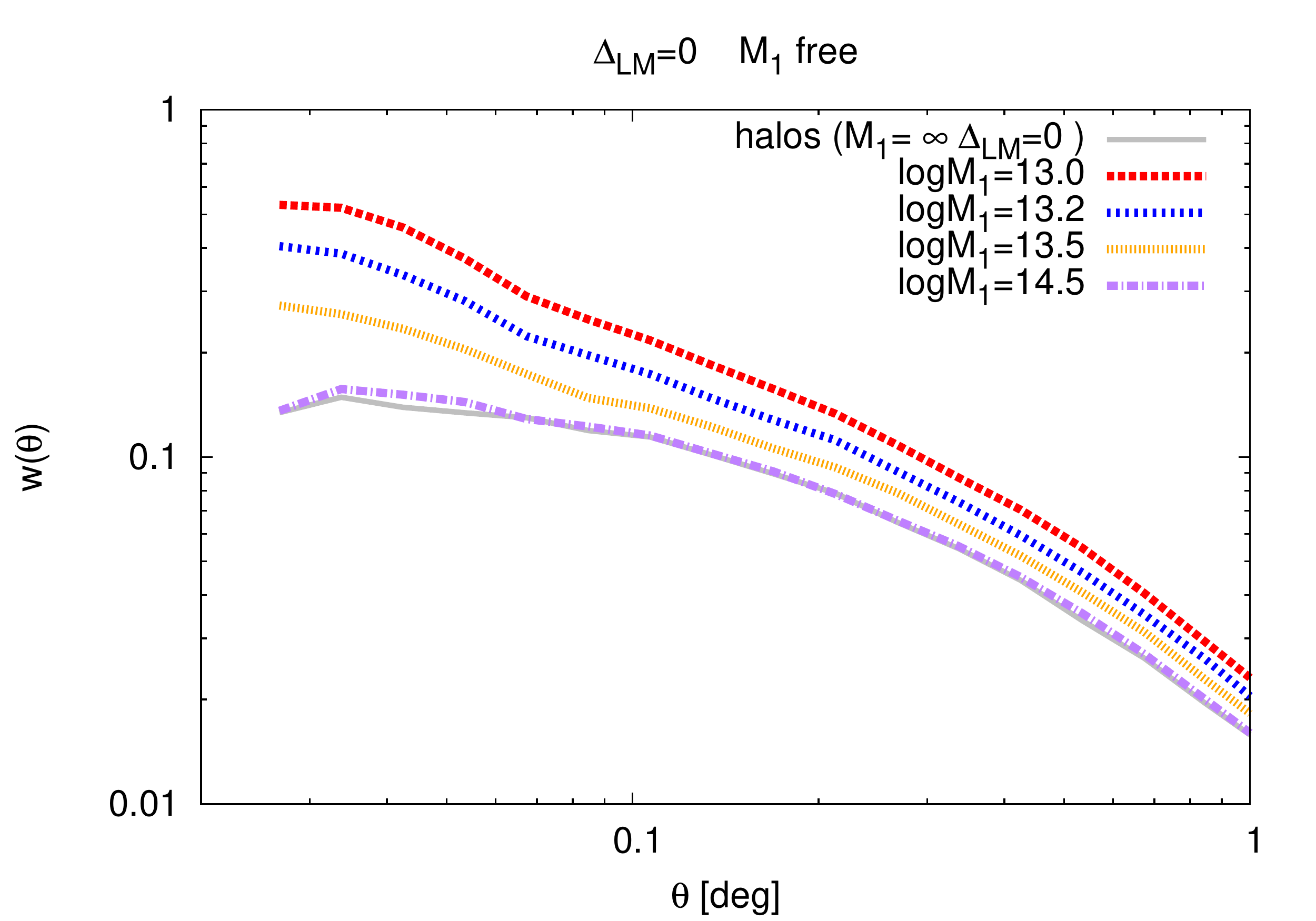}\includegraphics[height=0.5\linewidth,angle=270]{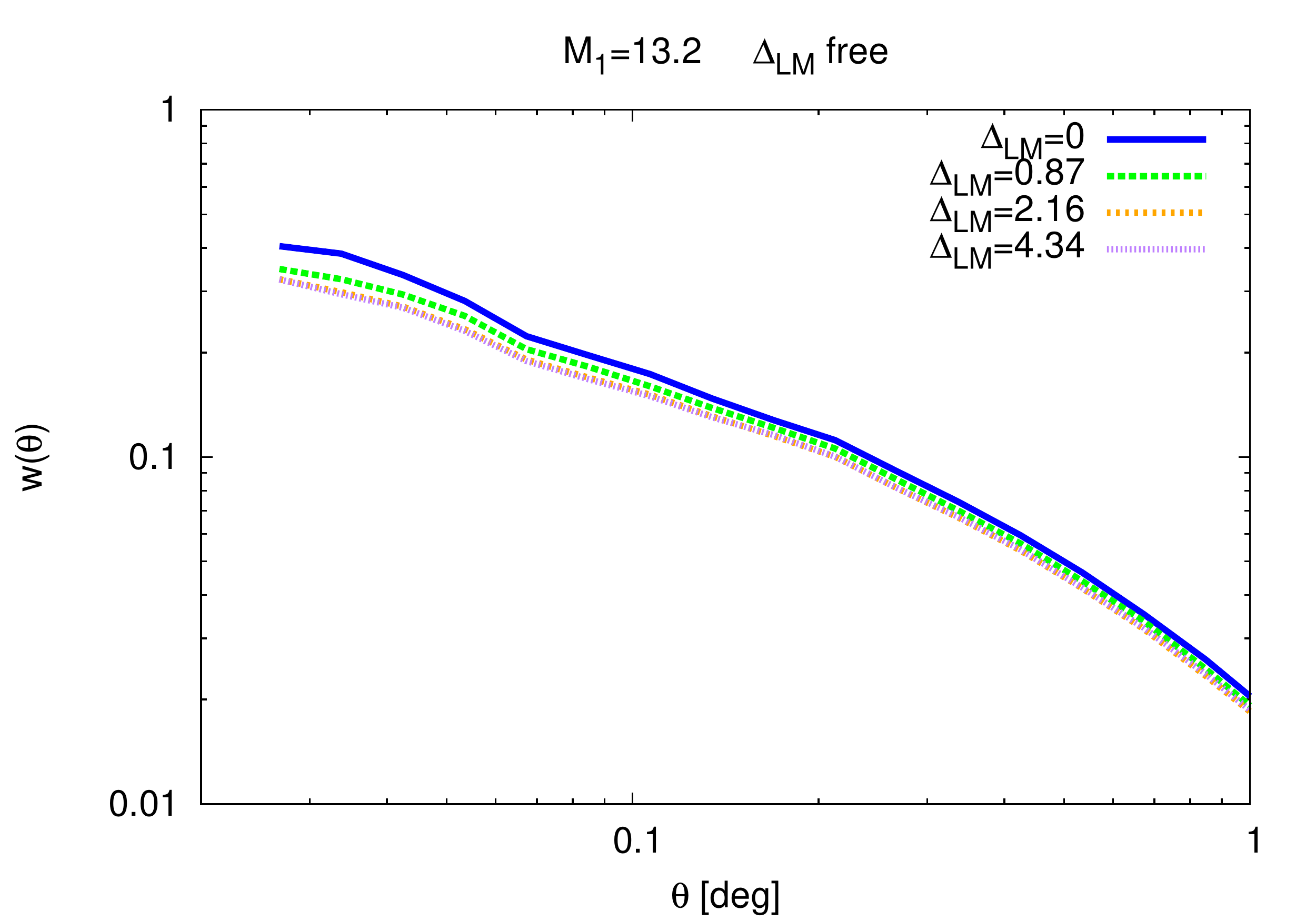}
  \caption{Effect on angular clustering in the redshift bin $0.6<z_{\rm ph}<0.65$ of the HOD parameters at fixed abundance and redshift distribution $n(z_{\rm sp},z_{\rm ph})$. Left: effect produced by $M_1$. Right: effect produced by $\Delta_{\rm LM}$}
  \label{fig:HOD}   
\end{figure*}

Central galaxies are placed at the center of the halo, whereas satellite galaxies are placed following a NFW \citep{NFW} profile.
The concentration (a parameter of the NFW profile) is determined from the mass by the mass-concentration relation given in \citet{concentration}. 
The velocities of the central galaxies are taken from the host halo, whereas the velocities of the satellite have an additional dispersion:

\begin{equation}
\begin{aligned}
 & v_{\rm sat\ i}=v_{\rm halo\ i} + \frac{1}{\sqrt{3}}\sigma_{v}(M_h)\cdot R^{\rm gauss}_{\mu=0\ \sigma=1} \ , \\
 & i=x,y,z\ ,
 \end{aligned}
\end{equation}
where $R^{\rm gauss}_{\mu=0\ \sigma=1}$ is a random number drawn from a Gaussian distribution with mean $\mu=0$ and standard deviation $\sigma=1$, and
$\sigma_{v}$ is the dispersion expected from the virial theorem:
\begin{equation}
\sigma_{v}(M_h)=	\sqrt{\frac{1}{5}\frac{GM_h}{R_{\rm vir}}} \ .
\end{equation}

In this case we use $R_{\rm vir}=R_{\rm 200,\ crit}$, since the mass definition used $M_h$ is that included inside this radius.

Following the concepts of Halo Abundance Matching, we assign a pseudo-luminosity $l_p$ 
to the galaxies, which represents the entire set of selection criteria (in our case \Eq{eq:sample}) compressed into a 1-dimensional criterion. This $l_p$ is not intended to represent a realistic luminosity. It is used to set the thresholds to match the abundance of the data in a simple way while accounting not only for the intrinsic Luminosity-Mass scatter, but also for the incompleteness of the sample. 
We model $l_p$ (in arbitrary scales) with a Gaussian scatter around the halo mass $M_h$ in logarithmic scales:

\begin{equation}
{\rm log}_{10}(l_p) = {\rm log}_{10}(M_h) + \Delta_{LM}\cdot R^{\rm gauss}_{\mu=0\ \sigma=1}
\label{eq:ham}
\end{equation}
where $\Delta_{LM}$ is a free parameter of the HAM model that controls the amount of scatter.

The abundance is then fixed by setting luminosity 
thresholds $l_p^{\rm th}(z_{\rm sp})$ that give us $n(z_{\rm sp})$ matching \Eq{eq:nz}. Note that $l_p^{\rm th}(z_{\rm sp})$ is implicitly another HOD parameter, since it will depend on the other two parameters ($M_1$ and $\Delta_{LM}$), but is not let free as it is defined by construction to match the abundance. 
Once these  HOD-HAM steps are complete, we generate a photometric redshift for each galaxy as explained in \Sec{sec:photoz}.

In \Fig{fig:HOD} we demonstrate the influence on clustering, under the assumption of fixed abundance and redshift distribution, of the 2 HOD parameters that we have introduced: $M_1$ and  $\Delta_{LM}$. We do so by studying the angular correlation function.

In order to create a galaxy catalogue identical to the halo catalogue, we would implement $M_1=\infty$, $\Delta_{LM}=0$. Deviations from those parameters control the clustering, as follows (\Fig{fig:HOD}): 

\begin{itemize}
\item $M_1$: By lowering this parameter, we oversample the most massive halos increasing the linear bias. It also introduces a 1-halo term that fades away as we increase $M_1$.

\item $\Delta_{LM}$: As we increase this value, lower mass halos enter into our selection and higher mass halos escape it, lowering the bias.

\end{itemize}

\begin{figure}
  \centering
   \includegraphics[height=1\linewidth,angle=270]{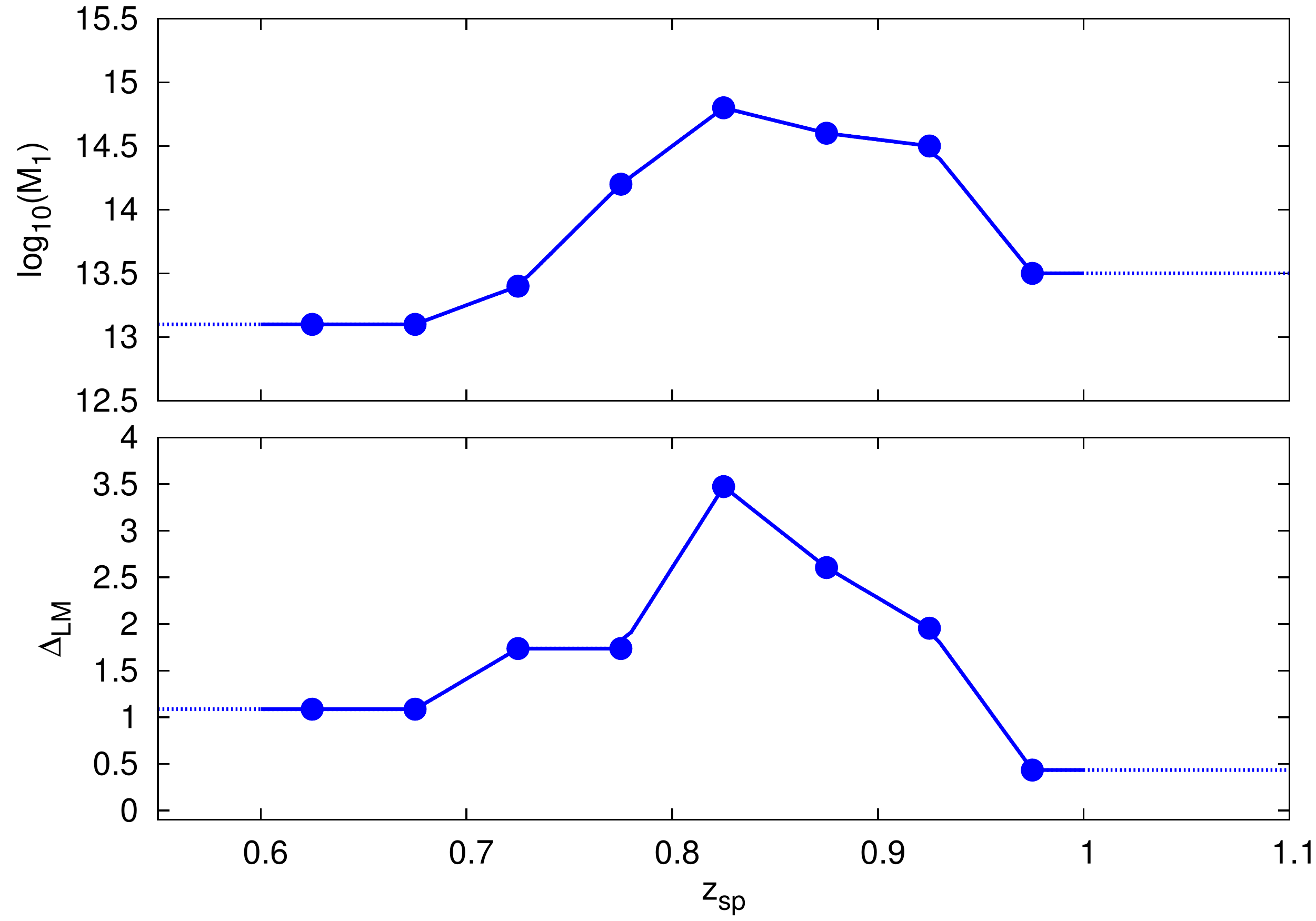}
  \caption{HOD parameter evolution with redshift $z_{\rm sp}$. Top: evolution of ${\rm log}M_1$ (in $M_{\sun}/h$). Bottom: evolution of $\Delta_{LM}$. 
  The evolution of the parameters is assumed smoothed, and they are linearly interpolating between the 8 pivots: the solid circles.
  Although the focus of the modelling and analysis it is the range $0.6<z<1$, there are contributions of galaxies with $z_{\rm sp}$ ranging from $0.45$ to $1.4$
  A flat evolution of the HOD parameters is assumed before and after the first and last pivot, respectively.}
  \label{fig:params}
\end{figure}

\begin{figure}
  \centering
  \includegraphics[width=1.0\linewidth]{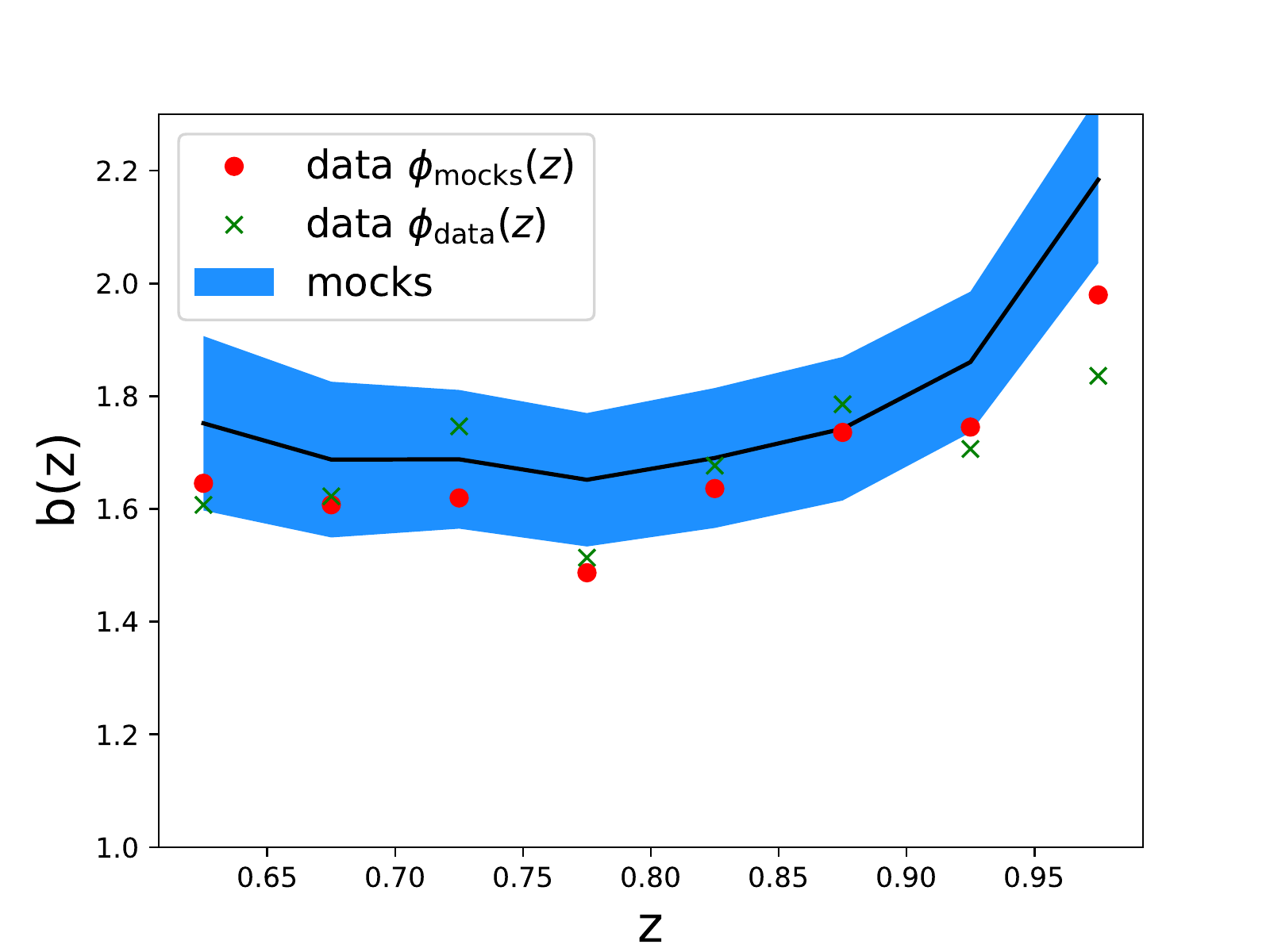}
  \caption{Bias evolution with redshift as measured from the mocks (shaded region representing the mean and standard deviation) and data. For all cases we fit the bias for the angular correlation in $z_{\rm ph}$-bins of width $\Delta z=0.05$. For the mocks we assume their own $\phi(z_{\rm sp})$ distribution when fitting $b$, whereas for the data we test the difference between assuming $\phi_{\rm mocks}(z_{\rm sp})$ or $\phi_{\rm data}(z_{\rm sp})$. At this stage we only consider the diagonal part of the covariance.}
  \label{fig:bias}
\end{figure}

We can find in the literature more complex $N_{\rm sat}(M_h)$ functions
but to  avoid huge degeneracies in the parameters, 
we choose to only have 2 free parameters $M_1$ and $\Delta_{LM}$. We do let them evolve with redshift in order to adapt for the clustering of the data, including selection effects.

We impose a smooth evolution of these parameters with redshift by making linear interpolations between 8 pivots 
at the center of each redshift bin,
and making the evolution flat beyond the center of the first and last redshift bin, since our measurements (in eight $z_{\rm ph}$-bins in $0.6<z_{\rm ph}<1.0$) will not be able to constrain the parameter evolution beyond these points. 
With these constraints, we find that a HOD parameter evolution as shown in \Fig{fig:params}, gives us a good match to the evolution of the amplitude of the clustering of the data. 
We have only explored large scales, since  we focus on BAO physics. This leaves some degeneracy 
between the two HOD parameters, which could be broken by including the small scales. We leave these studies for the next internal data release, which will include data from the first three years of the survey. 

Once the HOD parameters are fixed, we measure the bias of the mock catalogues by fitting the angular correlation to theoretical predictions (see how we compute the theoretical $w(\theta)$ in \Sec{sec:ang}). We fit correlations at angles that correspond to $20<r<60\ \mpch$ scales, different for each redshift bin. These measurements are meant to be a simple verification of the fit on the amplitude of the clustering, and hence we assume a simple diagonal covariance, measured from the mocks (unlike we will do in \Sec{sec:ang}). 
In \Fig{fig:bias} we show the bias evolution recovered from the mock catalogues, with the blue band representing the 1-$\sigma$ region computed as the standard deviation of best fit of each mock. 

We also present in \Fig{fig:bias} the bias measured from the data following the same procedure. In red we show the measured bias if we assume the same redshift distribution $\phi_i(z_{\rm sp})$ as in the mocks, whereas in green we show the measured bias assuming the $\phi_i(z_{\rm sp})$ estimated from the data itself.  
The main conclusion of this plot is that the amplitude of the clustering of the data (red circles) agrees within 1-$\sigma$ with the mocks. We notice a trend of the mocks having a slightly higher amplitude than the data, this effect is small compared to the uncertainties, but will be studied in more detail for future releases. 
This plot also shows us that the small differences between $\phi_{\rm mocks}$ and $\phi_{\rm data}$ in \Fig{fig:zdist} do not have a significant effect in the clustering, since all shifts are smaller than the $\sigma$ (and most of them being much smaller).

Looking at the mean halo mass of the catalogues we find that our Y1-BAO sample probes halo masses between 
$1.3 \times 10^{13} M_{\odot}/h$ and $2.1 \times 10^{13} M_{\odot}/h$ depending on the redshift range.

\section{Clustering Results} \label{sec:results}

In previous sections we designed a method to reproduce all the relevant properties of the Y1-BAO sample for clustering analysis. In this section we fix all the modelling and parameters and run $N_{\rm mocks}=1800$ realisations with different initial conditions. We analyse the clustering of all those mocks using different estimators in different spaces ($w_i(\theta)$, $\xi(r)$, $C_{l}$)), and their covariance matrices. We also compare those statistics with theoretical models, and measurements from the data.

\subsection{Theoretical modelling}
\label{sec:theory}

In this section we compare the clustering of the mock catalogues to several theoretical predictions. 
The baseline model we use is the linear theory with non-linear BAO damping, a Kaiser factor \citep{Kaiser} and linear bias: 

\begin{equation}
P(k,\mu) =  (b+\mu^2 f )^2 \big[ \big(P_{\rm lin}(k) -P_{\rm nw}(k)\big)e^{-k^2\Sigma^2_{\rm NL}} + P_{\rm nw}(k) \big].
\end{equation}

$P_{\rm lin}(k)$ is the linear power spectrum, and $P_{\rm nw}(k)$ it is a smoothed no BAO wiggle version of it obtained with the fitting formulas from  \citet{EnH}. We use the same linear power spectrum from \mice\ cosmology that we used for the 2LPT. The parameter $\Sigma_{\rm NL}$ represents the damping scale of the BAO due to non-linear evolution. We will discuss in Section \ref{sec:ang} \& \ref{sec:3D} its best-fit value. 

The details of the theoretical framework can be found in (\templates). But, going from $P(k,\mu)$ to
$\xi(r,\mu)$ can be summarised in three steps. First, decomposing the power spectrum in multipoles by convolving it with the Legendre Polynomial $L_l$: 

\begin{equation}
P_l(k) = \frac{2l+1}{2}\int_{-1}^1{\rm d}\mu P(k,\mu) L_l(\mu) .
\end{equation}

Second, Fourier transforming the multipoles using the Spherical Bessel functions $j_l$: 

\begin{equation}
\xi_l(s) = \frac{i^l}{2\pi ^2}\int {\rm d }k\ k^2 P_l(k)j_l (ks).
\end{equation}

And, finally, from the $\xi_l$ multipoles we can recover the full anisotropic correlation function: 

\begin{equation}
\xi(s,\mu)=\sum_l\xi_l (s) L_l(\mu) .
\label{eq:xismu}
\end{equation}

In the following sections we will see how to project $\xi(s,\mu)$ to obtain different estimators.

\subsection{Angular Clustering: $w(\theta)$}
\label{sec:ang}

\begin{figure*}
  \centering
   \includegraphics[height=1.0\linewidth, angle=270]{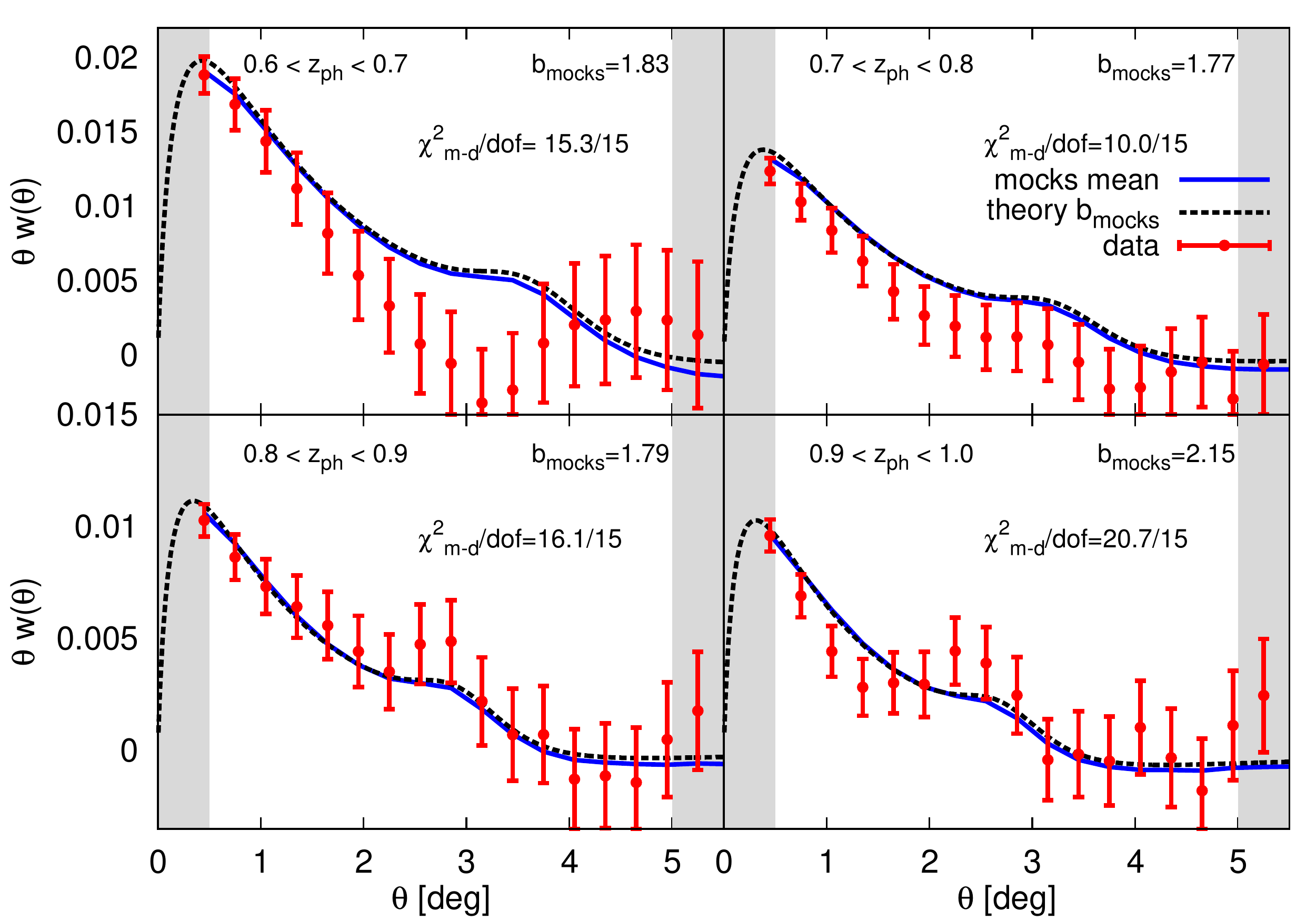}
  \caption{Angular clustering of mocks against data in four photometric redshift bins with width $\Delta z=0.1$. The red points with error bars represent the Y1-DES DNF-MOF data with the error bar computed as the standard deviation of the 1800 mocks. 
  The mean over the mock catalogues is represented by the blue solid line. We also show the theoretical curve that best fits the mocks. 
  We show the $\chi^2$ of the fit between the mocks and the data given by \Eq{eq:chi2} and the degrees of freedom (d.o.f., corresponding to the number of $\theta$-bins). Note that neighbouring $\theta$ bins are highly correlated. These $\chi^2$ show a good level of agreement in the clustering, this is one of the key ingredients to ensure that the mocks will give us the correct covariance matrices.
  }
  \label{fig:w4}
\end{figure*}

Here we study the angular correlation function of the final mock catalogues. We divide our catalogues in 4 redshift bins as we do in \main, and compute the correlation functions using the method given in \Sec{sec:cute}. We compare in \Fig{fig:w4} the mean correlation function $\bar{w}(\theta)$ of the mock catalogues
to the correlation from the data.  
The error bars attached to the data were computed as the square root of the diagonal of the covariance matrix, obtained from the mock catalogues with:

\begin{figure}
 \includegraphics[width=1\linewidth]{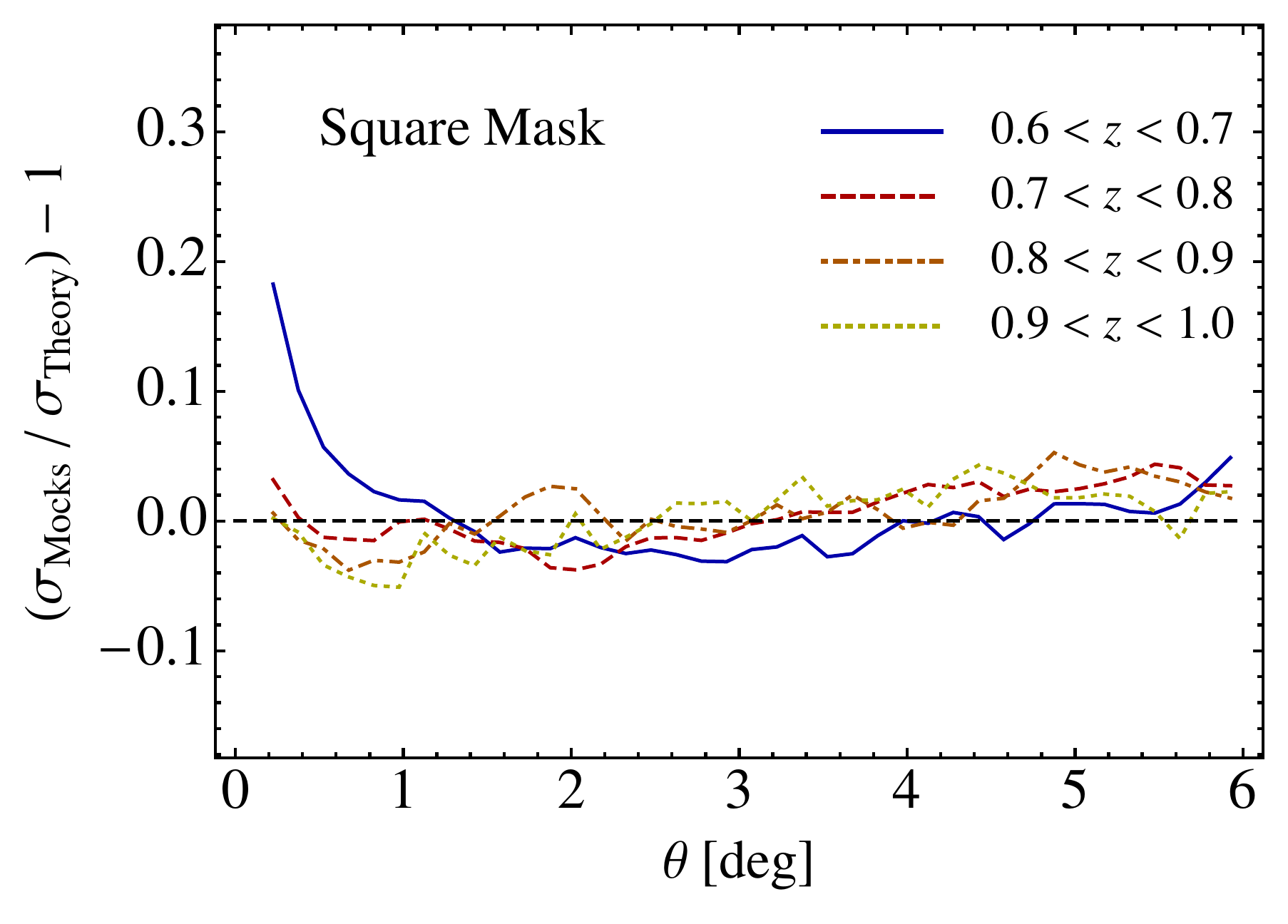}
 \includegraphics[width=1\linewidth]{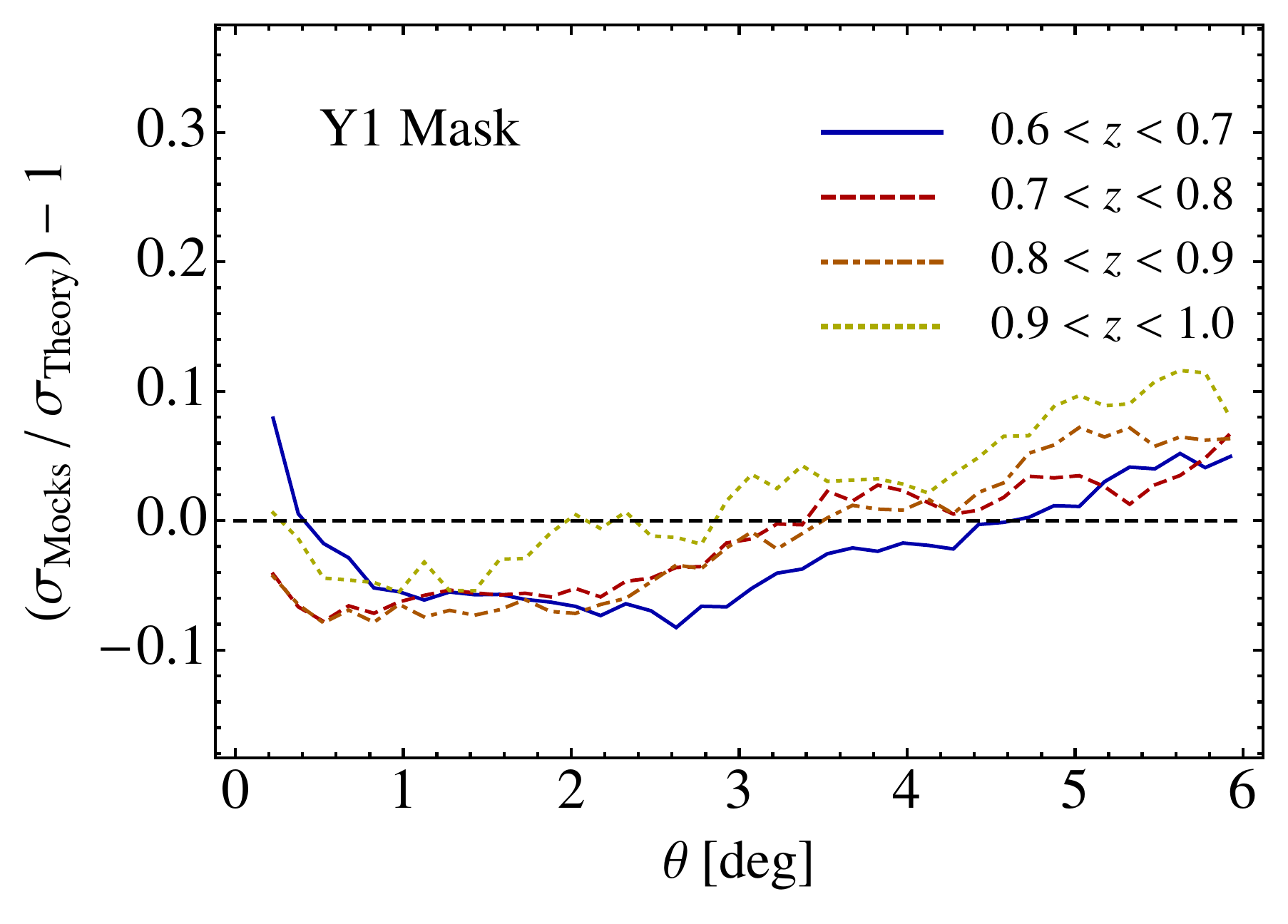}
 \includegraphics[width=1\linewidth]{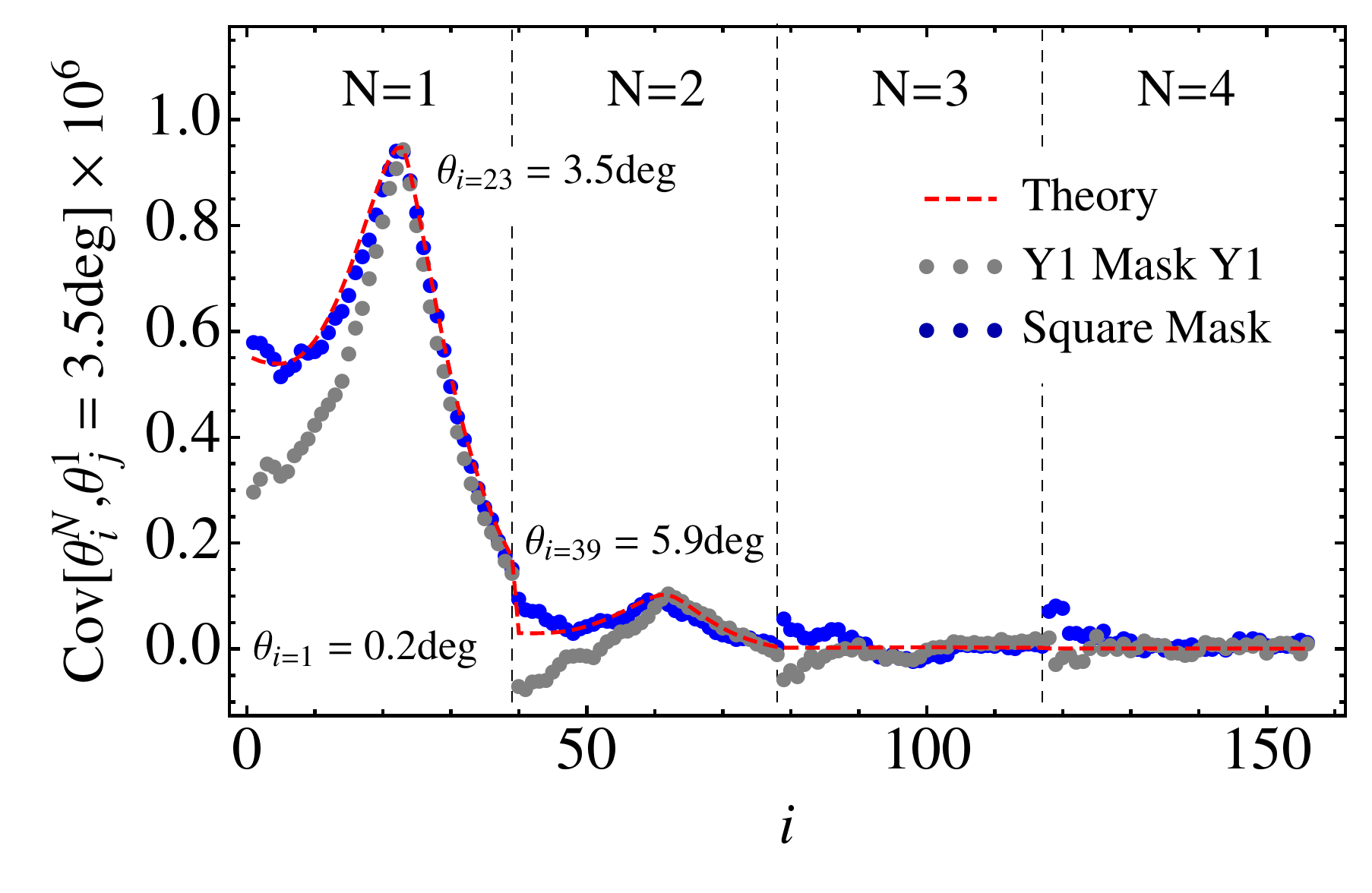}
 \caption{Covariance Matrices of the angular correlation function. Top: diagonal of the covariance matrices, theory vs `Square' mask mocks. Middle: diagonal of the covariance matrices, theory vs `Y1' mask mocks. Bottom: Column of the covariance matrix $C[\theta_i^N,\theta_j^M ]$ with fixed $\theta_j^1=3.5{\rm deg}$ as a function of $\theta_i^N$. The $i,j$ subscripts represent the index of the data vector (from 0 to 159, with $\theta$ ranging in $[0,6]{\rm deg}$ in $\Delta\theta=0.15{\rm deg}$ steps and 4 redshift bins), and the $N,M$ superscripts represent the index of the redshift bin. We compare theoretical predictions against Y1 mocks with `Square' and  `Y1' masks. In both the diagonal and off-diagonal cases the theoretical prediction lies closer to the `Square' mask mocks covariance.}
 \label{fig:cov}
\end{figure}

\begin{equation}
C_{ij} = \frac{1}{(N_{\rm mocks}-1)} \sum_{k=0}^{N_{\rm mocks}}  [\bar{w}(\theta_i) - w_k(\theta_i)]\cdot[\bar{w}(\theta_j) - w_k(\theta_j)].
\label{eq:cov}
\end{equation}

We compute the $\chi^2$ of the data with respect to the mean of the mocks using the covariance from the mocks using

\begin{equation}
\chi^2 = \sum_i \sum_j  [\bar{w}(\theta_i) - w_{\rm data}(\theta_i)]{C^{-1}}_{ij}[\bar{w}(\theta_j) - w_{\rm data}(\theta_j)],
\label{eq:chi2}
\end{equation}
and we display in \Fig{fig:w4} the goodness of the fit $\chi^2_{\rm red}=\chi^2/{d.o.f}$ between the mocks to the data for each $z_{\rm ph}$-bin. 
We find values near $1$, indicating a good fit to the data (with $p$-values of 0.430, 0.819, 0.375 and 0.147). Note the strong correlations between $\theta$ bins that move coherently up or down from realisation to realisation. This makes less intuitive the visual comparison (if one naively assumes a diagonal covariance matrix) between the curves to estimate the goodness of the fit: for example, the second bin has the best $\chi^2$, even if the data points appear to be systematically below the mocks line. 
We also bear in mind the cosmology from the mocks is not compatible with current cosmological constraints (e.g. \citealt{planck}), and this could introduce an extra $\chi^2$ contribution. However, the main difference due to cosmology would be the BAO position, and for the level of uncertainty that we have in a single $z_{\rm ph}$-bin (as opposed to combining the four, as we do in \main), this contribution is expected to be negligible.

Typically, we would need to correct the $\chi^2$ values by the factor in  
\citet{hartlap}, 
due to noise in the inverse of the covariance matrix  caused by having a finite number of mocks. 
We do include those factors, but find a negligible effect. 
This is because, for the results presented here, we do not consider the covariance between different $z_{\rm ph}$-bins, and, hence, our data vector is small compared to the large number of simulations used.

In \Fig{fig:w4} we also show theoretical angular correlation functions fitted to the mocks. 
The theoretical $w(\theta)$ can be computed by projecting $\xi(s,\mu)$ from \Eq{eq:xismu}, and taking into account the $z_{\rm sp}$-distributions within the bins (\Eq{eq:phi}): 

\begin{equation}
w(\theta)=\int dz_{\rm sp} \phi(z_{\rm sp}) \int dz'_{\rm sp} \phi(z'_{\rm sp}) \xi(r(z_{\rm sp},z'_{\rm sp},\theta),\mu(z_{\rm sp},z'_{\rm sp},\theta)).
\label{eq:w}
\end{equation}

We denote $r(z_{\rm sp},z'_{\rm sp},\theta)$ as the comoving distance between two galaxies, respectively at redshift $z$ and $z'$, and separated by a projected angle on the 
sky of $\theta$, and $\mu(z_{\rm sp},z'_{\rm sp},\theta)$ the cosine of the orientation of the pair of galaxies with respect to the line of sight. 

In \Sec{sec:theory} we left two parameters free: $b$ and $\Sigma_{\rm NL}$. The choice of the damping factor is discussed in \templates: we use a constant $\Sigma_{\rm NL}=5.2 \mpch$, which fits well the correlation functions of the mocks and it is within the theoretical expectations \citep{SnE07}. 
The linear bias $b$ is then fit for each redshift bin to the mean of the mocks, obtaining the values  written in \Fig{fig:w4}. We over-plot the theoretical $w(\theta)$ that best fits the mock catalogues, this best fit should not be confused with the best fit to the data presented in \main.

We now study the form of the covariance matrix obtained from the mock catalogues (\Eq{eq:cov}) and compare it to the theoretical model discussed in \citet{theory} and \templates.
For the mock covariance,
we used two sets of mocks in which the only difference is the mask we apply to obtain 8 mocks from the full sky catalogue. The one we denote as `Y1',  depicted in \Fig{fig:mask}, has the same 
footprint as the data and it is the standard one used for all the analysis throughout this paper. The `Square' mocks\footnote{Note that these square mocks are not the same as the ones in \comov, which had a simpler modelling} have a continuous mask with 
$\Delta {\rm ra} = \Delta {\rm sin}({\rm dec})=\sqrt{\Omega_{\rm sur}}$. 

We find that the theoretical predictions for the diagonal terms agree very well with the results from the `Square' mask mocks except for the lowest scales ($\theta<1\degree$). When applying the `Y1' mask we find that a tilt is introduced in the diagonal of the covariance matrix $\sigma(\theta)$.
Studying the non-diagonal covariance, we find again a good agreement between the `Square' mocks and the theory. In this case the effect of introducing the `Y1' mask
is to modify significantly the covariance at small scales. A more detailed analysis on the covariance matrices and their effects for BAO is discussed in \templates.

We conclude that the theoretical method agrees very well with the covariances from the mock catalogues if the geometry of the survey is ignored. But also that the geometry of the survey has a significant contribution to the covariance.

\subsection{3D clustering: $\xi_{\mu<0.8}(s_\perp)$}
\label{sec:3D}

\begin{figure}
\includegraphics[height=1\linewidth, angle=270]{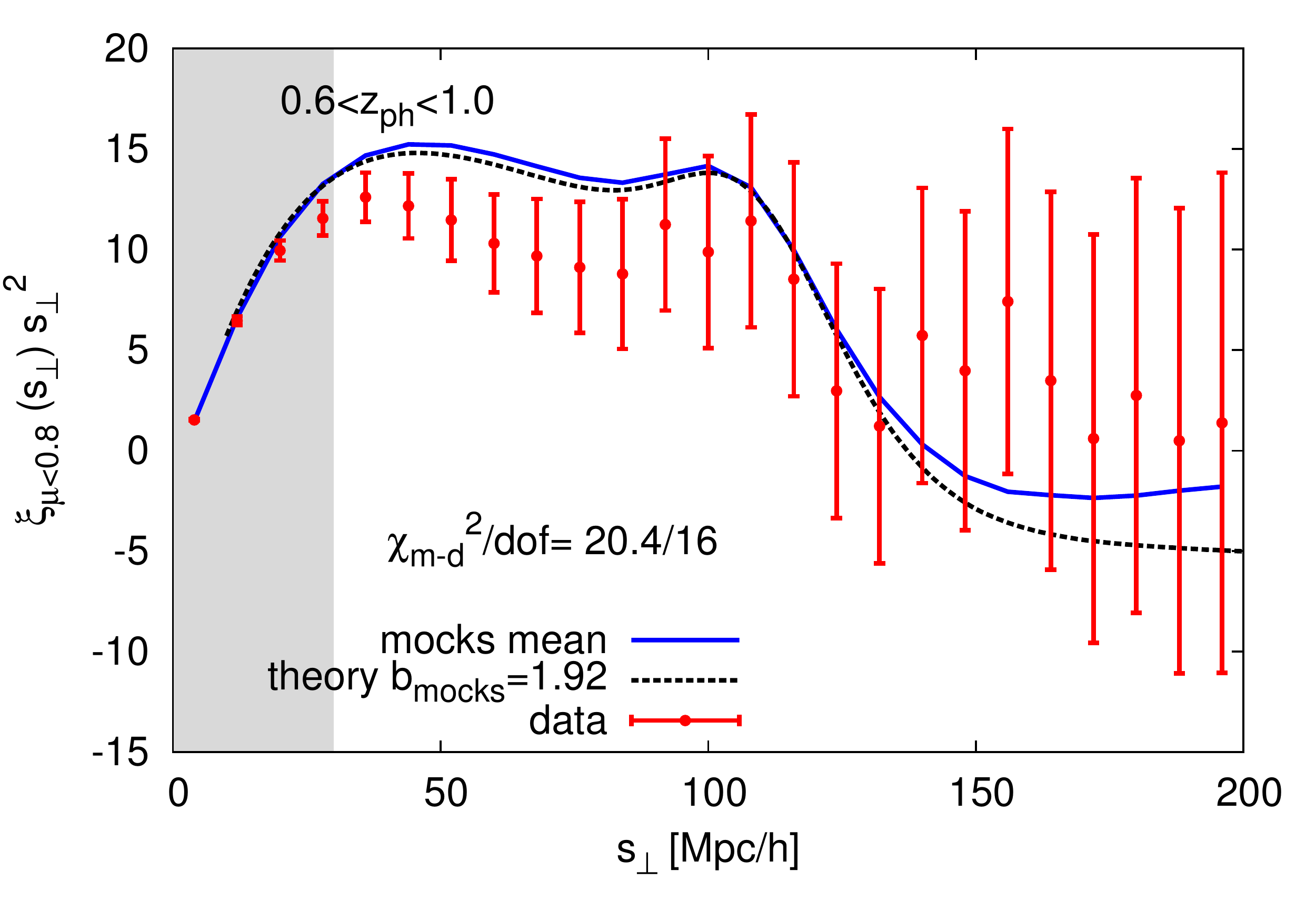}
\caption{3D correlation function using the estimator in \Eq{eq:3D}. The mean of the mocks is shown as a solid line, whereas the data is represented as points attached error bars coming from the diagonal of the mock covariance.}
\label{fig:3D}
\end{figure}

In photometric surveys most of the radial BAO information is lost due to redshift uncertainties. However, we showed in \comov, that also due to redshift uncertainties
the angular BAO information is spread to apparent $\mu>0$ modes. More specifically, most of the BAO information is spread nearly homogeneously in the range $0<\mu<0.8$. 
Hence, a nearly optimal way to study the BAO is analysing the following estimator (\comov): 

\begin{equation}
\xi_{\mu<0.8}(s_\perp) \equiv \frac{1}{0.8} \int_0^{0.8}d\mu\ \xi\big(\ s_\perp,\ s_\parallel(s_\perp,\mu)\ \big),
\label{eq:3D}
\end{equation}
being $s_\parallel$ ($s_\perp$) the projection of $s$ along (perpendicular to) the line of sight and 

\begin{equation}
s_\parallel (s_\perp,\mu) = s_\perp \sqrt{\frac{\mu^2}{1-\mu^2}}.
\end{equation}

Equivalently to \citet{FKP}, we apply inverse-covariance weighting to the galaxies to optimise the signal to noise of our measurements: 

\begin{equation}
w_{\rm FKP}(z) = \frac{b(z)D(z)}{1+n_{\rm eff}(z)P_{\rm lin}(k_{\rm eff},z=0)b^2(z)D^2(z)},
\end{equation}
where $n_{\rm eff}(z)$ is the effective number density accounting for the redshift uncertainty (see Eq.15 in \citealp{3DBAO}), and $k_{\rm eff}\approx0.12h/{\rm Mpc}$ the scale where the BAO informations is effectively coming from.

We compute the 3D correlation function $\xi(s_\perp,s_\parallel)$ of the data and mocks from pair-counts as explained in \Sec{sec:cute}, and integrating it using \Eq{eq:3D}. The results are shown in \Fig{fig:3D}, finding again a good agreement between mocks and data (with a $p$-value of 0.20). 
Again, the error-bars attached to the data come from the diagonal of the covariance from the mocks, and the $\chi^2$ were computed with the covariance from the mocks. 

From the theoretical point of view, we can project $\xi(s,\mu)$ from \Eq{eq:xismu} using \Eq{eq:3D}. For this we use a Gaussian uncertainty approximation, following the procedure in \comov. 
In this case we re-fit $\Sigma_{\rm NL}$, expecting this to capture part of the effect of the non-gaussian tails of the redshift uncertainties. We find a best fit value of $\Sigma_{\rm NL}=8 M{\rm pc}/h$. We compute the bias of the mocks $b_{\rm mocks}$ as the weighted average of the values obtained in the previous subsection (\Sec{sec:ang}). 
We plot the theoretical prediction for this bias, finding a good agreement that confirms the consistency of the mocks and theoretical framework we are working on.

\subsection{Clustering in Angular Harmonic space: $C_l$}
\label{sec:Cls}

In spectroscopic surveys, it has been shown that even though theoretically $\xi(r)$ and $P(k)$ carry the same information in different spaces, in reality, when dealing with finite volume, they can be provide complementary information \citep{BAO6}. 
In photometric surveys one usually deals with projected quantities. In this case, the complementary observables are the angular correlation $w(\theta)$ and its equivalent in harmonic space, the angular power spectrum $C_\ell$.

The galaxy number density contrast in a given redshift bin  $\delta_{{\rm gal}, i}(\hat{\bf n})$ can be decomposed into spherical harmonics $Y_{\ell m}$
as
\begin{equation}
\delta_{{\rm gal}, i} (\hat{\bf n}) = \sum_{\ell=0}^\infty \sum_{m=-\ell}^{\ell} a_{\ell m, i} Y_{\ell m}(\hat{\bf n})\,,
\label{eq:decomp}
\end{equation}
where $a_{\ell m}$ are the harmonic coefficients. The angular power spectrum $C_{\ell, i}$ is then defined via
\begin{equation}
\langle a_{\ell m, i} a_{\ell' m', i}^* \rangle \equiv \delta_{\ell \ell'}\delta_{mm'} C_{\ell, i}\,.
\label{eq:Cl}
\end{equation}

For data collected over the whole sky, an unbiased estimator of the angular power spectrum is simply the average of 
the $a_{\ell m}$ coefficients over all $m$ values:
\begin{equation}
\hat{C}_\ell = \frac{1}{2\ell+1} \sum_{m=-\ell}^{m=\ell} |a_{\ell m}|^2\,.
\end{equation}

When performing full-sky estimations, we compute the coefficients $a_{\ell m}$ from the pixelized density contrast maps using the {\tt anafast} routine within HEALPiX.

In the case of partial sky coverage, the pseudo-$C_\ell$ method \citep{Hivon:2001jp} is used to measure the $C_\ell$s. The measurement is performed with a binning of $\Delta \ell = 20$ using a resolution of $N_{\rm side} = 1024$.
A more detailed description of the methodology is found in \Cls.

\begin{figure*}
\includegraphics[height=0.5\linewidth]{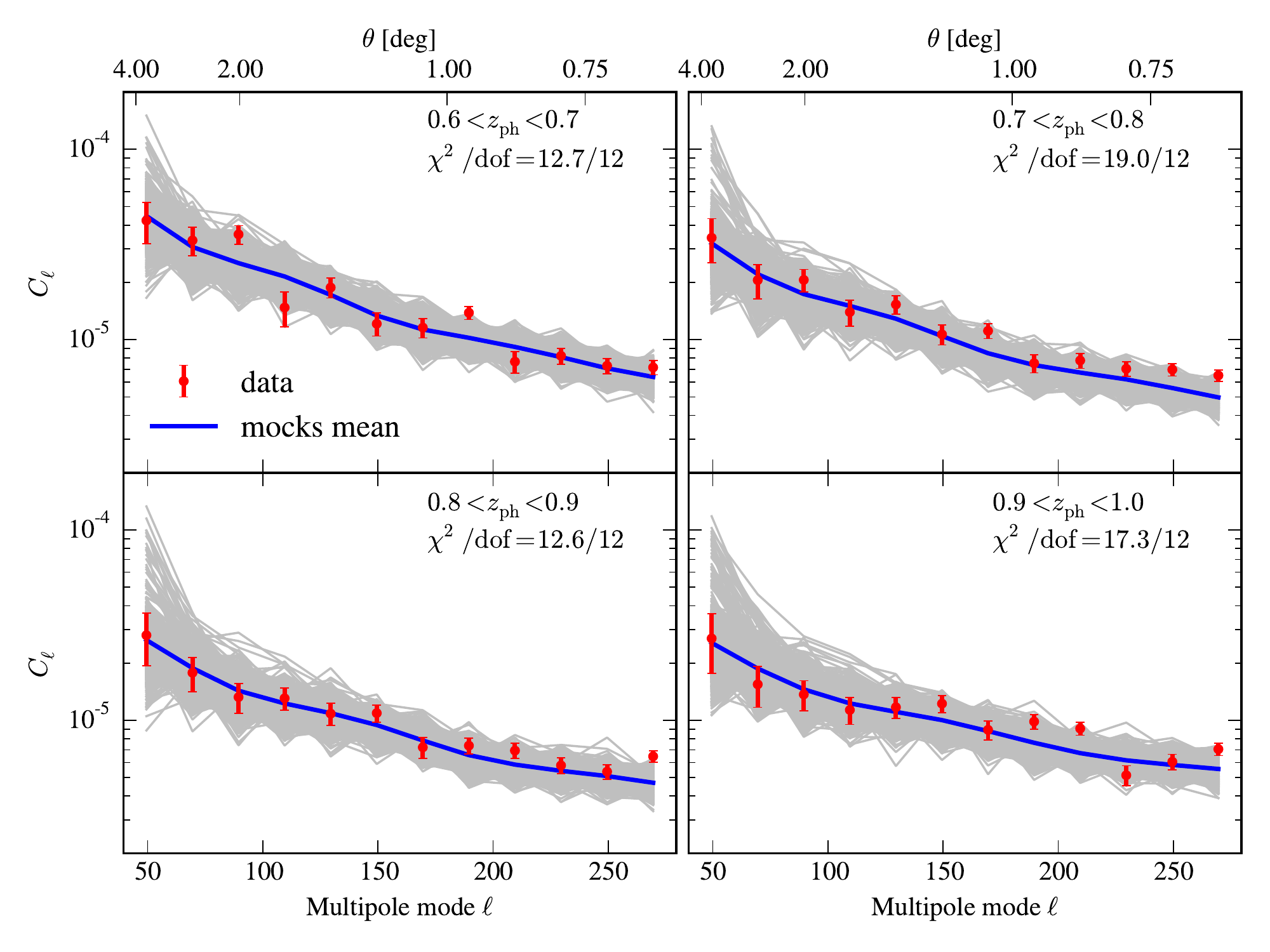}
\caption{Angular power spectrum for four $z_{\rm ph}$ bins for the 1800 mocks. In blue solid lines we show the mean of the mock catalogues, whereas the data are shown as points with error bars coming from the mock covariance. The results from all the individual realisation of the mocks are also shown in grey (forming nearly a shaded band). }
\label{fig:Cl}
\end{figure*}

Following the procedure of previous subsections, we compute the average and the covariance matrix for the angular power spectra in each redshift bin using the 1800 mocks. The measurement is also performed on Y1 data, with error bars estimated from the covariance matrix. The results in \Fig{fig:Cl} show that the angular power spectrum measured from the mocks are consistent with the measurements from data. For simplicity, we do not include the theoretical prediction here, but refer to \Cls, where we confirm that the bias values measured in \Sec{sec:ang} fit well the $C_\ell$ from the mocks.

\section{Summary and Conclusions}
\label{sec:conclusions}

In this paper we have designed and analysed a set of 1800 mock catalogues able to reproduce statistically the properties of the Dark Energy Survey Year-1 BAO sample. 

The three main properties reproduced are: 

\begin{itemize}
\item Sampled observational volume: $\{{\rm ra},{\rm dec},z_{\rm ph}\}$ (\Fig{fig:mask}).
\item Abundance of galaxies, redshift distribution and redshift uncertainty:  $n(z_{\rm ph},z_{\rm sp})$ (Figs. \ref{fig:nz}, \ref{fig:zdist}). 
\item Clustering as a function of redshift: $b(z_{\rm ph})$, $w_i(\theta)$, $\xi_{\mu<0.8}(s_\perp)$ and $C_l^i$ (Figs. \ref{fig:bias}, \ref{fig:w4}, \ref{fig:3D}, \ref{fig:Cl}).
\end{itemize}

Matching the properties listed above guarantees that our mock catalogues can correctly reproduce the clustering covariance of the data.

For the first time, we have presented a set of mock catalogues capable to reproduce simultaneously the clustering of a photometric sample together with an accurate description of redshift distribution and uncertainties. 

Throughout the paper we described in detail the way we design the galaxy catalogues, dividing the sections by the different processes/effects modelled (following the three bullet points above) and, in each section, building upon the modelling fixed in the sections before.

Once all the parameters and models have been fixed, catalogues are created following sequentially these steps: 

\begin{itemize}

\item Generate halo catalogues with \halogen\ at fixed redshifts in cubic volumes. (\Sec{sec:halogen})
\item Compose a full sky lightcone by superposing snapshots in redshift shells and using periodic conditions. (\Sec{sec:lightcone})
\item Add 1 central galaxy and $N_{\rm sat}$ satellite galaxies to each halo using \Eq{eq:hod}. (\Sec{sec:hod})
\item Assign a luminosity $l_p$ to each galaxy, based on the halo mass, using \Eq{eq:ham}. Then, select only the galaxies with $l_p>l_{\rm th}(z_{\rm sp})$  (\Sec{sec:hod}).
\item Draw a photometric redshift $z_{\rm ph}$ for each galaxy from the distribution $P(z_{\rm ph}|z_{\rm sp})$ in \Eq{eq:dbsk}. Then, select only galaxies in the range of interest $0.6<z_{\rm ph}<1$ (\Sec{sec:photoz}).
\item Apply the survey mask. (\Sec{sec:mask}) 

\end{itemize}

Some of the steps listed above involved
parameters that had to be adjusted to the data or to other simulations before starting producing the final batch of mock catalogues. The main fitted ingredients of the modelling are:

\begin{itemize}
\item \halogen\ parameters ($\alpha(M_h)$,$f_{\rm vel}(M_h)$) were tuned to reproduce the halo clustering
and velocity distributions of the reference $N$-Body simulation \mice\ as a function of mass and redshift. The input Halo Mass Function was also partially tuned to the simulation. 
\item We modelled $n(z_{\rm ph},z_{\rm sp})$ by fitting a \textit{double skewed} Gaussian to $\frac{\partial n}{\partial z_{\rm ph}}\Bigr|_{\substack{z_{\rm sp}}}$  (\Eq{eq:dbsk}) from the data for all $z_{\rm sp}$.
\item We explored the HOD parameter space $\{M_1(z_{\rm sp}),\Delta_{LM}(z_{\rm sp})\}$ to reproduce the bias evolution $b(z_{\rm sp})$ from the data. Simultaneously, the luminosity thresholds $l_{\rm th}(z_{\rm sp})$ need to be readjusted for each run in order to match the amplitude of $\frac{\partial n}{\partial z_{\rm ph}}\Bigr|_{\substack{z_{\rm sp}}}$.
\end{itemize}

This work is particularly relevant for the first BAO measurement with DES data presented in \main, where these mocks have been employed to run robustness tests, optimise the methodology, compute covariance matrices, and from them compute the best fit and uncertainty for angular scale of the BAO angular distance measurement $D_A$. 
It has also played key roles in other companion papers, e.g.: covariance for bias fits (\sample), study of methodology (\templates, \comov, \Cls) and photo-$z$ validation (\photoz).
This will set up a framework for the DES mock catalogue designs of the coming data releases, and can also valuable for many other photometric surveys studying the Large Scale Structure of the Universe.

\section*{Acknowledgements} \label{sec:Acknowledgements}

SA \& WJP acknowledge support from the UK Space Agency through grant ST/K00283X/1, and WJP acknowledges support from the European Research Council through grant {\it Darksurvey}, and the UK Science \& Technology Facilities Council through the consolidated grant ST/K0090X/1
SA thanks IFT computing facilities and support, that were necessary to run the simulations and fits.
SA thanks Alexander Knebe for the support during his PhD, when this project was started. 
NB acknowledges the use of University of Florida's supercomputer HiPerGator 2.0 as well as thanks the University of Florida's Research Computing staff. HC is supported by CNPq. ML is partially supported by FAPESP and CNPq. 
NK was financed by a FAPESP fellowship.
We thank the support of the Instituto Nacional de Ci\^encia
e Tecnologia (INCT) e-Universe (CNPq grant 465376/2014-2).

Funding for the DES Projects has been provided by the U.S. Department of Energy, the U.S. National Science Foundation, the Ministry of Science and Education of Spain, 
the Science and Technology Facilities Council of the United Kingdom, the Higher Education Funding Council for England, the National Center for Supercomputing 
Applications at the University of Illinois at Urbana-Champaign, the Kavli Institute of Cosmological Physics at the University of Chicago, 
the Center for Cosmology and Astro-Particle Physics at the Ohio State University,
the Mitchell Institute for Fundamental Physics and Astronomy at Texas A\&M University, Financiadora de Estudos e Projetos, 
Funda{\c c}{\~a}o Carlos Chagas Filho de Amparo {\`a} Pesquisa do Estado do Rio de Janeiro, Conselho Nacional de Desenvolvimento Cient{\'i}fico e Tecnol{\'o}gico and 
the Minist{\'e}rio da Ci{\^e}ncia, Tecnologia e Inova{\c c}{\~a}o, the Deutsche Forschungsgemeinschaft and the Collaborating Institutions in the Dark Energy Survey. 

The Collaborating Institutions are Argonne National Laboratory, the University of California at Santa Cruz, the University of Cambridge, Centro de Investigaciones Energ{\'e}ticas, 
Medioambientales y Tecnol{\'o}gicas-Madrid, the University of Chicago, University College London, the DES-Brazil Consortium, the University of Edinburgh, 
the Eidgen{\"o}ssische Technische Hochschule (ETH) Z{\"u}rich, 
Fermi National Accelerator Laboratory, the University of Illinois at Urbana-Champaign, the Institut de Ci{\`e}ncies de l'Espai (IEEC/CSIC), 
the Institut de F{\'i}sica d'Altes Energies, Lawrence Berkeley National Laboratory, the Ludwig-Maximilians Universit{\"a}t M{\"u}nchen and the associated Excellence Cluster Universe, 
the University of Michigan, the National Optical Astronomy Observatory, the University of Nottingham, The Ohio State University, the University of Pennsylvania, the University of Portsmouth, 
SLAC National Accelerator Laboratory, Stanford University, the University of Sussex, Texas A\&M University, and the OzDES Membership Consortium.

Based in part on observations at Cerro Tololo Inter-American Observatory, National Optical Astronomy Observatory, which is operated by the Association of 
Universities for Research in Astronomy (AURA) under a cooperative agreement with the National Science Foundation.

The DES data management system is supported by the National Science Foundation under Grant Numbers AST-1138766 and AST-1536171.
The DES participants from Spanish institutions are partially supported by MINECO under grants AYA2015-71825, ESP2015-66861, FPA2015-68048, SEV-2016-0588, SEV-2016-0597, and MDM-2015-0509, 
some of which include ERDF funds from the European Union. IFAE is partially funded by the CERCA program of the Generalitat de Catalunya.
Research leading to these results has received funding from the European Research
Council under the European Union's Seventh Framework Program (FP7/2007-2013) including ERC grant agreements 240672, 291329, and 306478.
We  acknowledge support from the Australian Research Council Centre of Excellence for All-sky Astrophysics (CAASTRO), through project number CE110001020.

This manuscript has been authored by Fermi Research Alliance, LLC under Contract No. DE-AC02-07CH11359 with the U.S. Department of Energy, Office of Science, Office of High Energy Physics. The United States Government retains and the publisher, by accepting the article for publication, acknowledges that the United States Government retains a non-exclusive, paid-up, irrevocable, world-wide license to publish or reproduce the published form of this manuscript, or allow others to do so, for United States Government purposes.


\bibliographystyle{mn2e} \label{sec:Bibliography}


\begin{thebibliography}{99} 

\bibitem[\protect\citeauthoryear{{DES Collaboration}}{{DES
  Collaboration}}{2017}]{Y1BAO}
{DES Collaboration}: {Alarcon} A., {Allam} S., {Annis} J., {Avila} S., Banerji {M}, Banik {N}, Bechtol {N}, G. M. Bernstein,
E. Bertin, D. Brooks, E. Buckley-Geer, D. L. Burke, H. Camacho, A. Carnero Rosell, et al. 2017, ArXiv:1712.06209

\bibitem[\protect\citeauthoryear{{Alonso}}{{Alonso}}{2012}]{cute}
{Alonso} D.,  2012, ArXiv:1210.1833

\bibitem[\protect\citeauthoryear{{Ata}, {Baumgarten}, {Bautista}, {Beutler},
  {Bizyaev}, {Blanton}, {Blazek}, {Bolton}, {Brinkmann}, {Brownstein},
  {Burtin}, {Chuang}, {Comparat}, {Dawson}, {de la Macorra} \& et al.}{{Ata}
  et~al.}{2017}]{BAO6}
{Ata} M.,  {Baumgarten} F.,  {Bautista} J.,  {Beutler} F.,  {Bizyaev} D.,
  {Blanton} M.~R.,  {Blazek} J.~A.,  {Bolton} A.~S.,  {Brinkmann} J.,
  {Brownstein} J.~R.,  {Burtin} E.,  {Chuang} C.-H.,  {Comparat} J.,  {Dawson}
  K.~S.,  {de la Macorra} A.,    et al. 2017, ArXiv:1705.06373

\bibitem[\protect\citeauthoryear{{Avila}, {Murray}, {Knebe}, {Power},
  {Robotham} \& {Garcia-Bellido}}{{Avila} et~al.}{2015}]{halogen}
{Avila} S.,  {Murray} S.~G.,  {Knebe} A.,  {Power} C.,  {Robotham} A.~S.~G.,
  {Garcia-Bellido} J.,  2015, \mnras, 450, 1856


\bibitem[\protect\citeauthoryear{}{{Bautista} et~al.}{2017}]{BAO7}
{Bautista} J.~E., {Vargas-Maga{\~n}a} M., {Dawson} K.~S.,
	{Percival} W.~J., {Brinkmann} J., {Brownstein} J., 
	{Camacho} B., {Comparat} J., {Gil-Mar{\'{\i}}n} H., 
	{Mueller} E.-M.,  {Newman} J.~A.,  {Prakash} A.,  {Ross} A.~J., 
	{Schneider} D.~P.,  {Seo} H.-J.,  {Tinker} J., {Tojeiro} R.,
	{Zhai} Z. and {Zhao} G.-B.,  arXiv:1712.08064

\bibitem[\protect\citeauthoryear{{Behroozi}, {Conroy} \& {Wechsler}}{{Behroozi}
  et~al.}{2010}]{HAM_2}
{Behroozi} P.~S.,  {Conroy} C.,    {Wechsler} R.~H.,  2010, \apj, 717, 379

\bibitem[\protect\citeauthoryear{{Ben{\'{\i}}tez}}{{Ben{\'{\i}}tez}}{2000}]{bpz1}
{Ben{\'{\i}}tez} N.,  2000, \apj, 536, 571

\bibitem[\protect\citeauthoryear{{Ben{\'{\i}}tez}, {Ford}, {Bouwens},
  {Menanteau}, {Blakeslee}, {Gronwall}, {Illingworth}, {Meurer}, {Broadhurst},
  {Clampin}, {Franx}, {Hartig}, {Magee}, {Sirianni}, {Ardila} \& et
  al.}{{Ben{\'{\i}}tez} et~al.}{2004}]{bpz2}
{Ben{\'{\i}}tez} N.,  {Ford} H.,  {Bouwens} R.,  {Menanteau} F.,  {Blakeslee}
  J.,  {Gronwall} C.,  {Illingworth} G.,  {Meurer} G.,  {Broadhurst} T.~J.,
  {Clampin} M.,  {Franx} M.,  {Hartig} G.~F.,  {Magee} D.,  {Sirianni} M.,
  {Ardila} D.~R.,    et al. 2004, \apjs, 150, 1

\bibitem[\protect\citeauthoryear{{Bertin} \& {Arnouts}}{{Bertin}
  \& {Arnouts}}{1996}]{MA}
{Bertin}, E. {Arnouts} S., 1996, \aap S, 117, 393

\bibitem[\protect\citeauthoryear{{Berlind} \& {Weinberg}}{{Berlind} \&
  {Weinberg}}{2002}]{HOD_2}
{Berlind} A.~A.,  {Weinberg} D.~H.,  2002, \apj, 575, 587

\bibitem[\protect\citeauthoryear{{Beutler}, {Blake}, {Colless}, {Jones},
  {Staveley-Smith}, {Campbell}, {Parker}, {Saunders} \& {Watson}}{{Beutler}
  et~al.}{2011}]{BAO3}
{Beutler} F.,  {Blake} C.,  {Colless} M.,  {Jones} D.~H.,  {Staveley-Smith} L.,
   {Campbell} L.,  {Parker} Q.,  {Saunders} W.,    {Watson} F.,  2011, \mnras,
  416, 3017

\bibitem[\protect\citeauthoryear{{Blake}, {Kazin}, {Beutler}, {Davis},
  {Parkinson}, {Brough}, {Colless}, {Contreras}, {Couch}, {Croom}, {Croton},
  {Drinkwater}, {Forster}, {Gilbank}, {Gladders} \& {Glazebrook}}{{Blake}
  et~al.}{2011}]{BAO4}
{Blake} C.,  {Kazin} E.~A.,  {Beutler} F.,  {Davis} T.~M.,  {Parkinson} D.,
  {Brough} S.,  {Colless} M.~.,  {Contreras} C.,  {Couch} W.,  {Croom} S.,
  {Croton} D.,  {Drinkwater} M.~J.,  {Forster} K.,  {Gilbank} D.,  {Gladders}
  M.,    {Glazebrook} et al.,  2011, \mnras, 418, 1707

\bibitem[\protect\citeauthoryear{{Bond}, {Cole}, {Efstathiou} \&
  {Kaiser}}{{Bond} et~al.}{1991}]{EPS}
{Bond} J.~R.,  {Cole} S.,  {Efstathiou} G.,    {Kaiser} N.,  1991, \apj, 379,
  440

\bibitem[\protect\citeauthoryear{{Bond} \& {Myers}}{{Bond} \&
  {Myers}}{1996}]{peakpatch}
{Bond} J.~R.,  {Myers} S.~T.,  1996, \apjs, 103, 1

\bibitem[\protect\citeauthoryear{{Bouchet}, {Colombi}, {Hivon} \&
  {Juszkiewicz}}{{Bouchet} et~al.}{1995}]{bouchet}
{Bouchet} F.~R.,  {Colombi} S.,  {Hivon} E.,    {Juszkiewicz} R.,  1995, \aap,
  296, 575

\bibitem[\protect\citeauthoryear{{Camacho}, et al. in preparation}{{Camacho}
  et~al.}{in preparation}]{Cls}
{Camacho} H.,  {et al.} ,  in preparation

\bibitem[\protect\citeauthoryear{{Carretero}, {Castander}, {Gazta{\~n}aga},
  {Crocce} \& {Fosalba}}{{Carretero} et~al.}{2015}]{MICE_HOD}
{Carretero} J.,  {Castander} F.~J.,  {Gazta{\~n}aga} E.,  {Crocce} M.,
  {Fosalba} P.,  2015, \mnras, 447, 646

\bibitem[\protect\citeauthoryear{{Chan} et al.}{{Chan}
  et~al.}{2017}]{templates}
{Chan} K.~C., Crocce M., Ross A. J., Avila S., Elvin-Poole J., Manera M., Percival W. J., Rosenfeld R., et al., ArXiv:1801.04390


\bibitem[\protect\citeauthoryear{{Chuang}, {Kitaura}, {Prada}, {Zhao} \&
  {Yepes}}{{Chuang} et~al.}{2015a}]{mocks3}
{Chuang} C.-H.,  {Kitaura} F.-S.,  {Prada} F.,  {Zhao} C.,    {Yepes} G.,
  2015, \mnras, 446, 2621

\bibitem[\protect\citeauthoryear{{Chuang}, {Zhao}, {Prada}, {Munari}, {Avila},
  {Izard}, {Kitaura}, {Manera}, {Monaco}, {Murray}, {Knebe}, {Sc{\'o}ccola},
  {Yepes}, {Garcia-Bellido} \& {Mar{\'{\i}}n}}{{Chuang} et~al.}{2015b}]{nifty}
{Chuang} C.-H.,  {Zhao} C.,  {Prada} F.,  {Munari} E.,  {Avila} S.,  {Izard}
  A.,  {Kitaura} F.-S.,  {Manera} M.,  {Monaco} P.,  {Murray} S.,  {Knebe} A.,
  {Sc{\'o}ccola} C.~G.,  {Yepes} G.,  {Garcia-Bellido} J.,    {Mar{\'{\i}}n}
  F.~A. et al.,  2015, \mnras, 452, 686

\bibitem[\protect\citeauthoryear{{Coles} \& {Jones}}{{Coles} \& {Jones}}{1991}]{lognormal}
{Coles} P., {Jones}, B., 1991, \mnras, 248, 1

\bibitem[\protect\citeauthoryear{{Cole}, {Percival}, {Peacock}, {Norberg},
  {Baugh}, {Frenk}, {Baldry}, {Bland-Hawthorn}, {Bridges}, {Cannon}, {Colless},
  {Collins}, {Couch}, {Cross}, {Dalton} \& {Eke}}{{Cole} et~al.}{2005}]{BAO1}
{Cole} S.,  {Percival} W.~J.,  {Peacock} J.~A.,  {Norberg} P.,  {Baugh} C.~M.,
  {Frenk} C.~S.,  {Baldry} I.,  {Bland-Hawthorn} J.,  {Bridges} T.,  {Cannon}
  R.,  {Colless} M.,  {Collins} C.,  {Couch} W.,  {Cross} N.~J.~G.,  {Dalton}
  G.,    {Eke} et al.,  2005, \mnras, 362, 505

\bibitem[\protect\citeauthoryear{{Conroy}, {Wechsler} \& {Kravtsov}}{{Conroy}
  et~al.}{2006}]{HAM_1}
{Conroy} C.,  {Wechsler} R.~H.,    {Kravtsov} A.~V.,  2006, \apj, 647, 201

\bibitem[\protect\citeauthoryear{{Cooray} \& {Sheth}}{{Cooray} \&
  {Sheth}}{2002}]{halomodel}
{Cooray} A.,  {Sheth} R.,  2002, \physrep, 372, 1

\bibitem[\protect\citeauthoryear{{Crocce}, {Ross}, {Sevilla-Noarbe}, {Gazta{\~n}aga}, {Elvin-Poole}, {Avila}, {Alarcon}, {Chan}, {Banik}, {Carretero}, {Sanchez} \& et al.}{{Crocce}
  et~al.}{2017}]{sample}
{Crocce} M., {Ross} A. J., {Sevilla-Noarbe} I., {Gazta{\~n}aga} E., {Elvin-Poole} J., {Avila} S., {Alarcon} A., {Chan} K. C., {Banik} N., {Carretero} J., {Sanchez} E. et al. et al.  2017, ArXiv:1712.06211


\bibitem[\protect\citeauthoryear{{Crocce}, {Cabr{\'e}} \&
  {Gazta{\~n}aga}}{{Crocce} et~al.}{2011}]{theory}
{Crocce} M.,  {Cabr{\'e}} A.,    {Gazta{\~n}aga} E.,  2011, \mnras, 414, 329

\bibitem[\protect\citeauthoryear{{Crocce}, {Castander}, {Gazta{\~n}aga},
  {Fosalba} \& {Carretero}}{{Crocce} et~al.}{2015}]{mice2}
{Crocce} M.,  {Castander} F.~J.,  {Gazta{\~n}aga} E.,  {Fosalba} P.,
  {Carretero} J.,  2015, \mnras, 453, 1513

\bibitem[\protect\citeauthoryear{{De Vicente}, {S{\'a}nchez} \&
  {Sevilla-Noarbe}}{{De Vicente} et~al.}{2016}]{DNF}
{De Vicente} J.,  {S{\'a}nchez} E.,    {Sevilla-Noarbe} I.,  2016, \mnras, 459,
  3078

\bibitem[\protect\citeauthoryear{{Diehl}, {Abbott}, {Annis}, {Armstrong},
  {Baruah}, {Bermeo}, {Bernstein}, {Beynon}, {Bruderer}, {Buckley-Geer},
  {Campbell}, {Capozzi}, {Carter}, {Casas}, {Clerkin} \& et al.}{{Diehl}
  et~al.}{2014}]{survey}
{Diehl} H.~T.,  {Abbott} T.~M.~C.,  {Annis} J.,  {Armstrong} R.,  {Baruah} L.,
  {Bermeo} A.,  {Bernstein} G.,  {Beynon} E.,  {Bruderer} C.,  {Buckley-Geer}
  E.~J.,  {Campbell} H.,  {Capozzi} D.,  {Carter} M.,  {Casas} R.,  {Clerkin}
  L.,    et al. 2014, Proc. SPIE 9149, 91490V

\bibitem[\protect\citeauthoryear{{Dodelson} \& {Schneider}}{{Dodelson} \&
  {Schneider}}{2013}]{cov1}
{Dodelson} S.,  {Schneider} M.~D.,  2013, \prd, 88, 063537

\bibitem[\protect\citeauthoryear{{Drlica-Wagner}, {Sevilla-Noarbe}, {Rykoff},
  {Gruendl}, {Yanny}, {Tucker}, {Hoyle}, {Carnero Rosell}, {Bernstein},
  {Bechtol}, {Becker}, {Benoit-Levy} \& {Bertin}}{{Drlica-Wagner}
  et~al.}{2017}]{gold}
{Drlica-Wagner} A.,  {Sevilla-Noarbe} I.,  {Rykoff} E.~S.,  {Gruendl} R.~A.,
  {Yanny} B.,  {Tucker} D.~L.,  {Hoyle} B.,  {Carnero Rosell} A.,  {Bernstein}
  G.~M.,  {Bechtol} K.,  {Becker} M.~R.,  {Benoit-Levy} A.,    {Bertin} et al.,
  2017, ArXiv:1708.01531

\bibitem[\protect\citeauthoryear{{Eisenstein} \& {Hu}}{{Eisenstein} \&
  {Hu}}{1998}]{EnH}
{Eisenstein} D.~J.,  {Hu} W.,  1998, \apj, 496, 605

\bibitem[\protect\citeauthoryear{{Eisenstein}, {Zehavi}, {Hogg}, {Scoccimarro},
  {Blanton}, {Nichol}, {Scranton}, {Seo}, {Tegmark}, {Zheng}, {Anderson},
  {Annis}, {Bahcall}, {Brinkmann} \& {Burles}}{{Eisenstein}
  et~al.}{2005}]{BAO2}
{Eisenstein} D.~J.,  {Zehavi} I.,  {Hogg} D.~W.,  {Scoccimarro} R.,  {Blanton}
  M.~R.,  {Nichol} R.~C.,  {Scranton} R.,  {Seo} H.-J.,  {Tegmark} M.,  {Zheng}
  Z.,  {Anderson} S.~F.,  {Annis} J.,  {Bahcall} N.,  {Brinkmann} J.,
  {Burles} et al.,  2005, \apj, 633, 560

\bibitem[\protect\citeauthoryear{{Favole}, {Comparat}, {Prada}, {Yepes},
  {Jullo}, {Niemiec}, {Kneib}, {Rodr{\'{\i}}guez-Torres}, {Klypin}, {Skibba},
  {McBridƒe}, {Eisenstein}, {Schlegel}, {Nuza}, {Chuang}, {Delubac}, T. and {Y{\`e}che}, C. and {Schneider}, D.~P.}{{Favole}
  et~al.}{2015}]{favole15}
{Favole} G.,  {Comparat} J.,  {Prada} F.,  {Yepes} G.,  {Jullo} E.,  {Niemiec}
  A.,  {Kneib} J.-P.,  {Rodr{\'{\i}}guez-Torres} S.~A.,  {Klypin} A.,  {Skibba}
  R.~A.,  {McBridƒe} C.~K.,  {Eisenstein} D.~J.,  {Schlegel} D.~J.,  {Nuza}
  S.~E.,    {Chuang} C.-H., {Delubac}, T. and {Y{\`e}che}, C. and {Schneider}, D.~P. 2015,
  \mnras, 461, 3421

\bibitem[\protect\citeauthoryear{{Feldman}, {Kaiser} \& {Peacock}}{{Feldman}
  et~al.}{1994}]{FKP}
{Feldman} H.~A.,  {Kaiser} N.,    {Peacock} J.~A.,  1994, \apj, 426, 23

  
  \bibitem[\protect\citeauthoryear{{Fosalba}, {Crocce}, {Gazta{\~n}aga} \&
  {Castander}}{{Fosalba} et~al.}{2015a}]{mice1}
{Fosalba} P.,  {Crocce} M.,  {Gazta{\~n}aga} E.,    {Castander} F.~J.,  2015,
  \mnras, 448, 2987

\bibitem[\protect\citeauthoryear{{Fosalba}, {Gazta{\~n}aga}, {Castander} \&
  {Crocce}}{{Fosalba} et~al.}{2015b}]{mice3}
{Fosalba} P.,  {Gazta{\~n}aga} E.,  {Castander} F.~J.,    {Crocce} M.,  2015,
  \mnras, 447, 1319

\bibitem[\protect\citeauthoryear{{Fosalba}, {Gazta{\~n}aga}, {Castander} \&
  {Manera}}{{Fosalba} et~al.}{2008}]{mice0}
{Fosalba} P.,  {Gazta{\~n}aga} E.,  {Castander} F.~J.,    {Manera} M.,  2008,
  \mnras, 391, 435

\bibitem[\protect\citeauthoryear{{Friedrich} \& {Eifler}}{{Friedrich} \&
  {Eifler}}{2017}]{cov5}
{Friedrich} O.,  {Eifler} T.,  2018, \mnras, 473, 4150

\bibitem[\protect\citeauthoryear{{Gazta{\~n}aga}, et al. in preparation}{{Gazta{\~n}aga}
  et~al.}{in preparation}]{Cls}
{Gazta{\~n}aga} E.,  {et al.} ,  in preparation


\bibitem[\protect\citeauthoryear{{G{\'o}rski}, {Hivon}, {Banday}, {Wandelt},
  {Hansen}, {Reinecke} \& {Bartelmann}}{{G{\'o}rski} et~al.}{2005}]{healpix}
{G{\'o}rski} K.~M.,  {Hivon} E.,  {Banday} A.~J.,  {Wandelt} B.~D.,  {Hansen}
  F.~K.,  {Reinecke} M.,    {Bartelmann} M.,  2005, \apj, 622, 759

\bibitem[\protect\citeauthoryear{{Guo}, {Zehavi} \& {Zheng}}{{Guo}
  et~al.}{2012}]{HOD_4}
{Guo} H.,  {Zehavi} I.,    {Zheng} Z.,  2012, \apj, 756, 127

\bibitem[\protect\citeauthoryear{{Guo}, {Zheng}, {Behroozi}, {Zehavi},
  {Chuang}, {Comparat}, {Favole}, {Gottloeber}, {Klypin}, {Prada},
  {Rodr{\'{\i}}guez-Torres}, {Weinberg} \& {Yepes}}{{Guo}
  et~al.}{2016}]{HOD_vs_HAM}
{Guo} H.,  {Zheng} Z.,  {Behroozi} P.~S.,  {Zehavi} I.,  {Chuang} C.-H.,
  {Comparat} J.,  {Favole} G.,  {Gottloeber} S.,  {Klypin} A.,  {Prada} F.,
  {Rodr{\'{\i}}guez-Torres} S.~A.,  {Weinberg} D.~H.,    {Yepes} G.,  2016,
  \mnras

\bibitem[\protect\citeauthoryear{{Hartlap}, {Simon} \& {Schneider}}{{Hartlap}
  et~al.}{2007}]{hartlap}
{Hartlap} J.,  {Simon} P.,    {Schneider} P.,  2007, \aap, 464, 399

\bibitem[\protect\citeauthoryear{{Hinshaw}, {Nolta}, {Bennett}, {Bean},
  {Dor{\'e}}, {Greason}, {Halpern}, {Hill}, {Jarosik}, {Kogut}, {Komatsu},
  {Limon}, {Odegard}, {Meyer}, {Page}, {Peiris}, {Spergel} \&
  {Tucker}}{{Hinshaw} et~al.}{2007}]{wmap3}
{Hinshaw} G.,  {Nolta} M.~R.,  {Bennett} C.~L.,  {Bean} R.,  {Dor{\'e}} O.,
  {Greason} M.~R.,  {Halpern} M.,  {Hill} R.~S.,  {Jarosik} N.,  {Kogut} A.,
  {Komatsu} E.,  {Limon} M.,  {Odegard} N.,  {Meyer} S.~S.,  {Page} L.,
  {Peiris} H.~V.,  {Spergel} D.~N.,    {Tucker} et al.,  2007, \apjs, 170, 288

\bibitem[\protect\citeauthoryear{Hivon, Gorski, Netterfield, Crill, Prunet \&
  Hansen}{Hivon et~al.}{2002}]{Hivon:2001jp}
Hivon E.,  Gorski K.~M.,  Netterfield C.~B.,  Crill B.~P.,  Prunet S.,
  Hansen F.,  2002, Astrophys. J., 567, 2

\bibitem[\protect\citeauthoryear{{Hoyle}, {Gruen}, {Bernstein}, {Rau}, {De
  Vicente}, {Hartley}, {Gazta{\~n}aga}, {DeRose}, {Troxel}, {Davis}, {Alarcon},
  {MacCrann}, {Prat}, {S{\'a}nchez}, {Sheldon} \& et al.}{{Hoyle}
  et~al.}{2017}]{photoz}
{Hoyle} B.,  {Gruen} D.,  {Bernstein} G.~M.,  {Rau} M.~M.,  {De Vicente} J.,
  {Hartley} W.~G.,  {Gazta{\~n}aga} E.,  {DeRose} J.,  {Troxel} M.~A.,  {Davis} C.,
   {Alarcon} A.,  {MacCrann} N.,  {Prat} J.,  {S{\'a}nchez} C.,  {Sheldon} E.,
    et al. 2017, ArXiv: 1708.01532

\bibitem[\protect\citeauthoryear{{Jing}, {Mo} \& {B{\"o}rner}}{{Jing}
  et~al.}{1998}]{HOD_0}
{Jing} Y.~P.,  {Mo} H.~J.,    {B{\"o}rner} G.,  1998, \apj, 494, 1

\bibitem[\protect\citeauthoryear{{Kaiser}}{{Kaiser}}{1984}]{peak_split}
{Kaiser} N.,  1984, \apjl, 284, L9

\bibitem[\protect\citeauthoryear{{Kaiser}}{{Kaiser}}{1987}]{Kaiser}
{Kaiser} N.,  1987, \mnras, 227, 1

\bibitem[\protect\citeauthoryear{{Kitaura}, {Rodr{\'{\i}}guez-Torres},
  {Chuang}, {Zhao}, {Prada}, {Gil-Mar{\'{\i}}n}, {Guo}, {Yepes}, {Klypin},
  {Sc{\'o}ccola} \& {Tinker}}{{Kitaura} et~al.}{2016}]{mocks2}
{Kitaura} F.-S.,  {Rodr{\'{\i}}guez-Torres} S.,  {Chuang} C.-H.,  {Zhao} C.,
  {Prada} F.,  {Gil-Mar{\'{\i}}n} H.,  {Guo} H.,  {Yepes} G.,  {Klypin} A.,
  {Sc{\'o}ccola} C.~G.,    {Tinker} J. et al.,  2016, \mnras, 456, 4156

\bibitem[\protect\citeauthoryear{{Klypin}, {Yepes}, {Gottl{\"o}ber}, {Prada} \&
  {He{\ss}}}{{Klypin} et~al.}{2016}]{concentration}
{Klypin} A.,  {Yepes} G.,  {Gottl{\"o}ber} S.,  {Prada} F.,    {He{\ss}} S.,
  2016, \mnras, 457, 4340

\bibitem[\protect\citeauthoryear{{Landy} \& {Szalay}}{{Landy} \&
  {Szalay}}{1993}]{LS}
{Landy} S.~D.,  {Szalay} A.~S.,  1993, \apj, 412, 64

\bibitem[\protect\citeauthoryear{{Manera}, {Scoccimarro} \&
  {Percival}}{{Manera} et~al.}{2013}]{mocks1}
{Manera} M.,  {Scoccimarro} R.,    {Percival} W.~J. et al., {Samushia}, L., {McBride} C.~K., 
{Ross} A.~J., {Sheth} R.~K., {White} M., {Reid} B.~A., {S{\'a}nchez}, A.~G.,
{de Putter} R., 	{Xu} X., {Berlind} A.~A., {Brinkmann} J., {Maraston} C.,
et al.,
2013, \mnras, 428, 1036

\bibitem[\protect\citeauthoryear{{Monaco}}{{Monaco}}{2016}]{monaco}
{Monaco} P.,  2016, Galaxies, 4, 53

\bibitem[\protect\citeauthoryear{{Monaco}, {Sefusatti}, {Borgani}, {Crocce},
  {Fosalba}, {Sheth} \& {Theuns}}{{Monaco} et~al.}{2013}]{pinocchio_2}
{Monaco} P.,  {Sefusatti} E.,  {Borgani} S.,  {Crocce} M.,  {Fosalba} P.,
  {Sheth} R.~K.,    {Theuns} T.,  2013, \mnras, 433, 2389

\bibitem[\protect\citeauthoryear{{Moutarde}, {Alimi}, {Bouchet}, {Pellat} \&
  {Ramani}}{{Moutarde} et~al.}{1991}]{2lpt_0}
{Moutarde} F.,  {Alimi} J.-M.,  {Bouchet} F.~R.,  {Pellat} R.,    {Ramani} A.,
  1991, \apj, 382, 377

\bibitem[\protect\citeauthoryear{{Murray}, {Power} \& {Robotham}}{{Murray}
  et~al.}{2013}]{hmfcalc}
{Murray} S.~G.,  {Power} C.,    {Robotham} A.~S.~G.,  2013, Astronomy and
  Computing, 3, 23

\bibitem[\protect\citeauthoryear{{Navarro}, {Frenk} \& {White}}{{Navarro}
  et~al.}{1996}]{NFW}
{Navarro} J.~F.,  {Frenk} C.~S.,    {White} S.~D.~M.,  1996, \apj, 462, 563

\bibitem[\protect\citeauthoryear{{Nuza}, {S{\'a}nchez}, {Prada}, {Klypin},
  {Schlegel}, {Gottl{\"o}ber}, {Montero-Dorta}, {Manera}, {McBride}, {Ross},
  {Angulo}, {Blanton}, {Bolton}, {Favole} \& {Samushia}}{{Nuza}
  et~al.}{2013}]{HAM_4}
{Nuza} S.~E.,  {S{\'a}nchez} A.~G.,  {Prada} F.,  {Klypin} A.,  {Schlegel}
  D.~J.,  {Gottl{\"o}ber} S.,  {Montero-Dorta} A.~D.,  {Manera} M.,  {McBride}
  C.~K.,  {Ross} A.~J.,  {Angulo} R.,  {Blanton} M.,  {Bolton} A.,  {Favole}
  G.,    {Samushia} 2013, \mnras, 432, 743

\bibitem[\protect\citeauthoryear{{Peacock} \& {Smith}}{{Peacock} \&
  {Smith}}{2000}]{HOD_1}
{Peacock} J.~A.,  {Smith} R.~E.,  2000, \mnras, 318, 1144

\bibitem[\protect\citeauthoryear{{Peebles} \& {Yu}}{{Peebles} \&
  {Yu}}{1970}]{thBAO1}
{Peebles} P.~J.~E.,  {Yu} J.~T.,  1970, \apj, 162, 815

\bibitem[\protect\citeauthoryear{{Planck Collaboration}, {Ade}, {Aghanim},
  {Arnaud}, {Ashdown}, {Aumont}, {Baccigalupi}, {Banday}, {Barreiro},
  {Bartlett} \& et al.}{{Planck Collaboration} et~al.}{2016}]{planck}
{Planck Collaboration} {Ade} P.~A.~R.,  {Aghanim} N.,  {Arnaud} M.,  {Ashdown}
  M.,  {Aumont} J.,  {Baccigalupi} C.,  {Banday} A.~J.,  {Barreiro} R.~B.,
  {Bartlett} J.~G.,    et al. 2016, \aap, 594, A13

\bibitem[\protect\citeauthoryear{{Pope} \& {Szapudi}}{{Pope} \&
  {Szapudi}}{2008}]{cov2}
{Pope} A.~C.,  {Szapudi} I.,  2008, \mnras, 389, 766

\bibitem[\protect\citeauthoryear{{Press} \& {Schechter}}{{Press} \&
  {Schechter}}{1974}]{PS}
{Press} W.~H.,  {Schechter} P.,  1974, \apj, 187, 425

\bibitem[\protect\citeauthoryear{{Rodr{\'{\i}}guez-Torres}, {Chuang}, {Prada},
  {Guo}, {Klypin}, {Behroozi}, {Hahn}, {Comparat}, {Yepes}, {Montero-Dorta},
  {Brownstein}, {Maraston} \& {McBride}}{{Rodr{\'{\i}}guez-Torres}
  et~al.}{2016}]{HOD_sergio}
{Rodr{\'{\i}}guez-Torres} S.~A.,  {Chuang} C.-H.,  {Prada} F.,  {Guo} H.,
  {Klypin} A.,  {Behroozi} P.,  {Hahn} C.~H.,  {Comparat} J.,  {Yepes} G.,
  {Montero-Dorta} A.~D.,  {Brownstein} J.~R.,  {Maraston} C.,    {McBride}
  C.~K. et~al.,  2016, \mnras, 460,1173
\bibitem[\protect\citeauthoryear{{Ross}, {Banik}, {Avila}, {Percival},
  {Dodelson}, {Garcia-Bellido}, {Crocce}, {Elvin-Poole}, {Giannantonio} \&
  {Sevilla-Noarbe}}{{Ross} et~al.}{2017a}]{3DBAO}
{Ross} A.~J.,  {Banik} N.,  {Avila} S.,  {Percival} W.~J.,  {Dodelson} S.,
  {Garcia-Bellido} J.,  {Crocce} M.,  {Elvin-Poole} J.,  {Giannantonio} T.,
  {Sevilla-Noarbe} I.,  2017, \mnras, 472, 4456

\bibitem[\protect\citeauthoryear{{Ross}, {Beutler}, {Chuang},
  {Pellejero-Ibanez}, {Seo}, {Vargas-Maga{\~n}a}, {Cuesta}, {Percival},
  {Burden}, {S{\'a}nchez}, {Grieb}, {Reid}, {Brownstein} \& {Dawson}}{{Ross}
  et~al.}{2017b}]{BAO5}
{Ross} A.~J.,  {Beutler} F.,  {Chuang} C.-H.,  {Pellejero-Ibanez} M.,  {Seo}
  H.-J.,  {Vargas-Maga{\~n}a} M.,  {Cuesta} A.~J.,  {Percival} W.~J.,  {Burden}
  A.,  {S{\'a}nchez} A.~G.,  {Grieb} J.~N.,  {Reid} B.,  {Brownstein} J.~R.,
  {Dawson} et al.,  2017, \mnras, 464, 1168

\bibitem[\protect\citeauthoryear{{}}{{Rozo} et al.}{2016}]{redmagic}
{Rozo}, E. and {Rykoff}, E.~S. and {Abate}, A. and {Bonnett}, C. and 
	{Crocce}, M. and {Davis}, C. and {Hoyle}, B. and {Leistedt}, B. and 
	{Peiris}, H.~V. and {Wechsler}, R.~H. and {Abbott}, T. and {Abdalla}, F.~B. and 
	{Banerji}, M. and {Bauer}, A.~H. and {Benoit-L{\'e}vy}, A. and 
	{Bernstein}, G.~M. and {Bertin}, E. and {Brooks}, D. and {Buckley-Geer}, E. and 
	{Burke}, D.~L. and {Capozzi}, D. and {Rosell}, A.~C. and {Carollo}, D. and 
	{Kind}, M.~C. and {Carretero}, J. and {Castander}, F.~J. et al.  2016, \mnras, 461, 1431

\bibitem[\protect\citeauthoryear{}{{Schneider et al.}}{2011}]{schneider}
{Schneider} M.~D., Cole, S., Frenk, C. S., Szapudi, I.  2011, \apj, 737, 11


\bibitem[\protect\citeauthoryear{{Scoccimarro}}{{Scoccimarro}}{2000}]{cov4}
{Scoccimarro} R.,  2000, \apj, 544, 597

\bibitem[\protect\citeauthoryear{{Scoccimarro} \& {Sheth}}{{Scoccimarro} \&
  {Sheth}}{2002}]{pthalos}
{Scoccimarro} R.,  {Sheth} R.~K.,  2002, \mnras, 329, 629

\bibitem[\protect\citeauthoryear{{Seo} \& {Eisenstein}}{{Seo} \&
  {Eisenstein}}{2007}]{SnE07}
{Seo} H.-J.,  {Eisenstein} D.~J.,  2007, \apj, 665, 14

\bibitem[\protect\citeauthoryear{{Skibba} \& {Sheth}}{{Skibba} \&
  {Sheth}}{2009}]{Skibba2009}
{Skibba} R.~A.,  {Sheth} R.~K.,  2009, \mnras, 392, 1080

\bibitem[\protect\citeauthoryear{{Sunyaev} \& {Zeldovich}}{{Sunyaev} \&
  {Zeldovich}}{1970}]{thBAO2}
{Sunyaev} R.~A.,  {Zeldovich} Y.~B.,  1970, \apss, 7, 3

\bibitem[\protect\citeauthoryear{{Tassev}, {Zaldarriaga} \&
  {Eisenstein}}{{Tassev} et~al.}{2013}]{cola}
{Tassev} S.,  {Zaldarriaga} M.,    {Eisenstein} D.~J.,  2013, Journal of
  Cosmology and Astroparticle Physics, 6, 36

\bibitem[\protect\citeauthoryear{{Taylor} \& {Joachimi}}{{Taylor} \&
  {Joachimi}}{2014}]{cov3}
{Taylor} A.,  {Joachimi} B.,  2014, \mnras, 442, 2728

\bibitem[\protect\citeauthoryear{{Trujillo-Gomez}, {Klypin}, {Primack} \&
  {Romanowsky}}{{Trujillo-Gomez} et~al.}{2011}]{HAM_3}
{Trujillo-Gomez} S.,  {Klypin} A.,  {Primack} J.,    {Romanowsky} A.~J.,  2011,
  \apj, 742, 16

\bibitem[\protect\citeauthoryear{{Watson}, {Iliev}, {D'Aloisio}, {Knebe},
  {Shapiro} \& {Yepes}}{{Watson} et~al.}{2013}]{watson}
{Watson} W.~A.,  {Iliev} I.~T.,  {D'Aloisio} A.,  {Knebe} A.,  {Shapiro} P.~R.,
     {Yepes} G.,  2013, \mnras, 433, 1230

\bibitem[\protect\citeauthoryear{{White}, {Tinker} \& {McBride}}{{White}
  et~al.}{2014}]{qpm}
{White} M.,  {Tinker} J.~L.,    {McBride} C.~K.,  2014, \mnras, 437, 2594

\bibitem[\protect\citeauthoryear{{Zehavi}, {Zheng}, {Weinberg}, {Blanton},
  {Bahcall}, {Berlind}, {Brinkmann}, {Frieman}, {Gunn}, {Lupton}, {Nichol},
  {Percival}, {Schneider}, {Skibba}, {Strauss}, {Tegmark} \& {York}}{{Zehavi}
  et~al.}{2011}]{HOD_zehavi}
{Zehavi} I.,  {Zheng} Z.,  {Weinberg} D.~H.,  {Blanton} M.~R.,  {Bahcall}
  N.~A.,  {Berlind} A.~A.,  {Brinkmann} J.,  {Frieman} J.~A.,  {Gunn} J.~E.,
  {Lupton} R.~H.,  {Nichol} R.~C.,  {Percival} W.~J.,  {Schneider} D.~P.,
  {Skibba} R.~A.,  {Strauss} M.~A.,  {Tegmark} M.,    {York} D.~G.,  2011,
  \apj, 736, 59

\bibitem[\protect\citeauthoryear{{Zel'dovich}}{{Zel'dovich}}{1970}]{ZA}
{Zel'dovich} Y.~B.,  1970, \aap, 5, 84


\bibitem[\protect\citeauthoryear{{Zheng}, {Berlind}, {Weinberg}, {Benson},
  {Baugh}, {Cole}, {Dav{\'e}}, {Frenk}, {Katz} \& {Lacey}}{{Zheng}
  et~al.}{2005}]{HOD_3}
{Zheng} Z.,  {Berlind} A.~A.,  {Weinberg} D.~H.,  {Benson} A.~J.,  {Baugh}
  C.~M.,  {Cole} S.,  {Dav{\'e}} R.,  {Frenk} C.~S.,  {Katz} N.,    {Lacey}
  C.~G.,  2005, \apj, 633, 791




\end{thebibliography}

\label{lastpage}
\
\end{document}